\documentclass[a4paper,10pt,notitlepage]{article}
\usepackage[a4paper, left=3.5cm, right=3.5cm, top=2cm]{geometry}
\usepackage{amsfonts,amssymb,amsmath,amsthm,hyperref,colonequals}
\usepackage{caption}
\usepackage{cite}
\usepackage{color}
\usepackage{tabularx}
\usepackage{graphicx}
\usepackage{bbm}
\usepackage[export]{adjustbox}
\usepackage{shuffle}
\usepackage[nice]{nicefrac} 
\usepackage{url}
\usepackage{mathtools}
\usepackage{epsfig}
\usepackage{physics}
\usepackage[utf8]{inputenc} 
\theoremstyle{definition}
\newcommand{\bea}{\begin{eqnarray}}
\newcommand{\eea}{\end{eqnarray}}
\newcommand{\be}{\begin{equation}}
\newcommand{\ee}{\end{equation}}

\newcommand*\pFqskip{8mu}
\catcode`,\active
\newcommand*\pFq{\begingroup
        \catcode`\,\active
        \def ,{\mskip\pFqskip\relax}%
        \dopFq
}
\catcode`\,12
\def\dopFq#1#2#3#4#5{%
        {}_{#1}F_{#2}\biggl[\genfrac..{0pt}{}{#3}{#4};#5\biggr]%
        \endgroup
}

\begin{document}

\thispagestyle{empty}
\setcounter{page}{0}
\begin{flushright}\footnotesize
\vspace{0.5cm}
\end{flushright}
\setcounter{footnote}{0}

\begin{center}
{\LARGE{
\textbf{Entropy production in inflation from spectator loops} 
}}

\bigskip
\medskip

{\sc \large
Pavel Friedrich$^*$ and Tomislav Prokopec$^\ddagger$ }\\[5mm]

{\it Institute for Theoretical Physics, Spinoza Institute and the
Center for Extreme Matter and Emergent Phenomena (EMME$\Phi$),\\ 
Utrecht University, Buys Ballot Building,
Princetonplein 5, 3584 CC Utrecht, The Netherlands
}\\[5mm]
\let\thefootnote\relax\footnotetext{\\$^*$Electronic address: \texttt{p.friedrich@uu.nl}\\
$^\ddagger$Electronic address: \texttt{t.prokopec@uu.nl}
}

\end{center}
\begin{abstract}
Perturbations in cosmic microwave background (CMB) photons and large scale structure of the universe
are sourced primarily by the curvature perturbation which is widely believed to be produced during inflation.
In this paper we present a 2-field inflationary model in which the inflaton couples bi-quadratically to a spectator field. We show that the spectator induces a rapid growth of the momentum of the curvature perturbation and the associated Gaussian van Neumann entropy during inflation such that the initial conditions at the end of inflation are substantially different from the standard ones.
Consequently, one ought to reconsider the kinetic equations describing evolution of the photon, dark matter and baryonic fluids in radiation and matter eras and take account of the fact that 
the curvature perturbation and its canonical momentum are two {\it a priory} independent stochastic fields.
We also briefly analyze possible imprints on the CMB temperature fluctuations from the
more general inflationary scenario which contains light spectator fields coupled to the inflaton.
\end{abstract}
 

\newpage
\setcounter{page}{1}


\tableofcontents
\addtolength{\baselineskip}{5pt}

\section{Overview and motivation}

It is a remarkable fact that all of the modern cosmic microwave background (CMB) data, together with various large scale structure 
(LSS) probes, can be described by a class of simple cosmological models containing just six parameters \cite{Aghanim:2015xee, Aghanim:2018eyx}. 
Two of these parameters - the amplitude ($A_s$)
and spectral slope ($n_s-1$) of the curvature spectrum - are primordial in origin, while four - the Hubble parameter today ($H_0$) (or equivalently
the angular scale of the first acoustic peak ($\theta=r_*/D_A$)), 
the reionization optical depth ($\tau_r$), (relative) baryonic density ($\Omega_b$) and cold dark matter density ($\Omega_c$)
-- are late time observables.
Since the simplest `vanilla' cosmological model assumes a spatially flat universe ($\Omega_\kappa=0$),
the dark energy density $\Omega_{de}=\Omega_\Lambda$ is not an independent parameter, {\it i.e.} $\Omega_{de}=1-\Omega_b-\Omega_c$.
For more details we refer to~\cite{Akrami:2018odb,Aghanim:2018eyx}.

Cosmological models have been tested for various other features, that include various probes of isotropy and homogeneity, 
statistical Gaussianity (the amplitude of primordial bispectrum and trispectrum), the amplitude and slope of tensor perturbations,
but for all of these only upper bounds exists, albeit there is a statistically weak evidence supporting some of the 
probes that indicate deviation from statistical isotropy or Gaussianity~\cite{Akrami:2019bkn}.

Another interesting class of features is encoded in isocurvature modes (see {\it e.g.}
 Ref.~\cite{Byrnes:2006fr}). Even though there are many potential physical degrees of freedom which 
can play the role of isocurvature modes, there is no strong evidence in the data that would suggest that any of these contribute dominantly
to the CMB photon temperature fluctuations. Indeed, \cite{Akrami:2018odb} has looked for 
traces of cold dark matter density isocurvature (CDI), neutrino density isocurvature (NDI) and neutrino velocity isocurvature (NVI) modes
in the data, and places upper limits on the relative amount of CDI, NDI and NVI of 
 $2.5 \%$, $7.4 \%$, and $6.8 \%$, respectively, at the scale of $k = 0.002\, \text{Mpc}^{-1}$. 
Signatures that are analogous to isocurvature modes are produced by topological defects and therefore similar upper bounds 
can be placed on the contribution of various classes of topological defects (which include cosmic strings, monopoles and textures) 
to the observed spectrum~\cite{Ade:2013xla}.

In this paper we study an idea with similar effects, namely how spectator fields during inflation decohere the Gaussian density matrix of the curvature perturbation on super-Hubble scales by means of quantum loop interactions.\footnote{  While this work was nearing completion, we became aware that a similar problem had recently been addressed by \cite{Hollowood:2017bil}.} 
This decoherence is manifested as an increase of entropy during inflation and can produce similar signals as isocurvature modes and topological defects in the effective CMB temperature fluctuations. This is so because isocurvature modes tend to produce peaks which are out-of phase with the adiabatic mode, and therefore tend to wash out the coherent CMB oscillations.
Let us be a bit more precise about the last statement and recap the form of the effective photon temperature fluctuation $\Delta\hat{ T}$ in momentum space before recombination in a simple approximation which we review in Appendix  \ref{appphoton},
\be
\Delta\hat{ T} \big(\vec{k}, \eta \big) \approx   \frac{1}{2} \hat{\Psi} (\eta_{\text{cmb}},\vec{k}) \cos  \big[ k r_s (\eta)  \big]+ 2 \frac{{\Psi}^{\prime}(\eta_{\text{cmb}},\vec{k}) }{k c_s(\eta_{\text{cmb}}) } \sin \big[ k r_s (\eta) \big]  \, . \label{effTempPre}
\ee
Here, $c_s(\eta)$ denotes the speed of sound and the sound horizon $r_s(\eta)$ is its integral over conformal time. The stochastic variable $\hat{\Psi} (\eta_{\text{cmb}},\vec{k})$ is the gauge invariant perturbation of the trace of the spatial metric at conformal time $\eta =  \eta_{\text{cmb}}$ within the radiation era some time before recombination such that it is observable in the CMB. Its derivative in conformal time, $\hat{\Psi}^{\prime} (\eta_{\text{cmb}},\vec{k})$, is an {\it{a priori} } stochastically independent variable. 
We can conclude that coherent CMB oscillations are possible if the stochastic operators $\hat{\Psi} (\eta_{\text{cmb}},\vec{k})$ and $\hat{\Psi}^{\prime} (\eta_{\text{cmb}},\vec{k})$ are linearly related (which induces a phase-shift) or either of them is much smaller than the other. As we pointed out above, Planck data is mostly consistent with coherent CMB oscillation such that the standard case is to discard  the initial time-derivative of the gravitational potential and consider only the adiabatic mode whose associated operator is conserved on super-Hubble scales. Still, the constraints to wash out the CMB oscillation reside in the range of percent so its worth studying mechanisms that can contribute to it. This allows us to either target those effects by precision cosmology or to rule them out.
We remind ourselves in Appendix \ref{curvlin} that the linear dynamics of single-field inflation on super-Hubble scales effectively decreases the number of independent stochastic operators to the aforementioned adiabatic mode. 
Thus, one way of obtaining a non-vanishing and stochastic independent time-derivative of the initial gravitational potential in \eqref{effTempPre} is to work with non-trivial background trajectories in multi-field inflationary model, leading to the aforementioned isocurvature modes whose stochastic independence can be traced back to independent quantum fluctuations whose presence is guaranteed by vacuum expectation values of the additional fields.

We obtain a significant amount of decoherence at the end of inflation by going beyond the tree-level analysis and relying purely on {\it interactions} of the inflaton perturbation $\varphi$ with a {\it spectator field} $\chi$ that has a zero expectation value. 
 We chose such a simple model because the inflaton coupling to the spectator field is controlled 
by a separate coupling constant, which is independent on the loop counting parameter of quantum gravity, 
$\kappa^2 H^2 \sim H^2/M_{\rm P}^2 \sim 10^{-12}$ (here $H$ is the inflationary Hubble parameter and 
$M_{\rm P}\simeq 2.4\times 10^{18}~{\rm GeV}$ is the reduced Planck mass), which governs the strength of 
interactions in the inflation sector. Moreover, since the spectator field does not acquire an expectation value, it is invariant under coordinate transformations to first order in perturbations. 
Thus, if we express corrections to the inflaton propagator in terms of the gauge invariant curvature perturbation $\mathcal{R}$ and take corrections to the inflaton expectation value $\overline{\phi}$ into account, our results are to first order in perturbations gauge invariant and we may compare them to the tree-level analysis at the end of inflation.

The effect of quantum corrections to the power spectrum of the curvature perturbation has been studied in \cite{Weinberg:2005vy,Weinberg:2006ac} with the conclusion that loop corrections on super-Hubble scales can at most be enhanced as powers of logarithms of the scale factor. However, the power spectrum of the $\mathcal{R} \mathcal{R}$-correlator remains approximately frozen due to the coupling constant suppression and the limit on how long inflation lasts. In this paper, we reconsider these observation with a concrete calculation in the above mentioned model involving spectator fields. The model consists of two canonical scalar fields on locally de Sitter background that interact {\it via} a cubic interaction which is derived by expanding a bi-quadratic action around the vev of the inflaton. While the interactions with the spectator indeed produce logarithmic corrections to the comoving curvature perturbations, the corrections to the canonical momentum of the comoving curvature perturbations grow exponentially in time (inverse power in conformal time) and may induce considerable fluctuations. The question whether these field excitations are stochastically independent can be answered by calculating the Gaussian part of the von Neumann entropy $S_{\text{vN}}$ associated to $\hat{\mathcal{R}}$ and $\hat{\mathcal{\pi}}_{\mathcal{R}}$, which is conveniently represented in momentum space by,
\be
S_{\text{vN}}\big[ \mathcal{R},\mathcal{\pi}_{\mathcal{R}}\big] 
   = \frac12\sum_{\vec k}s_{\text{vN}}(\eta,k)
\,,\quad
s_{\text{vN}}=\frac{\Delta_{ \mathcal{R} } +1}{2} \log\frac{\Delta_{ \mathcal{R} } +1}{2} -\frac{\Delta_{ \mathcal{R} } -1}{2} \log \frac{\Delta_{ \mathcal{R} } -1}{2} \, , \label{entropy}
\ee
which depends on the Gaussian invariant $\Delta_{ \mathcal{R} }^2$ (see e.g. \cite{Koksma:2010zi}),
\be
\Delta_{ \mathcal{R} }^2(\eta, k)=4\left[  \Delta_{\mathcal{R}\mathcal{R}}(\eta,k) \Delta_{\pi_\mathcal{R} \pi_\mathcal{R}}(\eta,k) - \Delta_{ \mathcal{R}\pi_\mathcal{R}}^2(\eta,k)\right]\, ,
\ee
where $\Delta_{\mathcal{R}\mathcal{R}}$,  $\Delta_{\pi_\mathcal{R} \pi_\mathcal{R}}$ and
$\Delta_{ \mathcal{R}\pi_\mathcal{R}}$ are the equal-time momentum space two-point functions.
 The Gaussian invariant $\Delta_{ \mathcal{R} }^2$ is identical to {\it one} for linearly evolved fields prepared in
 a pure Gaussian initial state (an important example of which is 
  the Bunch-Davies vacuum) and thus yields {\it zero} Gaussian von Neumann entropy. 
  A large Gaussian invariant on the other hand would indicate a big uncertainty in the phase-space 
which is spanned by the operators $\hat{\mathcal{R}}$ and $\hat{\mathcal{\pi}}_{\mathcal{R}}$.

In order to see how quantum interactions with spectators during inflation influence the CMB, we relate the gauge-invariant gravitational potential $\hat{\Psi}$ shortly before the end of inflation to the gauge-invariant curvature perturbation $\hat{\mathcal{R}}$ and evolve it to the radiation era where we assume a simple scenario in which we switch of the interactions after inflation. 
In Appendix \ref{appphoton} we review that equation \eqref{effTempPre} then takes the following form,
\begin{multline}
\Delta\hat{ T} \big(\eta,\vec{k} \big) \approx   \frac{1}{2} \Big[  \frac{2}{3} \hat{\mathcal{R}}(\eta_e,\vec{k} ) - \frac{a^3(\eta_e)}{a^3(\eta_{\text{cmb}})} \frac{ H }{2 M_p^2k^2a(\eta_e)}\hat{\pi}_{\mathcal{R}} (\eta_e,\vec{k})  \Big] \cos  \big[ k r_s(\eta) \big]\\+\frac{6 H}{k c_s(\eta_{\text{cmb}})}  \frac{ a^4(\eta_e)}{a^4(\eta_{\text{cmb}})}  \Big[ \frac{ H}{2 M_p^2k^2}\hat{\pi}_{\mathcal{R}} (\eta_e,\vec{k}) + {a(\eta_e)}\hat{\mathcal{R}}(\eta_e,\vec{k} ) \Big] \sin \big[ k r_s (\eta) \big]  \, , \label{effTempCurv}
\end{multline}
where the parameter $H$ is the Hubble scale at beginning of inflation and the argument $\eta_e$ is some time shortly before the end of inflation such that the slow-roll parameter $\epsilon = \mathcal{H}^2 - \mathcal{H}^{\prime} 
$, with $\mathcal{H}a = a^{\prime}$, is still small,  $\epsilon(\eta_e) \ll 1$. 
We see that the main contribution $\propto \hat{\pi}_{\mathcal{R}}$  in \eqref{effTempCurv} could wash out the Sakharov oscillations if it was able to balance the heavy suppression by the pre factor $\propto a^{-4}$, which is for initially small amplitudes only possible if $\hat{\pi}_{\mathcal{R}}$ was growing during inflation. As we review in Appendix \ref{curvlin}, linear single-field inflation yields the following relation on super-Hubble scales in slow roll regime,
\be
\hat{\pi}_{\mathcal{R}}^{\text{(lin)}}(\eta_e,\vec{k}) = -   \frac{2 M_p^2a(\eta_e) \epsilon(\eta_e) }{ H} \Big[\hat{\mathcal{R}}(\eta_e,\vec{k}) + \mathcal{O}\big( k \eta_e \big) \Big]\,,
\ee
such that stochastic independent off-peak contribution in \eqref{effTempCurv} can safely be neglected.
However, in models in which the inflaton couples to other matter fields with unsuppressed couplings (a notable example being Higgs inflation),
there is no reason to {\it a priory} expect that the standard tree level results apply and thus spectator fields without vevs might still contribute to stochastic independent modes.

 While this work is inspired by the large literature on decoherence and classicalization of cosmological 
 perturbations~\cite{Brandenberger:1992sr,Brandenberger:1992jh,Prokopec:1992ia,Polarski:1995jg,Lesgourgues:1996jc,Brandenberger:1990bx,Kiefer:1998qe,Prokopec:2006fc},
it also differs from it in important aspects. 
In contrast with the effective approaches based on studying the approximate evolution of the reduced density 
matrix~\cite{Zurek:2003zz},
we use standard perturbative methods of the quantum field 
theory~\cite{Calzetta:1995ea,Koksma:2010zi,Koksma:2009wa,Koksma:2011dy,Prokopec:2012xv}.
Furthermore, we identify the late time (CMB) observables that can be used 
to quantify the amount of decoherence in the curvature perturbation 
(expressed through the Gaussian part of the von Neumann entropy)
that occurs during inflation and subsequent epochs, while most of the existing works 
base their analysis on standard criteria for classicalization 
often used in condensed matter systems, such as the diagonalization rate of the reduced density 
matrix in a suitably chosen pointer basis.
While early works~\cite{Brandenberger:1992sr,Brandenberger:1992jh,Prokopec:1992ia,Polarski:1995jg,Lesgourgues:1996jc,Brandenberger:1990bx,Kiefer:1998qe,Prokopec:2006fc} used the late time observer's inability to get a complete access to the state of cosmological perturbations 
as the principal source of decoherence and classicalization
(the so-called `decoherence without decoherence'), later works used more realistic settings, in which 
(dissipative) interactions among quantum fields during (or after) inflation is the principal cause for decoherence. 
The interactions considered range from self-interactions of the inflaton 
field~\cite{Lombardo:2005iz,Martineau:2006ki,Nelson:2016kjm,Nelson:2017pmc}, interactions with 
gravitational waves~\cite{Calzetta:1995ys,Ye:2018kty},
interactions with other scalar fields~\cite{Rostami:2017akw,Liu:2016aaf,Martin:2018zbe,Martin:2018lin}, 
as well as interactions with massive fermionic fields~\cite{Boyanovsky:2018soy}. 

Encouraged by the result of \cite{Prokopec:2006fc} we decided to investigate the effect of one-loop interactions between the spectator and the inflaton where the fields interact bi-quadratically. When this work was nearing completion, we became aware that a similar problem was addressed in \cite{Hollowood:2017bil} based on the density matrix formalism developed in \cite{Boyanovsky:2015tba,Boyanovsky:2015jen}. While the authors of references \cite{Hollowood:2017bil,Boyanovsky:2015jen,Boyanovsky:2015tba} start from a cubic interaction and make use of the density matrix formalism, we start from a bi-quadratic interaction which provides a stable theory for a positive coupling.  By expanding around the inflaton condensate, we also obtain an effective cubic vertex which turns out to yield the dominate contributions to decoherence. However, we approach the problem differently by providing a one loop evaluation of the inflaton propagator  $\Delta_{\varphi  \varphi}(\eta, \eta^{\prime},k) $ from which we can fully reconstruct the Gaussian part of the density matrix. 

 The paper is organized as follows. In section~\ref{entropyResults} we explain the model set up and how to relate the various two-point functions. In the follow-up section~\ref{2PF}, we present the main steps 
in the calculation, including renormalization, the solution of the equation of motion for the statistical propogator, symmetry properties and the super-Hubble limit.
In section \ref{discus} we come back to the implications of our results and discuss extensions of the presented analysis. Moreover, we make a comparison with the findings of references \cite{Hollowood:2017bil,Boyanovsky:2015jen,Boyanovsky:2015tba}. Some important technical details of the calculations are presented in 
several appendices. 
\par
We work in natural units in which $c=\hbar=1$ and with the metric tensor with a mostly plus signature, $(-,+,+,+)$.



\section{Growing curvature momentum from quantum interactions \label{entropyResults}}
\label{Growing curvature momentum from quantum interactions}

 Coupling of the comoving curvature perturbation to other fields can be mediated not only via tree level processes,
but can be also studied at the quantum (loop) level. 
Take  a simple two scalar field inflationary model
that interact via a bi-quadratic interaction term,
\begin{equation}
 S[\phi,\chi] =S_{\text{EH}}+ \int d^D x \sqrt{-g}\left(-\frac12 g^{\mu\nu}(\partial_\mu\phi)(\partial_\nu\phi)
                       - \frac12 g^{\mu\nu}(\partial_\mu\chi)(\partial_\nu\chi)-V(\phi,\chi)\right)
\,,
\label{action for 2 fields}
\end{equation}
where $S_{\text{EH}}$ is the Einstein-Hilbert action, 
\begin{equation}
  S_{\text{EH}} = \frac{1}{16\pi G}\int d^D x \sqrt{-g}R
  \,,
\label{Einstein Hilbert}
\end{equation}
$D$ is the number of space-time dimensions, $R=R[g_{\mu\nu}]$ is the Ricci curvature scalar,
the field $\phi$ is the inflaton with the perturbation,
\begin{equation}
 \hat\varphi=\hat \phi-\overline\phi
 \,,\quad  \overline\phi(t)=\langle\hat \phi(x)\rangle
\label{inflaton perturbation}
\end{equation} 
the field $\chi$ is a spectator with a vanishing expectation value, $\langle\hat \chi\rangle=0$ 
and  the potential $V(\phi,\chi)$ reads,
\begin{equation}
V(\phi, \chi) = \frac{m^2_{\phi}}{2}  \phi^2 + \frac{m_\chi^2}{2}\chi^2 + \frac{g}{4}\chi^2\phi^2
\,,
\label{potential}
\end{equation}
where both fields are assumed to be light,
\begin{equation}
H\gg m_\chi \, , m_\phi \,. 
\label{potential:2}
\end{equation}

We are interested in studying the dynamics of the metric and field perturbations 
around a cosmological background,
with the metric tensor (in the plasma restframe) given by, 
\begin{equation}
 \overline{g}_{\mu\nu}
  ={\rm diag}\left(-\overline N^{\,2}(t),\underbrace{a^2(t), \cdots , a^2(t)}_{D-1\rm\ times}\right)
\,, \quad
\overline{g}={\rm det}\left[\overline{g}_{\mu\nu}\right] = \overline N^{\,2}a^{2(D-1)}
\,,
\label{background metric}
\end{equation}
where $\overline N(t)$ is the lapse function and $a(t)$ is the scale factor. 
While it would be of interest to study both the dynamics of the quantum gravitational and quantum 
scalar perturbations, for simplicity in this work we limit ourselves to studying the dynamics of 
the scalar curvature perturbation induced by its bi-quadratic interaction term 
given in Eq.~(\ref{potential}). 
This process is controled by the coupling constant $g$ which is generally 
different from the gravitational coupling constant $\kappa = 1/\sqrt{16\pi G}$, where $G$ 
denotes the Newton constant, and therefore can be separately studied. 
To show that, in what follows we recall some of the basics of the quantum perturbative gravity in inflationary 
space-times.

The theory~(\ref{action for 2 fields}) has two dynamical scalar degrees of freedom, which in the 
comoving gauge, in which $\varphi=0$, are the scalar metric perturbation $\psi= -{\rm Tr}[\delta g_{ij}]/(6a^2)$
and the isocurvature field, $\chi$, and one transverse, traceless tensor perturbation, 
$h_{ij}=\delta g_{ij}/a^2$, with $\delta_{ij}h_{ij}=0=\partial_i h_{ij}$. In addition, there are constraint degrees 
of freedom: one scalar and one transverse vector degree of freedom, 
namely the lapse function $N(x)$ and the shift vector $N_i(x)$ 
(with $\partial_i N_i=0$). Since one can choose a gauge in which the lapse and shift decouple from the 
dynamical degrees of freedom, one can ignore them~\cite{Prokopec:2012ug,Prokopec:2013zya}. 

The dynamics of the linear scalar cosmological perturbations is governed 
by the well-known  Mukhanov-Sasaki action~\cite{Brandenberger:1993zc}. When written for  the curvature  perturbation $\mathcal{R}$, the action reads~\cite{Brandenberger:1993zc,Malik:2008im,Weenink:2010rr}:
\begin{equation}
 S^{(2)}_s[\mathcal{R}] =  \int d^Dx \overline{N}a^{D-1}2\epsilon M_{\rm P}^2\left[\frac12\dot{\mathcal{R}}^2
                - \frac1{2a^2}(\partial_i\mathcal{R})^2\right]
\,,\qquad \epsilon = - \frac{\dot H}{H^2}
\,,\qquad M_{\rm P}^2 \equiv \frac{1}{8\pi G}
\,,
\label{quadratic scalar action R}
\end{equation}
and the quadratic action for the tensor perturbations, $h_{ij}=\delta g_{ij}/a^2$,
which in the traceless and transverse gauge ($\delta_{ij}h_{ij}=0=\partial_i h_{ij}$) reduces to,  
\begin{equation}
 S^{(2)}_t = \frac{M_{\rm P}^2}{8} \int d^Dx \overline{N}a^{D-1}\left[\dot h_{ij}^2
                - \frac1{a^2}(\partial_l h_{ij})^2\right]
\,,
\label{quadratic tensor action}
\end{equation}
where a {\it dot} signifies a reparametrization invariant derivative with respect to time, 
$\dot{X}\equiv \overline{N}^{-1}\partial_tX$. Note that both actions~(\ref{quadratic scalar action R}) 
and~(\ref{quadratic tensor action}) are manifestly gauge invariant, as they are  
written for the gauge invariant curvature perturvation $\mathcal R$ and gauge invariant tensor perturbation 
$h_{ij}$. If one fixes a gauge completely, one can easily get the corresponding gauge fixed action 
from~(\ref{quadratic scalar action R}). For example, 
in the comoving gauge ($\varphi=0$), in which $\mathcal R\rightarrow \psi$, the action for $\psi$ identical in form 
as the action~(\ref{quadratic scalar action R}) for $\mathcal R$; in the zero-curvature gauge ($\psi=0$),
the action for $\varphi$ is obtained by exacting the replacement, 
$\mathcal R\rightarrow \varphi/(\sqrt{2\epsilon}M_{\rm P})$ in~(\ref{quadratic scalar action R}),  
\begin{equation}
 S^{(2)}_s[\varphi] =  \int\! d^Dx \overline{N}a^{D-1}\!\left[\frac12\dot{\varphi}^2
                \!-\! \frac1{2}\left(\frac{\partial_i\varphi}{a}\right)^2\!
                 \!+\!\frac14\left(\frac{(a^{D-1}\dot\epsilon)^\cdot}{a^{D-1}\epsilon}
                       \!-\!\frac12\frac{\dot\epsilon^2}{\epsilon^2}\right)\!\varphi^2\right]
\,,
\label{quadratic scalar action}
\end{equation}
such that the linear dynamics of the inflaton perturbation corresponds to that of a harmonic oscillator with a time 
dependent frequency. Since $\chi$ remains invariant to first order under gauge transformations, 
the quadratic action for $\chi$ is by itself gauge invariant, 
\begin{equation}
 S^{(2)}_s[\chi] =  \int\! d^Dx \overline{N}a^{D-1}\!\left[\frac12\dot{\chi}^2
                \!-\! \frac1{2}\left(\frac{\partial_i\chi}{a}\right)^2
                 \!-\!\frac12\left(m_\chi^2 \!+\! \frac{g}{2}\overline\phi^2\right)\chi^2\right]
\,.
\label{quadratic spectator action}
\end{equation}
In addition, there are two physical constraint fields - the lapse and (transverse) shift function, 
but they decouple from the dynamical degrees of freedom $\mathcal{R}$ and $h_{ij}$.
While this decoupling is clearly evident (from the Helmholz decomposition) 
at the linear order in the perturbations, one has to work harder 
to show that it also works at higher order in perturbations~\cite{Prokopec:2012ug,Prokopec:2013zya}.
In fact, there  are gauges in which the constraint fields can play an important role~\cite{Prokopec:2010be}.
The leading order actions~(\ref{quadratic tensor action}--\ref{quadratic spectator action})
 are supplemented by the higher order actions 
describing cubic, quartic and higher order interactions~\cite{Maldacena:2002vr,Prokopec:2012ug,Prokopec:2013zya}.
Generically, while all gravitational interactions are suoppressed by powers of the gravitational coupling constant $\kappa = 1/\sqrt{16\pi G}$, 
the interactions involving the scalar curvature perturbation are in addition suppressed by powers of the slow roll parameters,
$\epsilon=-\dot H/H^2$ and/or its derivatives (no such suppression occurs in the tensor interactions). 
However, that does not mean that scalar loops are suppressed when compared with the tensor loops, 
since the scalar curvature propagator is
enhanced by a factor $\sim 1/\epsilon$ when compared with the tensor propagator, thus nullfying the 
slow-roll vertex suppression. The result is that, quite generically, each gravitational loop contributes as, 
$\sim \kappa^2 H^2\sim H^2/M_{\rm P}^2 $. In addition, 
Weinberg's theorem~\cite{Weinberg:2005vy, Weinberg:2006ac}
allows for a secular enhancement in the form of powers the number of e-foldings, $\mathcal{N}=\ln(a)$.
Since not much is known about such secular enhancements of the gravitational loops
(most notably because the problem of gauge dependence of gravitational loops is {\it not} well 
understood~\cite{Miao:2018bol,Miao:2017feh}), 
for the sake of simplicity we neglect them in what follows. 

From Eq.~(\ref{quadratic spectator action}) we see that the inflaton condensate $\overline\phi\sim HM_{\rm P}/m_\phi$ generates a mass for the spectator 
field $\chi$   of the order, 
\begin{equation}
\delta m_\chi^2 =\frac{g}{2} \overline\phi^{\,2}\sim g  H^2 \frac{M_{\rm P}^2}{m_\phi^2}
\,.
\end{equation}
Since light scalar field fluctuations grow during inflation, their effect on the inflaton 
fluctuation will be larger than from a heavy scalar field. Demanding that $\chi$ remains light during inflation,
$\delta m_\chi^2\ll H^2$, 
leads to the following condition on the coupling constant,
\be 
0< g \lesssim \frac{m^2_{\phi}}{M_{\rm P}^2} \sim 10^{-12} 
\, .
\label{gCondition}
\ee
Let us first consider the tadpole contribution to the expectation value of the inflaton field $\overline{\phi}$, 
which contributes to the inflaton equation of motion as, 
\begin{equation}
(\Box -m_\phi^2)\overline{\phi} = \frac{g}{2}\overline{\phi} i\Delta_\chi(x;x)\, .
\label{tadpole}
\end{equation}
This ought to renormalized by the non-minimal coupling counterterm, $\int d^Dx\Big(-\frac12 \delta \xi R\bar\phi^2\Big)$.
According to \eqref{potential:2}, we assume that the coincident scalar propagator is that of the massless scalar in de Sitter space. The finite part of the coincident propagator is given by
\be
i\Delta_\chi(x;x)_{\rm fin}\simeq [H^2/(4\pi^2)]\ln(a)\, ,
\ee 
which exhibits a secular growth and modifies the inflaton mass by
$\delta m_\phi^2 = [gH^2/(8\pi^2)]\ln(a)\ll m_\phi^2$ by a negligibly small amount. Moreover, this contribution changes 
the expansion rate and slow roll parameters, but by a small amount.
These corrections are important for maintaining gauge invariance of the corrected comoving curvature perturbation at linear order. The reason is that the inflaton vev enters the definition of the curvature perturbation and its corrections are of the similar order as the non-local self-mass corrections.
However, local terms will not induce dissipative effects that could affect the entropy of cosmological perturbations~\cite{Weenink:2011dd} and they are negligible for the canonical momentum of the comoving curvature perturbation and correlators thereof, as we will see explicitly later on.

The interaction between the inflaton and spectator fields is governed by 
the and generates cubic and quartic interactions, whose actions are, 
\begin{eqnarray}
 S^{(3)}_s[\varphi,\chi] &=&  \int\! d^Dx \overline{N}a^{D-1}\!\left(-\frac{h}{2}\varphi\chi^2\right)
 \,,\quad h = g\overline\phi
 \,,
\label{cubic interaction}\\
S^{(4)}_s[\varphi,\chi] &=&  \int\! d^Dx \overline{N}a^{D-1}\!\left(-\frac{g}{4}\varphi^2\chi^2\right)
\,.
\label{quartic interaction}
\end{eqnarray}
Let us first make a rough comparison of the effects induced by these two interactions 
on the dynamics of the inflaton perturbation. 

The one-loop $\mathcal{O}(g)$ contribution 
generated by the quartic interaction~(\ref{quartic interaction}) will (upon renormalization) generate a time 
dependent mass term for the inflaton fluctuations, $\delta m_\phi^2 = (g/2)i\Delta_\varphi(x;x)$,
where $i\Delta_\varphi(x;x)=\langle\hat\varphi(x)^2\rangle$ denotes the coincident two-point function
for the inflaton perturbation) and thus 
will not generate any entropy or any other dissipative effects in the scalar sector of the 
theory. 

\begin{figure}[!htb]
\minipage{0.48\textwidth}
  \includegraphics[width=\linewidth]{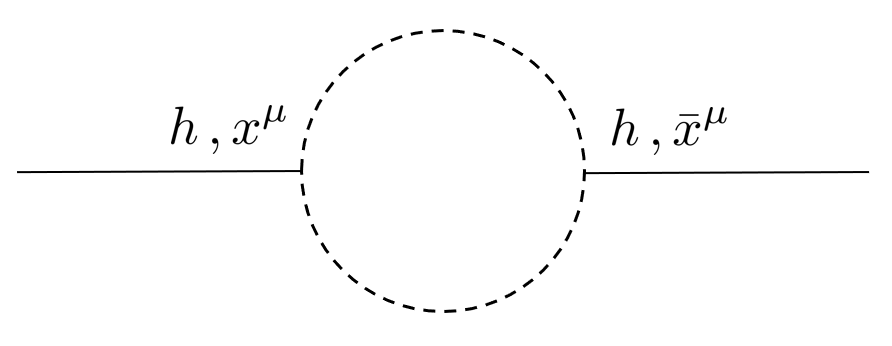}
  \caption{The one-loop Feynman diagram for
    the inflaton two-point function (solid lines)  generated by the cubic interaction in~(\ref{cubic interaction}). 
    The spectator field $\chi$ (dashed lines) runs in the loop. 
    The vertex coupling strength 
    is  $h=g\overline\phi$.}\label{fig:loop1}
\endminipage \hfill
\minipage{0.48\textwidth}
  \includegraphics[width=\linewidth]{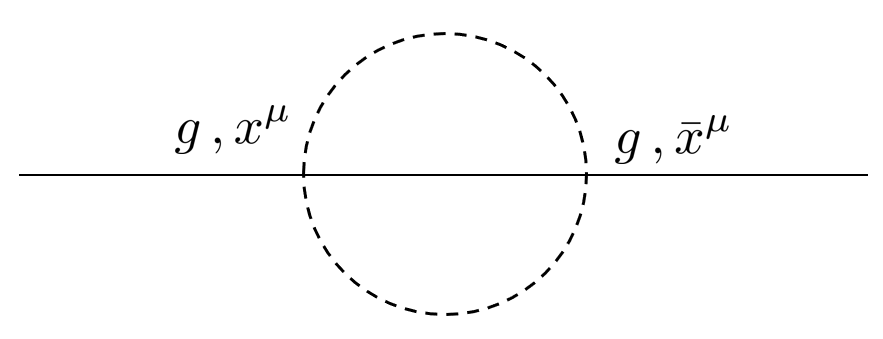}
  \caption{The two-loop diagram generated by the quartic interaction in~(\ref{quartic interaction})
    with the spectator $\chi$ (dashed lines) and inflaton (solid lines) running in the loops.
    The vertex coupling strength 
    is $g$.\\}\label{fig:loop2} 
\endminipage
\end{figure}

Next, at order $g^2$ there are two contributions: 
the one-loop contribution in figure~\ref{fig:loop1} which is generated by the cubic action~(\ref{cubic interaction})
and the two-loop contribution in figure~\ref{fig:loop2} generated by the quartic interaction~(\ref{quartic interaction}).
Since we are primarily interested in super-Hubble fluctuations, we shall compare the size of these two 
diagrams for super-Hubble distances, $\|\vec x-\vec x^{\,\prime}\|\gg 1/H$ and at equal time, $t=t'$.
It is not hard to see that the ratio of the two-loop to the one-loop contribution scales roughly as,
\begin{equation}
\frac{i\Delta_\varphi(t,\vec x;t,\vec x^{\,\prime})}{\overline\phi(t)^2}
\sim \frac{m_\phi^2 \ln(a)}{M_{\rm P}^2}\ll 1
\,,
\label{ratio of two to one loop}
\end{equation}  
where we made use of $i\Delta_\varphi(t,\vec x;t,\vec x^{\,\prime})\sim H^2\ln(a)$,
$\overline\phi\sim HM_{\rm P}/m_\phi$ and $m_\phi\ll H$ (in the above estimate, factors of order one such 
as powers of $\pi$ have been neglected). This means that the principal 
diagram that contributes (in a dissipative manner) to the dynamics of the inflaton perturbation,
and therefore also to the curvature perturbation, is the one-loop diagram in figure~\ref{fig:loop1}.
%
%

In what follows we shall compare the size of the one loop spectator diagram with that of the 
quantum gravitational loops. From Eq.~(\ref{resDeltaF}) we see that the ratio of the one-loop to the tree level 
Hadamard function is of the order $\delta F_\varphi/F_{\varphi,dS}\sim (h^2/H^2)\ln^3(a)$ , 
which ought to be compared with the corresponding quantum gravitational contribution, 
$\kappa^2 H^2\ln^{n_g}(a) \sim (H^2/M_{\rm P}^2)\ln^{n_g}(a)$, where $n_g$ as an unspecified positive 
integer which parametrizes our ignorance of the quantum gravitational loops. 
Upon dividing the two contributions we get, 
\begin{equation}
  \frac{(g^2\overline\phi^2/H^2)\ln^3(a)}{\kappa^2H^2\ln^{n_g}(a)} 
   \lesssim \frac{m^2_{\phi}}{H^2}[\ln(a)]^{3-n_g} 
\,,
\label{comparison g and kappa}
\end{equation}
Knowing the secular terms can be crucial, since each power of $\ln(a)$ produces an enhancement 
by a factor $\sim 10^2$, and that can be detrimental for determining whether the quantum gravitational or 
spectator contributions in~(\ref{comparison g and kappa}) dominate. 
From the estimate in~(\ref{comparison g and kappa})
 we see that the condition that $\chi$ remains light in inflation implies that the contribution 
from the spectator loop can be comparable to the quantum gravitational loops. 
This means that, before  one makes any definite conclusion
concerning the strength of decoherence during inflation, one also ought to investigate the effect of 
the quantum gravitational loops. In fact, there have been several attempts to do precisely 
that~\cite{Nelson:2016kjm,Nelson:2017pmc,Rostami:2017akw,Gong:2019yyz,Liu:2016aaf}.
In addition, a lot of work has been invested into a much easier set of problems, namely into studying
how the inflaton coupling with the other quantum fields (scalar, fermionic or vector) induces decoherence in 
the inflaton sector~\cite{Brandenberger:1990bx,Prokopec:2006fc,Liu:2016aaf,Boyanovsky:2018soy}.
While the earlier works considered simple models with bilinear couplings~\cite{Brandenberger:1990bx,Prokopec:2006fc} (since these couplings are non-dissipative, they are not true interactions), more recent works 
studied true interactions~\cite{Liu:2016aaf,Boyanovsky:2018soy}. These type of studies are much easier, since 
the hardest problem -- the problem of gauge dependence -- is absent in these studies. 

While these attempts represent important first steps, 
it is fair to say that no definite answer to that question has been given as yet.
The principal reason is that none of the existing works has seriously addressed the issue of gauge (in-)dependence,
neither have the authors performed a complete quantum calculation which must include: 
(a) a complete set of Feynman rules, with all relevant vertices and propagators included 
(currently there exists no propagator for 
that encompasses the dynamics of both scalar and tensor perturbations in inflation); 
(b) a complete calculation of the one-loop diagrams
that includes (preferably dimensional) regularization and renormalization, with 
the notable exception of Refs.~\cite{Boyanovsky:2015tba, Hollowood:2017bil}, where 
normal ordering was used to renormalize the self-mass; (c) a study of how
the inflaton two-point function gets modified by the one-loop quantum fluctuations, which also includes
a detailed analysis of how it depends on the choice of gauge.
Before we have good understanding of all of these steps and problems, we cannot say anything definite 
regarding the importance of the quantum gravitational loops for the evolution of 
cosmological perturbations. 

As a final remark, we point out that, because the spectator loop is controlled by a different coupling constant ($g$) 
from that governing the quantum gravitational loops ($\kappa$), one can unambiguously separate the two.
In other words, the quantum gravitational loops cannot cancel or compensate the effects of the 
spectator loop studied in this work.

In principle we could include slow-roll corrections in our study. 
However, including them would significantly complicate the spectator propagator, and thus also the whole 
calculation. Therefore, for simplicity, we shall consider a nearly de Sitter inflation,
in which the effects due to slow roll corrections are negligibly small. We point out that 
the spectator field is very different from the inflaton in that taking the limit $\epsilon\rightarrow 0$ 
in the scalar sector of the graviton is a delicate one, because the curvature propagator is in that limit enhanced
as $\propto 1/\epsilon$, {\it cf.} the action for the curvature perturbation~(\ref{quadratic scalar action}). 
No such enhancement is present in the spectator sector of the theory, implying that 
there is no subtlety involved in taking the limit $\epsilon\rightarrow 0$. 
Moreover, the tensor-to-scalar ratio $r\simeq 16\epsilon\leq 0.065$ 
is known to be small, implying that $\epsilon<1/200$, such that taking the limit $\epsilon\rightarrow 0$ 
should give reasonably accurate answers.  
Next, the spectral slope of the curvature perturbation is also quite small, 
$n_s-1\simeq -0.035=-2\epsilon -\epsilon_2 \approx -\epsilon_2$, and it is controlled by 
the second slow roll parameter $\epsilon_2=\dot\epsilon/(\epsilon H)\simeq 0.035$. 
This near scale invariance of the scalar perturbation also tells us that approximating 
the tree level equation for the inflaton perturbation by that of a massless scalar, $\Box \varphi= 0 $, 
constitutes a reasonably  accurate approximation,
where $\Box =g^{\mu\nu}\nabla_\mu\nabla_\nu$ is the d'Almbertian operator.

With these remarks in mind, we can now proceed to the calculation of the 
Hadamard function induced by the one-loop diagram shown in figure~\ref{fig:loop1}. 
The calculation will be done entirely on spatially flat sections of de Sitter space (Poincar\'e patch), 
in which the scale factor in conformal time $d\eta=dt/a$ reads, 
\begin{equation}
 a(\eta) = -\frac{1}{H\eta}\,,\qquad (\eta<0)
\,.
\label{scale factor on de Sitter}
\end{equation}
The relevant action is simply, 
\begin{equation}
S\big[\varphi,\chi\big] \approx  \int d^4 x \sqrt{-\overline{g}_{{dS}}}\left(-\frac12 \overline{g}^{\mu\nu}_{{dS}}(\partial_\mu\varphi)(\partial_\nu\varphi)                       - \frac12 \overline{g}^{\mu\nu}_{{dS}}(\partial_\mu\chi)(\partial_\nu\chi)- \frac{h}{2} \varphi \chi^2 \right)
\,,
\label{action for 2 fields:dS}
\end{equation}
with a de Sitter background metric $\overline{g}_{\mu \nu}^{{dS}}$. The free theory is solved in momentum space, with $k=\|\vec k\,\|$, for each field by the Bunch-Davies vacuum whose  positive ($+$)
and negative ($-$) frequency mode functions are given by
\be
u^{\pm}_{{dS}} (\eta, k) =  \frac{H}{\sqrt{2k^3}} (1\pm i k \eta)e^{\mp i k \eta}\, . \label{modfunc}
\ee
In Appendix \ref{defAndConv}, we give the definition of the Wightman functions $\Delta_{\varphi}^{\mp \pm}$ as well as the spectral (causal) two-point function $ \Delta^c_{\varphi}$ and the Hadamard (statistical) two-point function $F_{\varphi }$ in momentum space. For the Bunch-Davies vacuum they read 
\begin{align}
i \Delta_{\varphi , {dS}}^{\mp \pm}(\eta, \eta^{\prime},k) =  \frac{H^2}{2 k^3}(1\pm i k \eta) (1 \mp i k \eta^{\prime})e^{\mp i k (\eta-\eta^{\prime} )} \, , \label{BD2PF}
\end{align}
\be
 \Delta^c_{\varphi , {dS}}(\eta, \eta^{\prime},k) =  \frac{H^2}{ k^3} \Big[k(\eta-\eta^{\prime}) \cos \big[ k (\eta - \eta^{\prime}) \big]-(1 + k^2 \eta \eta^{\prime}) \sin \big[ k (\eta - \eta^{\prime} ) \big] \Big] \, ,\label{BDDelta}
\ee
\be
 F_{\varphi, {dS}}(\eta, \eta^{\prime},k) = \frac{H^2}{2 k^3} \Big[ (1 + k^2 \eta \eta^{\prime})\cos \big[ k (\eta - \eta^{\prime}) \big]+k(\eta-\eta^{\prime}) \sin \big[ k (\eta - \eta^{\prime} ) \big] \Big] \, .\label{BDF}
\ee
In the following section \ref{2PF}, we compute the one-loop correction to the statistical propagator 
in the super-Hubble limit as,
\begin{multline}
\delta F_{ \varphi}(\eta ,\eta^{\prime},k)=  \big[ F_{ \varphi} -F_{\varphi ,{ {dS}}} \big] (\eta ,\eta^{\prime},k)\\ = \frac{h^2}{2^6 3^3  k^3 \pi^2} \Bigg\lbrace 
6 \Big[4 \log \left(\frac{H}{2 k}\right)-4 \gamma_E +5 \Big]
 \log (-2 k \eta  ) \log (-2 k\eta^{\prime} )\\
  -\Big[(106-48 \gamma_E ) \log \left(\frac{H}{2 k}\right)-18 \log \left(\frac{\mu }{k}\right)+36 \gamma_E  (\gamma_E -3)+\pi ^2+\frac{208}{3}\Big] 
 \log( 4 k^2 \eta \eta^{\prime}  )\\
+ \Big[
 12 \log \left(\frac{H}{2 k}\right)-5 \Big]\Big[\log^2(-2 k \eta )+\log^2(-2 k\eta^{\prime})  \Big]\\
+4 \Big[
 \log^3 (-2 k \eta  )+ \log^3(-2 k\eta^{\prime})\Big]+ 
\mathcal{O}\big( k\eta, k \eta^{\prime} \big)
 \Bigg\rbrace\, , \label{resDeltaF}
\end{multline}
where the parameter $\mu$ is the renormalization scale and the quantity 
$\gamma_E =-\psi(1)\approx 0.577216 $ is Euler's constant, where $\psi(z)=(d/dz)\ln(\Gamma(z))$ 
is the digamma function (not to be confused with the spatial scalar metric perturbation $\psi$).
We now have to express these results in terms of the comoving curvature perturbation which we achieve in a first approximation by using linear relations.
The comoving curvature perturbation $\mathcal{R}$ and its canonical momentum $\pi_{\mathcal{R}}$ read to linear order  in zero curvature gauge $\psi=0$,
\begin{eqnarray}
\mathcal{R} &\equiv& \psi + \frac{H}{\dot{\phi}} \varphi  \; \longrightarrow \;  \frac{H}{\dot{\phi}} \varphi = \frac{1}{\sqrt{2 \epsilon}}\frac{\varphi}{M_p}\, , \\
\pi_{\mathcal{R}} &\equiv& {2 a^2 M_p^2 \epsilon} \partial_{\eta} \mathcal{R}  \; \longrightarrow \;   {\sqrt{2  \epsilon} M_p a^2} \Big[ \partial_{\eta}\varphi - (\partial_{\eta} \epsilon)\frac{\varphi }{2  \epsilon} \Big]  \, .
\label{momentum of curvature perturbation}
\end{eqnarray}
This procedure gives results that are gauge invariant to first order in coordinate gauge transformations if the one-loop corrections discussed above are consistently taken into account.
However, our primary goal is to calculate the entropy increase from the dissipative part of the spectator loop in figure~\ref{fig:loop1},
which is controlled by the coupling constant $g$ and which is different -- and thus independent --
 from the gravitational coupling  $\kappa=\sqrt{16\pi G}$. This observation provides evidence 
 that our final result for the entropy is gauge independent. 

Using the linear relations~(\ref{momentum of curvature perturbation}) 
we can express the statistical two-point functions of the comoving curvature perturbation and its canonical momentum in terms of the inflaton correlator to linear order as
\begin{eqnarray}
\Delta_{\mathcal{R} \mathcal{R}}(\eta,k)&
\equiv& F_{\mathcal{R} }(\eta, \eta,k)  =\frac{1}{2 \epsilon M_p^2} F_{ \varphi}(\eta, \eta,k)\, ,\\
\Delta_{\mathcal{R} \pi_{\mathcal{R}} }(\eta,k)&
\equiv& {2 a^2 M_p^2 \epsilon} \partial_{\eta^{\prime}} F_{\mathcal{R} }(\eta, \eta^{\prime},k)\Big|_{\eta = \eta^{\prime}}=  a^2\Big[ \frac{1}{2}  \partial_{\eta}  -  \frac{(\partial_{\eta} \epsilon) }{2 \epsilon} \Big] F_{ \varphi}(\eta,\eta,k)\, ,\\
\Delta_{\pi_{\mathcal{R}} \pi_{\mathcal{R}} }(\eta,k)&
\equiv& \nonumber({2 a^2 M_p^2 \epsilon})^2 \partial_{\eta}\partial_{\eta^{\prime}} F_{\mathcal{R} }(\eta, \eta^{\prime},k)\Big|_{\eta = \eta^{\prime}} \\ &=& 2 \epsilon M_p^2 a^4  \Big[ \partial_{\eta} \partial_{\eta^{\prime}} F_{ \varphi}(\eta, \eta^{\prime},k)\Big|_{\eta = \eta^{\prime}}  - \frac{(\partial_{\eta} \epsilon)}{2\epsilon} \partial_{\eta}  F_{ \varphi}(\eta, \eta,k)  + \frac{(\partial_{\eta} \epsilon)^2}{4\epsilon^2}  F_{ \varphi}(\eta, \eta,k)\Big] \, .
\end{eqnarray}
Thus, shortly before the end of inflation at $\eta = \eta_e$ such that the slow-roll parameter $\epsilon(\eta_e) $ is still small and to leading order  a constant, we have the following leading order corrections to the comoving curvature correlators on super-Hubble scales $|k\eta_e| \ll 1$,
\begin{eqnarray}
\Delta_{\mathcal{R} \mathcal{R}}(\eta_e,k) &\approx& \frac{H^2}{4   M_p^2 k^3  \epsilon(\eta_e) }\Big[1 +\frac{h^2}{108 \pi^2 H^2} \Big[ \log^3(-2 k\eta_e) + \mathcal{O}\big(\log^2(-2 k\eta_e) \big)\Big]+ \mathcal{O}\big(\epsilon(\eta_e), k\eta_e \big)\Big]
, 
\quad\;
\label{RR correlator}
\\
\Delta_{\mathcal{R} \pi_{\mathcal{R}} }(\eta_e,k) &\approx& -\frac{H a(\eta_e) }{2 k  }\Big[1 +\frac{a^2(\eta_e) h^2}{72 \pi^2 k^2} \Big[ \log^2(-2 k\eta_e) + \mathcal{O}\big(\log(-2 k\eta_e) \big)\Big]+ \mathcal{O}\big(\epsilon(\eta_e), k\eta_e \big)\Big]
\, ,
\label{Rpi correlator}\\
\Delta_{\pi_{\mathcal{R}} \pi_{\mathcal{R}} }(\eta_e,k) &\approx& k M_p^2  a^2(\eta_e)   \epsilon(\eta_e) \Big[1 + \frac{h^2 a^4(\eta_e) H^4}{36  \pi^2 H^2 k^4} \Big[ \log \left(\frac{H}{2 k}\right)+\frac{5}{4}- \gamma_E  \Big] + \mathcal{O}\big(\epsilon(\eta_e), k\eta_e \big)\Big]
\, .
\label{pipi correlator}
\end{eqnarray}

We note that our result satisfies Weinberg's theorem, since the $\Delta_{\mathcal{R} \mathcal{R}}$ correlator in \eqref{RR correlator} receives only logarithmic corrections
in time multiplying a constant
\be
\propto h^2H^{-2} = g^2 \bar{\phi}^2 H^{-2} \sim g^2 M_P^2 m^{-2} \lesssim 10^{-12} \, .
\ee
The one-loop corrections to $\epsilon$ are also small as argued below \eqref{tadpole}. 
Although corrections to  $\Delta_{\mathcal{R} \mathcal{R}}$ are negligible, the corrections to $\Delta_{{\cal R}\pi_{\cal R}}$ and 
$\Delta_{\pi_{\cal R}\pi_{\cal R}}$, which are induced by dissipative effects, can become very large since they multiply powers of the scale factor.

In order to study the physical implications at the end of inflation on supper-Hubble scales, we will rescale 
$\pi_{\mathcal{R}}$ by its linear relation to the gauge invariant gravitational potential    \eqref{general formula for psi R},
\be
\Psi =  -    \frac{   \mathcal{H}  }{ 2 M_p^2k^2a^2 }\pi_{\mathcal{R}}\, .
\ee
We quantify possibly large corrections of the $\pi_{\mathcal{R}} \pi_{\mathcal{R}} $-correlator to the tree-level result $\overline{\Delta}_{\pi_{\mathcal{R}} \pi_{\mathcal{R}}}$  by the ratio
\begin{multline}
\Delta_{\text{infl}} \equiv \frac{{H}}{2 M_p^2 k^2 a(\eta_e)}\left|\frac{ \Delta_{\pi_{\mathcal{R}} \pi_{\mathcal{R}} }- \overline{\Delta}_{\pi_{\mathcal{R}} \pi_{\mathcal{R}}}}{\Delta_{\mathcal{R} \mathcal{R}}}\right|^{1/2} 
\\\approx  \frac{ \epsilon(\eta_e)h}{6 \pi H}  \frac{a^2(\eta_e) H^2}{k^2}\left| \log \left(\frac{H}{ k}\right)\right|^{1/2} \lesssim 10^{-12}\frac{ \epsilon(\eta_e)}{6 \pi }  \frac{a^2(\eta_e) H^2}{k^2}\left| \log \left(\frac{H}{ k}\right)\right|^{1/2}\, , \label{pipiInf}
\end{multline}
where we kept only the dominant logarithmic contribution and substituted the estimate for the coupling constant $h = g \bar{\phi}  $ from \eqref{gCondition}. We note that  the quantity $\Delta_{\text{inf}}$ in \eqref{pipiInf} is of order one after, 
\be
 N_{\text{dec}}
  \approx \frac12\log\left[\frac{6 \pi H}{\epsilon(\eta_e)h\left| \log (H/k)\right|^{1/2}}\right] 
   \gtrsim 20 
 \label{Ndeco}
\ee
e-folds  the mode $k$ spends on super-Hubble scales.
This marks the time at which quantum corrections dominate the tree-level result for the $\pi_{\mathcal{R}}\pi_{\mathcal{R}}$-correlator and 
the decoherence sets in. In fact, the time scale~(\ref{Ndeco})  is a couple of e-folds 
longer than the decoherence time associated with the growth of entropy, which is controlled 
by the time at which the momentum-momentum correlator~(\ref{pipi correlator})
becomes loop dominated, 
$N_{\rm entropy}\simeq \frac12\log\left[\frac{6\pi H}{h|\log(H/k)|^{1/2}}\right]
  =N_{\text{dec}}-\frac12\log(1/\epsilon(\eta_e))$.
Furthermore, the decoherence time-scale $ N_{\text{dec}}$ in~(\ref{Ndeco}) differs essentially 
from the breakdown-time of standard perturbation theory which is governed by 
the perturbativity time associated with the $\mathcal{RR}$-correlator~(\ref{RR correlator}),~\footnote{The standard estimate for the perturbativity time is larger, $ N_{\text{pert}}\sim 10^{13}$ e-folds, and it is based on the assumption that there are only two powers of the logarithms in 
the $\mathcal{RR}$-correlator~(\ref{RR correlator}). However, the detailed calculation performed 
in this work shows that there are in fact three powers of the logarithm, thus shortening significantly 
$N_{\text{pert}}$.}
\be
 N_{\text{pert}}
  \approx \left[\frac{108\pi^2 H^2}{h^2}\right] ^\frac13 \gtrsim 10^9
\,, 
 \label{Npert}
\ee
which is a much larger time scale because the correlators entering loop calculations grow only logarithmically 
with the scale factor.
We can also quantify possibly large corrections of the ${\mathcal{R}} \pi_{\mathcal{R}} $-correlator to the tree-level result  by the ratio
\begin{multline}
\theta_{\text{infl}} \equiv  \frac{{H}}{2 M_p^2 k^2 a(\eta_e)}
\left|
\frac{\Delta_{ {\mathcal{R}} \pi_{\mathcal{R}} }-\overline{\Delta}_{ {\mathcal{R}} \pi_{\mathcal{R}}} }
{\Delta_{\mathcal{R} \mathcal{R}}}\right| 
\\
\approx \epsilon(\eta_e)\frac{a^2(\eta_e) h^2}{ 72\pi^2 k^2}  \log^2(-2 k\eta_e)  \lesssim 10^{-24}\epsilon(\eta_e)\frac{a^2(\eta_e) H^2}{ 72 \pi^2 k^2} \Big| \log^2\Big(\frac{H a(\eta_e)}{k}\Big) \Big| \, .
\qquad\qquad\qquad
\label{RpiInf}
\end{multline}
From \eqref{pipiInf} and \eqref{RpiInf}, we see an enhancement of  the $\hat\pi_{\mathcal{R}}$ operator by the factor $a^2(\eta_e)H^2/k^2$ at the end of inflation.
The source of this amplification, however, lies in the vacuum quantum uncertainty of the spectator field $\chi$ which is coupled to the inflaton via the interaction term $\overline{\phi} \varphi \chi^2$. Since the quantum fluctuations of the spectator is independent of the inflaton quantum fluctuations they will lead to an independent, amplified late-time stochastic source. We can make the latter statement quantitative by invoking the Gaussian entropy of the corrected two-point functions.
Since we used linear relations as a first approximation, the Gaussian invariant associated with the comoving curvature perturbation is identical to  the Gaussian invariant associated with the inflaton perturbation,
\begin{eqnarray}
\frac{\Delta^2_{\mathcal{R}}(\eta, k)}{4 } \nonumber &=& \Delta_{\mathcal{R} \mathcal{R}}(\eta, k) \Delta_{ \pi_{\tiny{\mathcal{R}}} \pi_{\tiny{\mathcal{R}}}}(\eta, k) - \Delta_{\mathcal{R} \pi_{\tiny{\mathcal{R}}}}^2(\eta, k)\\ &=& a^4\Big[  F_{ \varphi}(\eta, \eta,k)  \partial_{\eta} \partial_{\eta^{\prime}} F_{ \varphi}(\eta, \eta^{\prime},k)\Big|_{\eta = \eta^{\prime}} -  \frac{1}{4} ( \partial_{\eta}  F_{ \varphi}(\eta, \eta,k) )^2\Big] =\frac{\Delta^2_{\varphi}(\eta, k)}{4 }\, .
\end{eqnarray}
The Gaussian invariant $\Delta^2_{\varphi}$ of the inflaton perturbation $\varphi$ and hence of the comoving curbature perturbation $\Delta^2_{\mathcal{R}}$  is given by 
\be
\frac{\Delta_{ \varphi }^2(\eta, k)}{4a^4} =\frac{\Delta_{ \mathcal{R} }^2(\eta, k)}{4a^4} =   F_{\varphi}(\eta, \eta^{\prime},k) \partial_{\eta}\partial_{\eta^{\prime}} F_{\varphi}(\eta, \eta^{\prime},k) - \Big[ \partial_{\eta^{\prime}} F_{\varphi}(\eta, \eta^{\prime},k) \Big]^2 \Bigg|_{\eta^{\prime} = \eta}\, ,
\label{gaussian invariant}
\ee 
which can be used to calculate the Gaussian part of the von Neumann entropy
\be
S_{\text{vN}}\big[ \mathcal{\mathcal{R}}  \big] = \frac{\Delta_{ \mathcal{\mathcal{R}} } +1}{2} \log\frac{\Delta_{ \mathcal{\mathcal{R}} } +1}{2} -\frac{\Delta_{ \mathcal{\mathcal{R}} } -1}{2} \log \frac{\Delta_{ \mathcal{\mathcal{R}} } -1}{2}
  = S_{\text{vN}}\big[ \mathcal{\varphi} \big]
 \, . \label{entropyPhi}
\ee
The last equality follows from the fact that $\mathcal{R}$ and $\varphi$ are related 
by a (time dependent) rescaling, and since the von Neuman entropy is expressed in terms of 
the Gaussian invariant of the  state $\Delta_\varphi^2$, it cannot depend on a linear field redefinition.
This is one way to understand why local mass corrections  changing the vev of the inflaton via \eqref{tadpole} do not contribute to the entropy.

The mode functions of the non-interacting theory in the Bunch-Davies vacuum yield a Gaussian invariant that is identical to one and hence result zero von Neumann entropy. The same reasoning holds for the spectator field $\chi$ which we also prepare in the Bunch-Davies vacuum.
Thus, the Bunch-Davies vacuum for the fields $\varphi$ and $\chi$ represents a state with minimal uncertainty which is solely due to the quantum nature of the theory.
However, once interactions are taken into account the Gaussian invariant and hence the entropy get perturbatively corrected\begin{multline}
\delta \Big[\frac{ \Delta_{\varphi}^2}{4 a^4} \Big]= \delta \Big[ F_{\varphi}(\eta, \eta) \partial_{\eta} \partial_{\eta^{\prime}} F_{\varphi}(\eta ,\eta^{\prime}) -  \big[ \partial_{\eta^{\prime}}  F_{\varphi}(\eta ,\eta^{\prime})\big]^2 \Big]\Bigg|_{\eta^{\prime} = \eta}
\\
= \Big[  F_{\varphi, {dS}}(\eta, \eta) \partial_{\eta} \partial_{\eta^{\prime}} \delta F_{\varphi}(\eta ,\eta^{\prime}) + \delta F_{\varphi}(\eta, \eta) \partial_{\eta} \partial_{\eta^{\prime}}  F_{\varphi, {dS}}(\eta ,\eta^{\prime}) \\-2  \big[\partial_{\eta^{\prime}}   F_{\varphi, {dS}}(\eta ,\eta^{\prime})\big] \partial_{\eta^{\prime}}  \delta F_{\varphi}(\eta ,\eta^{\prime})\Big] \Bigg|_{\eta^{\prime} = \eta} \\
=   \frac{H^2}{2 k} \Big[(1 + k^2 \eta^2) \partial_{k\eta} \partial_{k\eta^{\prime}} \delta F_{\varphi}(\eta ,\eta^{\prime}) + \delta F_{\varphi}(\eta, \eta) k^2 \eta^2 - 2   \eta\partial_{\eta^{\prime}}   \delta F_{\varphi}(\eta ,\eta^{\prime})\Big]\Bigg|_{\eta^{\prime} = \eta}\,. 
\end{multline}

The correction to the Gaussian invariant of the inflaton perturbation is to leading order in the super-Hubble limit given by
\be
\delta \Big[\frac{ \Delta_{\varphi}^2}{4 } \Big] = \frac{1}{9 \pi^2} \frac{h^2}{H^2} \left(\frac{ H a }{2 k }\right)^6 \Big[4 \log \left(\frac{H}{2 k}\right)+5-4 \gamma_E  + 
\mathcal{O}\big( k\eta \big)\Big] \, . 
\label{gaussPert}
\ee
This expression is greater than zero for $H > 2k$, which is amply satisfied for the scales we will be interested in. We conclude that cubic interactions in inflation  of the type $g\overline{\phi}\varphi \chi^2$ lead do a growth of the Gaussian invariant $\Delta_{\varphi}^2$ by a factor of $a^6$ on super-Hubble scales and correspondingly to a growth of the Gaussian entropy. This growth is to leading order due to the quantum loop corrected $\pi_{\mathcal{R}}\pi_{\mathcal{R}}$-correlator which grows much faster than the correction to the $\mathcal{R}\mathcal{R}$-correlator.
This leads to two conclusions. First, the $\pi_{\mathcal{R}}\pi_{\mathcal{R}}$-correlator in \eqref{pipi correlator}  which was calculated with dissipative corrections is linearly gauge invariant. This follows from the fact that entropy production results only from dissipative effects \cite{Weenink:2011dd} and the statement that the entropy (or the associated Gaussian invariant \eqref{gaussian invariant}) are to linear order gauge invariant. 
The second conclusion is that we can view the quantum loop corrected operators $\hat{\mathcal{R}}$ and $\hat{\mathcal{\pi}}_{\mathcal{R}}$ as stochastically independent at the end of inflation, in contrast to the tree-level result \eqref{solPiSingleField}.

\begin{figure}[!htb]
\minipage{0.32\textwidth}
  \includegraphics[width=\linewidth]{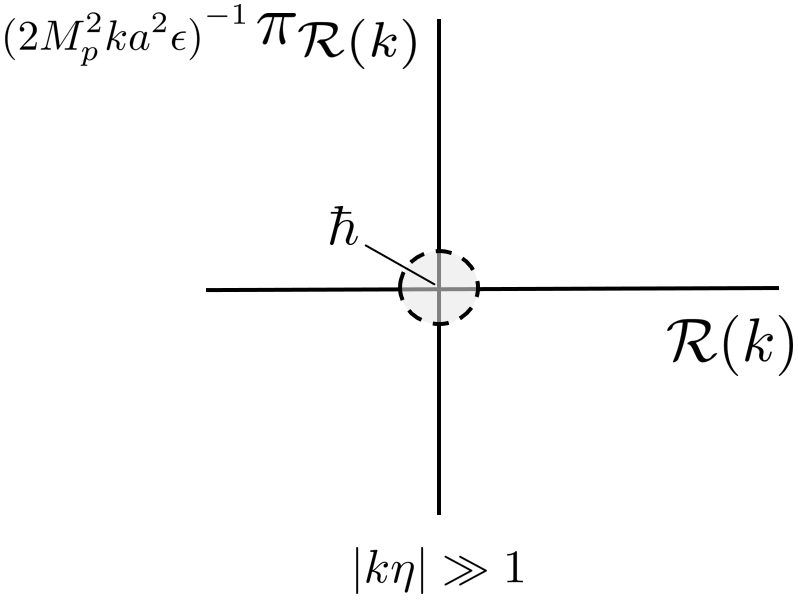}
  
  \caption{Phase diagram for mode k early in the sub-Hubble regime. The rescaling for the momentum $\pi_{\mathcal{R}}$ follows from initial conditions of the linear evolution \eqref{solInflationRPi} at early times.\\\\\\\\}\label{fig:snap1}
\endminipage\hfill
\minipage{0.32\textwidth}
  \includegraphics[width=\linewidth]{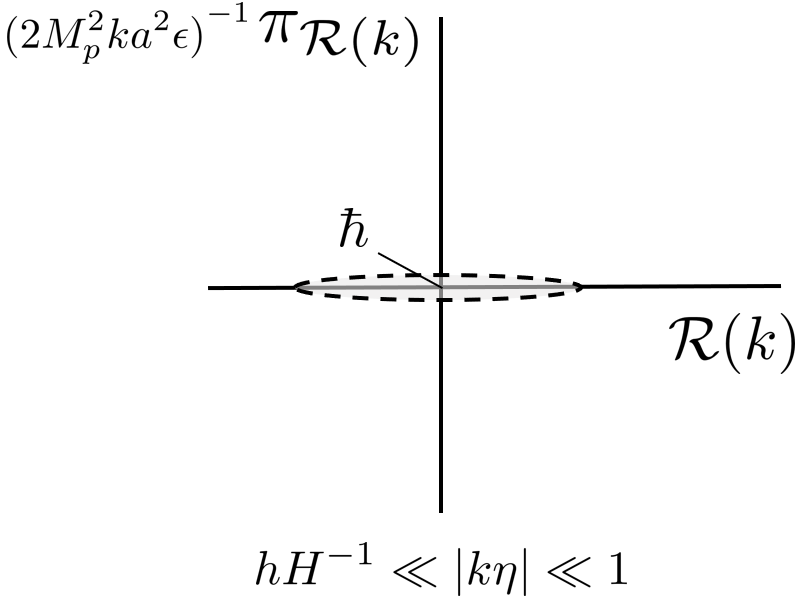}
  \caption{Phase diagram for mode k at intermediate times such that k is super-Hubble but quantum loop corrections are still neglible. The semi-minor is enlarged to be visible and is substantially smaller than the one in figure \ref{fig:snap3}. \\\\}\label{fig:snap2}
\endminipage\hfill
\minipage{0.32\textwidth}%
  \includegraphics[width=\linewidth]{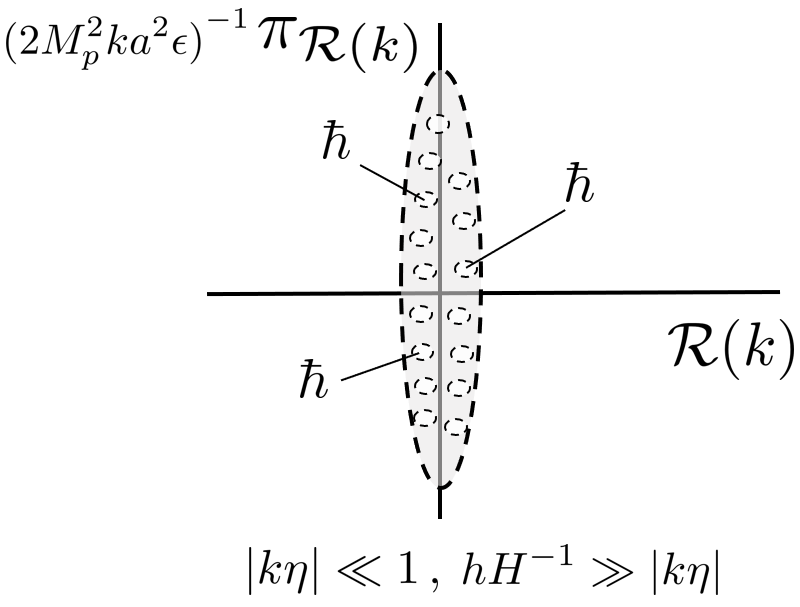}
  \caption{Phase diagram for mode k which is super-Hubble at late times where quantum loop corrections balance the suppresion from the small coupling constant.  Note 
  that the axes in this figure are compressed, which was necessary as the surface area of this state 
is very large when measured in units of $\hbar$.
}\label{fig:snap3}
\vskip -0.2cm
\endminipage
\end{figure}

Let us visualize this statement by three snapshots of a phase-space diagram associated to $\mathcal{R}(\vec{k})$ and $\pi_{\mathcal{R}}(\vec{k})$ for a given mode $\vec{k}$. The first snapshot in figure \ref{fig:snap1} is taken while the mode is deep in the sub-Hubble regime where it is governed by tree-level level dynamics due to the smallness of the coupling constant. 
The state is then approximately in its adiabatic, Gaussian vacuum, indicated by 
the circle on the phase space diagram, representing the set of points of equal probability amplitude. 
In an intermediate step in snapshot in figure \ref{fig:snap2}, the mode becomes super-Hubble but the enhancement due to the factor of $k^{-2}a^2H^2$ in \eqref{pipiInf} is still too small to compensate the small coupling $h H^{-1}$. This phase is thus still dominated by the linear analysis and results in the usual squeezed state \cite{Albrecht:1992kf}. In the final snapshot in figure  \ref{fig:snap3} at the end of inflation, more precisely, for all modes that have evolved for $\gtrsim 20$ e-folds on super-Hubble scales,
{\it cf.} the estimate~\eqref{Ndeco}. For these modes,
the enhancement of the $\pi_{\mathcal{R}}\pi_{\mathcal{R}}$-correlator due to the factor $k^{-2}a^2H^2$ in \eqref{pipiInf} is now big enough to overcome the suppression of the small coupling $h H^{-1}$. The state is still squeezed, but now mostly in the momentum direction.
\par
A tempting question to ask is how the enhanced $\pi_{\mathcal{R}}$-operator at the end of inflation affects the effective temperature perturbation. In order to answer this question we still have to map these correlators to a time deep in the radiation era $\eta_{\text{cmb}} \approx 10^{-1} \eta_{\text{rec}}$, some time before recombination at $\eta = \eta_{\text{rec}}$. As a first attempt, we pick the simplest possible scenario and assume that the comoving curvature perturbation $\mathcal{R}$ and the gauge invariant gravitational potential $\Psi$ will not be further affected on super-Hubble scales during the transition to radiation such that we can make use of standard linear relations. We review this process in Appendix \ref{appphoton}. The effective photon temperature perturbation relevant for the CMB at $\eta_{\text{cmb}}$ (which is a conformal time early enough from the decoupling time such
that the linear collisionless evolution still applies) may then be expressed according to \eqref{finalCMBTemp} in terms the comoving curvature perturbation just before the end of inflation at $\eta_{e}$ as follows,
\begin{multline}
\Delta\hat{ T} \big(\eta,\vec{k} \big) \approx   \frac{1}{2} \Big[  \frac{2}{3} \hat{\mathcal{R}}(\eta_e,\vec{k} ) - \frac{a^3(\eta_e)}{a^3(\eta_{\text{cmb}})} \frac{ H }{2 M_p^2k^2a(\eta_e)}\hat{\pi}_{\mathcal{R}} (\eta_e,\vec{k})  \Big] \cos  \big[ k r_s(\eta) \big]\\+ \frac{ 6 H}{k c_s(\eta_{\text{cmb}})}  \frac{ a^4(\eta_e)}{a^4(\eta_{\text{cmb}})}  \Big[ \frac{ H}{2 M_p^2k^2}\hat{\pi}_{\mathcal{R}} (\eta_e,\vec{k}) + {a(\eta_e)}\hat{\mathcal{R}}(\eta_e,\vec{k} ) \Big] \sin \big[ k r_s (\eta) \big]  
\, . 
\label{effTemp2}
\end{multline}
We already know that the tree-level contribution to the sine term in equation \eqref{effTemp2} is insignificant in this scenario.
Let us thus define here another quantity that allows us to measure the relative amplitude of orthogonal oscillations in \eqref{effTemp2} if we assume the quantum contributions to the $\pi_{\mathcal{R}}$ operator to be dominant,
\begin{eqnarray}
\frac{\Delta_{\text{sin}}}{\Delta_{\text{cos}}} \equiv \frac{a^4(\eta_e)}{a^4(\eta_{\text{cmb}})} \frac{18 H a(\eta_e)}{  k c_s(\eta_{\text{cmb}})}  \Delta_{\text{infl}} \sim
\frac{h}{H}\frac{a^4(\eta_e)}{a^4(\eta_{\text{cmb}})} 
\frac{ 3\epsilon(\eta_e)}{ \pi  }  \frac{a^3(\eta_e) H^3}{k^3c_s(\eta_{\text{cmb}})} \Big| \log \left(\frac{H}{ k}\right) \Big|\, .
\label{momentum contribution to CMB}
\end{eqnarray}
Putting in the estimate for our coupling constant $h$ from~(\ref{gCondition}) we get
\be
\frac{\Delta_{\text{sin}}}{\Delta_{\text{cos}}}
\lesssim 10^{-12}\frac{a(\eta_e)}{a(\eta_{\text{cmb}})} 
\frac{ 3\epsilon(\eta_e)}{ \pi  }  \frac{\mathcal{H}^3(\eta_{\text{cmb}})}{k^3c_s(\eta_{\text{cmb}})} \Big| \log \left(\frac{H}{ k}\right) \Big|\ll 1\, .\label{momentum contribution to CMB:2}
\ee
It is thus not sufficient to have quantum loop enhancements of the $\pi_{\mathcal{R}}\pi_{\mathcal{R}}$-correlator only during inflation since the linear evolution throughout radiation suppresses it such that at the times of CMB it again becomes small. It is a natural question to ask whether quantum corrections during radiation will hinder this decay in a way that is similar the to quantum corrected processes that take place during inflation and we leave this for future studies.

\section{Kadanoff-Baym equation for the statistical propagator \label{2PF}}
\subsection{Effective action}

In this section, we lay out in some detail how we calculate the quantum loop correction to the statistical propagator of the inflaton perturbation that we present in \eqref{resDeltaF}.
We will perform this calculation in the Schwinger-Keldysh formalism for which the first step is to write down the 2-particle-irreducible (2PI) effective action \cite{jackiw}. We will work with an accuracy of a two-loop effective action,
where dissipative effects can occur. The two-particle irreducible (2PI) effective action 
corresponding to the tree-level action~\eqref{action for 2 fields:dS}
 can be written in the two-loop approximation as,
\begin{equation}
\Gamma[i\Delta^{cd}_{\varphi},i\Delta^{cd}_{\chi}]
 =\Gamma_0[i\Delta^{cd}_{\varphi},i\Delta^{cd}_{\chi}]+\Gamma_1[i\Delta^{cd}_{\varphi},i\Delta^{cd}_{\chi}]+\Gamma_2[i\Delta^{cd}_{\varphi},i\Delta^{cd}_{\chi}]\, , \quad c\, ,d = \pm\, ,
\end{equation}
where the three constituent functional are given by
\begin{align}
\Gamma_0[i\Delta^{cd}_{\varphi},i\Delta^{cd}_{\chi}]
=& \frac12\int d^Dxd^Dx'\sqrt{-\overline{g}_{{dS}}(x)}\bigg(\nonumber
\sum_{c,d=\pm} \overline{\Box}^{{dS}}_x\delta^D(x-x')c\delta^{cd}i\Delta^{dc}_{\varphi}(x',x)\\
&+\sum_{c,d=\pm}\overline{\Box}^{{dS}}_x \delta^D(x-x')c\delta^{cd}i\Delta^{dc}_{\chi}(x',x)\bigg)\, ,\\
\Gamma_1[i\Delta^{cd}_{\varphi},i\Delta^{cd}_{\chi}]=&-\frac{i}{2}\Tr[\log(i\Delta^{cd}_{\varphi}(x;x'))]-\frac{i}{2}\Tr[\log(i\Delta^{cd}_{\chi}(x;x'))]\, ,\\
\Gamma_2[i\Delta^{cd}_{\varphi},i\Delta^{cd}_{\chi}]=&\int d^Dxd^Dx'\sqrt{-\overline{g}_{{dS}}(x)}\sqrt{-\overline{g}_{{dS}}(x^{\prime})}\sum_{c,d=\pm}cd\frac{ih^2}{4}\qty(i\Delta^{cd}_{\chi}(x,x'))^2i\Delta^{cd}_{\varphi}(x,x')\,,
\end{align}
and the elements of the Keldysh propagators $i\Delta^{cd}_{\varphi, \chi}$ may be identified in terms of the statistical and spectral two-point functions,
\begin{eqnarray}
i\Delta^{\mp \pm}_{\varphi,\chi}(x,x^{\prime}) &=& F_{\varphi, \chi}  (x,x^{\prime}) \pm \frac{1}{2} i\Delta^c_{\varphi , \chi}(x,x^{\prime})\, , \\
 i\Delta^{\pm \pm}_{\varphi,\chi}(x,x^{\prime}) &=& F_{\varphi, \chi}(x,x^{\prime}) 
      \pm \frac{1}{2} \text{sign} \big[ x^0 - (x^0)^{\prime}\big] i\Delta^c_{\varphi , \chi}(x,x^{\prime})\, .
\end{eqnarray}
Applying the variational principle yields the following equations of motion
\begin{align}
\overline{\Box}^{{dS}}_xi\Delta^{ab}_{\varphi}(x;x^{\prime \prime})=\frac{a \,  \delta^{ab}i \delta^D(x-x^{\prime \prime})}{\sqrt{-\overline{g}_{{dS}}(x)}}\label{geneom}
+ \int d^Dx^{\prime }\sqrt{-\overline{g}_{{dS}}(x^{\prime })}\sum_{c=\pm}  c \,i  M^{ac}_{\varphi}(x,x^{\prime })i\Delta^{cb}_{\varphi}(x^{\prime },x^{\prime \prime})\, ,
\end{align}
\begin{align}
\overline{\Box}^{{dS}}_{x^{\prime \prime}}i\Delta^{ab}_{\varphi}(x;x^{\prime \prime})=\frac{a \,  \delta^{ab}i \delta^D(x-x^{\prime \prime})}{\sqrt{-\overline{g}_{{dS}}(x)}}\label{geneomPP}
+ \int d^Dx^{\prime }\sqrt{-\overline{g}_{{dS}}(x^{\prime })}\sum_{c=\pm}c \,i \Delta^{ac}_{\varphi}(x,x^{\prime }) i  M^{cb}_{\varphi}(x^{\prime},x^{\prime\prime })\, ,
\end{align}
where the corresponding self-masses $iM^{ab}_{\varphi}(x,x^{\prime})$ read
\begin{equation}
iM^{ab}_{\varphi}(x,x^{\prime})=-\frac{ih^2}{2}\qty(i\Delta^{cd}_{\chi}(x,x'))^2\, .
\end{equation}

\subsection{Renormalizing the self-mass}

We attempt to solve equation \eqref{geneom} by using the expression for the free propagators in the Bunch-Davies vacuum,
\begin{equation}
iM^{ab}_{\varphi}(x,x^{\prime})=-\frac{ih^2}{2}\qty(i\Delta^{ab}_{\chi}(x,x'))^2 \approx -\frac{ih^2}{2}\qty(i\Delta^{ab}_{dS}(x,x'))^2 
\label{selfMassIni}\, .
\end{equation}
The self-masses \eqref{selfMassIni} are products of distributions that have local contributions $\propto \delta^D (x, x^{\prime})$
which would yield indefinite answers when integrated against a test function. The singularities can be isolated by differential, dimensional regularization in position space where they takes the form, 
$\propto(D-4)^{-1} \delta^D (x,x^{\prime})$  (and/or derivatives thereof). 
We renormalize the self-mass \eqref{selfMassIni} 
by adding suitable local counterterms to the effective action which can be used to subtract these divergent contributions,  yielding eventually finite answers in the limit $D\rightarrow 4$. 

Let is first write down the de Sitter Feynman propagator in position space in $D$ space-time dimensions which has been computed in terms of the quantity 
\begin{align}
y \equiv y_{++}\, ,
\end{align}
where in de Sitter invariant length functions
\begin{align}
y_{ab} =  a a^{\prime}  H^2  \Delta x_{ab}^2  
= a(\eta) a(\eta^{\prime} )  H^2  \Delta x_{ab}^2\big(\eta\!-\!\eta^{\prime}, \vec x \!-\!\vec x^{\,\prime}\big) 
= \frac{\Delta x_{ab}^2\big(\eta\!-\!\eta^{\prime},\vec x \!-\!\vec x^{\,\prime}\big)  }{\eta \eta^{\prime}}\, 
\end{align}
can be expressed with the Lorentz invariant length functions
\begin{align}
\Delta x^2_{\pm \pm} &= - \big(| \eta \!-\! \eta^{\prime} | \mp i \varepsilon \big)^2 + \| \vec{x} \!-\! \vec{x}^{\,\prime} \|^2\, , \\
\Delta x^2_{\pm \mp} &= - \big(\eta \!-\! \eta^{\prime}  \pm i \varepsilon \big)^2 + \| \vec{x} \!-\! \vec{x}^{\,\prime} \|^2\, .
\end{align}
The de Sitter propagator in position space has has been given by \cite{Onemli:2004mb},
\begin{multline}
i\Delta^{++}_{dS} = \frac{H^{D-2}}{(4 \pi)^{D/2}} \Bigg[- \sum_{n=0}^{\infty} \frac{1}{n - \frac{D}{2} +1 } \frac{\Gamma\big[ n+ \frac{D}{2}\big]}{\Gamma\big[ n+ 1\big]} \Big( \frac{y}{4} \Big)^{n-\frac{D}{2}+1}    \\ 
- \frac{\Gamma\big[ D-1\big]}{\Gamma\big[ \frac{D}{2}\big]} \pi \cot \big[ \pi \frac{D}{2} \big]
 + \sum_{n=1}^{\infty} \frac{1}{n} \frac{\Gamma\big[ n+ {D} -1\big]}{\Gamma\big[ n+ \frac{D}{2}\big]}\Big( \frac{y}{4} \Big)^{n} 
 + \frac{\Gamma\big[ D-1\big]}{\Gamma\big[ \frac{D}{2}\big]} \log \big[ a a^{\prime}\big]
\Bigg] \, , \label{deSitterWood}
\end{multline}
where we use in this section the notation $a^{\prime} = a(\eta^{\prime})$ and it should be clear from the context whether a prime denotes a time derivative or refers to a coordinate. 
We can expand expression \eqref{deSitterWood} around $D=4$ and get
 \be
i\Delta^{++}_{dS} = \frac{H^{D-2}}{(4 \pi)^{D/2}} \Big[
 {   \Gamma \Big[\frac{D-2}{2} \Big]}\Big( {\frac{y}{4}}\Big)^{1-\frac{D}{2}}- 2 \log \Big[
 \frac{\sqrt{e} \,  y }{4 a a^{\prime}} \Big] 
\Big]  + \mathcal{O} \big( D-4 \big) \, .
\ee
Taking the square leads to
 \be
\Big(i\Delta^{++}_{dS} \Big)^2 = \frac{H^{2D-4}}{(4 \pi)^{D}} \Big[
{  \Gamma^2 \Big[\frac{D-2}{2} \Big]} \Big( {\frac{y}{4}}\Big)^{2-{D}} 
-{\frac{16}{y}}  \log \Big[
 \frac{\sqrt{e} \,  y }{4 a a^{\prime}} \Big] 
+ 4 \log^2 \Big[
 \frac{\sqrt{e} \,  y }{4 a a^{\prime}} \Big] 
\Big]  + \mathcal{O} \big( D-4 \big) \, , 
\ee
and we note that the non-integrable piece of the self-mass is contained in the first term $\propto y^{2-D}$.
Let us simplify the notation and denote the de Sitter d'Alembert operator as\footnote{We would like to remark that due to symmetry reasons we may use in the following derivations also derivatives acting on primed coordinates
\begin{align}
\frac{\square^{\prime}}{H^2}= (\eta^{\prime})^2 \Big[-\frac{\partial^2}{\partial (\eta^{\prime})^2} + \frac{D-2}{\eta^{\prime}} \frac{\partial}{\partial \eta^{\prime}} + \delta^{ij} \frac{\partial^2}{\partial (x^i)^{\prime} \partial (x^j)^{\prime}} \Big] \, .
\end{align}}
\begin{align}
\frac{\square}{H^2}\equiv \frac{\overline{\Box}^{{dS}}}{H^2} = \eta^2 \Big[-\frac{\partial^2}{\partial \eta^2} + \frac{D-2}{\eta} \frac{\partial}{\partial \eta} + \delta^{ij} \frac{\partial^2}{\partial x^i \partial x^j} \Big] \, .
\end{align}
We will make use of two relations that were established in \cite{Prokopec:2008gw},
\begin{multline}
\Big( {\frac{y_{\pm \pm}}{4}}\Big)^{2-{D}} = \Big[\frac{2}{(D-3)(D-4)} \frac{\square}{H^2} -\frac{D(D-2)}{2 (D-3)(D-4)}   + \frac{D-6}{2(D-3)} \Big]\Big( {\frac{y_{\pm \pm}}{4}}\Big)^{3-{D}} \\
-\Big[ \frac{2}{(D-3)(D-4)} \frac{\square}{H^2} - \frac{D(D-2)}{2 (D-3)(D-4)} \Big]\Big( {\frac{y_{\pm \pm}}{4}}\Big)^{1-(D/2)} \\ \pm \frac{2(4 \pi)^{D/2}}{(D-3)(D-4)\Gamma\big[\frac{D}{2}-1 \big]} \frac{i \delta^D (x-x^{\prime})}{(Ha)^D}\, ,\label{deSitterRel}
\end{multline}
as well as
\be
 \frac{\square}{H^2}\Big( {\frac{y_{\pm \pm}}{4}}\Big)^{1-(D/2)} = \pm \frac{(4 \pi)^{D/2}}{\Gamma\big[\frac{D}{2}-1 \big]} \frac{i \delta^D (x-x^{\prime})}{(Ha)^D} + \frac{D(D-2)}{4}\Big( {\frac{y_{\pm \pm}}{4}}\Big)^{1-(D/2)}\, .
\ee
Let us introduce the renormalization parameter $\mu$ with energy dimension one.
We can rewrite \eqref{deSitterRel} by adding a $\mu$-dependent term that vanishes on $D=4$ in such a way that the divergence in the self-mass may be removed with a mass counter term in the action $\propto (D-4)^{-1}\mu^{D-4} a^{-D} \delta^D \big(x-x^{\prime} \big)$. Moreover, we use
\be
\Big( {\frac{y_{\pm \pm}}{4}}\Big)^{3-{D}} = \Big( {\frac{y_{\pm \pm}}{4}}\Big)^{1-(D/2)}\Big[1- \frac{D-4}{2}\log \big[  y_{\pm \pm} \big] + \mathcal{O}\big[(D-4)^2 \big] \Big]\, ,
\ee
and expand the non-singular terms in \eqref{deSitterRel},
\begin{multline}
\Big( {\frac{y_{\pm \pm}}{4}}\Big)^{2-{D}} =
 \pm \frac{2 (4 \pi )^{D/2} }{(D-3)(D-4)\Gamma\big[\frac{D}{2}-1\big] } \Big(  \frac{ \mu}{H}\Big) ^{D-4}  \frac{i \delta^D \big(x-x^{\prime} \big)}{(Ha)^D}\\
-  \frac{\square}{H^2} \Big(\frac{4}{y_{\pm \pm}}\log \Big[\frac{\mu^2 y_{\pm \pm}}{H^2} \Big] \Big) - \frac{4}{y_{\pm \pm}} \Big(2 \log \Big[ \frac{\mu^2 y_{\pm \pm} }{H^2}\Big] -1 \Big)  + \mathcal{O}(D-4)\, ,
\end{multline}
which leads to
\begin{multline}
\Big(i\Delta^{\pm \pm}_{dS} \Big)^2 = 
\pm   \frac{2  \Gamma\big[\frac{D}{2}-1\big]{ \mu} ^{D-4}  }{(4 \pi )^{D/2}(D-3)(D-4)} 
 \frac{i \delta^D \big(x\!-\!x^{\prime} \big)}{a^D} \\
 -  \frac{H^{2D-4}}{(4 \pi)^{D}}  \Bigg[  \frac{\square}{H^2} \Big(\frac{4}{y_{\pm \pm}}\log \Big[\frac{\mu^2 y_{\pm \pm}}{H^2} \Big] \Big) - \frac{4}{y_{\pm \pm}} \Big(2 \log \Big[ \frac{\mu^2 y_{\pm \pm} }{H^2}\Big]  -1\Big)  \\
+  \frac{16}{y_{\pm \pm}}  \log \Big[
 \frac{\sqrt{e} \,  y }{4 a a^{\prime}} \Big] 
- 4 \log^2 \Big[
 \frac{\sqrt{e} \,  y_{\pm \pm} }{4 a a^{\prime}} \Big] 
\Bigg]  + \mathcal{O} \big( D-4 \big) \, .
\label{some equation}
\end{multline} 
The divergent local contribution in the first line of~(\ref{some equation}) yields a divergent contribution to
 the self-mass~(\ref{selfMassIni}), 
\begin{equation}
\left(i M^{cd}_{\varphi}(x,x')\right)_{\rm div}
  =h^2\frac{\Gamma\big[\frac{D}{2}-1\big]{ \mu} ^{D-4}  }{(4 \pi )^{D/2}(D-3)(D-4)} 
 \frac{\delta^D \big(x\!-\!x^{\prime} \big)}{a^D} c\delta^{cd}
 \,,
\label{counterterm self-mass}
\end{equation}
which can be removed by adding the following counterterm action,~\footnote{The corresponding 
counterterm action in the one-particle irreducible formalism is local in the fields, 
\begin{equation}
S_{\rm ct}^{\rm 1PI} = \int d^Dx a^D \left(-\frac12\delta m^2 \sum_{c=\pm}c[\varphi^{c}(x)]^2\right)
\,.
\label{counterterm:1PI}
\end{equation}
} 
\begin{equation}
S_{\rm ct} = 
    \int d^Dx a^D \left(-\frac12\delta m^2\sum_{c,d=\pm}c\delta^{cd}i\Delta_\varphi^{cd}(x,x)\right)
\,,
\label{counterterm}
\end{equation}
where $\delta m^2$ is proportional to the inflaton condensate squared, 
\begin{equation}
\delta m^2 = -g^2\overline\phi^{\,2}
                       \frac{\Gamma\left(\frac{D}{2}\!-\!1\right)\mu^{D-4}}{(4\pi)^{D/2}(D\!-\!3)(D\!-\!4)}
\,,
\label{divergent mass}
\end{equation}
and diverges as $\propto 1/(D-4)$.
Clearly, the counterterm~(\ref{counterterm}) is the divergent mass counterterm of the 2PI formalism.
It is easy to check that varying the action~(\ref{counterterm}) and adding it to the equations of 
motion~(\ref{geneom}--\ref{geneomPP}) removes 
the divergent parts of the self-masses. The resulting renormalized self-mass $i M^{++}_{\phi, \text{ren}}$ is,
\begin{multline}
i M^{++}_{\varphi, \text{ren}}(x,x') =  \frac{i h^2}{2} \frac{H^{4}}{(4 \pi)^{4}} \Bigg[
    \frac{\square}{H^2} \Big(\frac{4}{y}\log \Big[\frac{\mu^2 y}{H^2} \Big] \Big) \!-\! \frac{4}{y} \Big(2 \log \Big[ \frac{\mu^2 y}{H^2}\Big]  \!-\!1\Big) 
\!+\! {\frac{16}{y}}\log \Big[
 \frac{\sqrt{e} \,  y }{4 a a^{\prime}} \Big] 
\!-\! 4 \log^2 \Big[
 \frac{\sqrt{e} \,  y }{4 a a^{\prime}} \Big] 
\Bigg] 
 .
\label{renormalized self mass}
\end{multline}
The other renormalized self-masses, $i M^{ab}_{\varphi, \text{ren}}(x,x')$ ($a,b=\pm$), are obtained
simply by replacing $y(x,x')=y_{++}(x,x')$ in~(\ref{renormalized self mass}) by $y_{ab}(x,x')$.

\subsection{Self-mass in momentum space}

Ultimately, we will be interested in the Wigner transform of the spatially dependent piece of the self-mass. 
This may be conveniently achieved by extracting d'Alembert's operators and dropping homogeneous 
(momentum independent) contributions. 
If the d'Alembertian in de Sitter space-time is acting on non-singular functions 
(not containing $y^{-1}$), we have,
\begin{align}
\frac{\square}{H^2} f(y) = (4-y) y f^{\prime \prime} (y) + 4(2-y)f^{\prime}(y)\, ,
\end{align}
which gives the identities
\begin{eqnarray}
\frac{1}{y} &=& \frac{1}{4} \frac{\square}{H^2} \log \big( y\big) + \frac{3}{4}\, ,\\
\frac{\log \big(y \big)}{y}   &=&
  \frac{1}{8} \frac{\square}{H^2} \Big[ \log^2 \big(y\big) - 2  \log \big(y\big)  \Big]
  + \frac{3}{4} \log \big( y \big) 	
- \frac{1}{2}\,.
\end{eqnarray}
These identities allow us to rewrite the self-mass~(\ref{renormalized self mass}) as,
\begin{multline}
\label{inhomSelfMass}
i M^{++}_{\varphi, \text{ren}}(x, x^{\prime}) =  \frac{i h^2}{2} \frac{H^{4}}{(4 \pi)^{4}} \Bigg\lbrace
       \frac{\square^2}{H^4} \Bigg[  \frac{1}{2}\log^2 \Big(\frac{y}{4}\Big) +   \log \Big[ \frac{4 \mu^2 }{ e H^2} \Big] \log \Big(\frac{y}{4}\Big) \Bigg]	
    \\
 +    2 \frac{\square}{H^2} \Bigg[ \frac{1}{2}\log^2 \Big(\frac{y}{4}\Big) 
+  \log \Big[
 \frac{ e \,  H^2 }{ 4 \mu^2} \Big]  \log \Big( \frac{y}{4}\Big) \Bigg] +2\Big[  1-  2\log (a a^{\prime})\Big]\frac{\square}{H^2} \log \Big( \frac{y}{4}\Big)
  \\+ 2 \Big[1 	
 + 4 \log (a a^{\prime}) \Big] \log \Big( \frac{ y}{4} \Big)  
 - 4 \log^2 \Big( \frac{ y}{4} \Big) 
\Bigg\rbrace + \text{hom} \, .
\,,
\end{multline}
where {hom.} encode spatially homogeneous ($y$ independent) contributions, which are of no importance for 
this study.  
At this stage, we would like to emphasize, that the expression for the self-mass \eqref{inhomSelfMass} could have also been written with the de Sitter d'Alembertian operators acting on the primed space-time coordinates.
We now perform the spatial Wigner transform of the self-mass~(\ref{inhomSelfMass}) according to, 
\begin{align}
i M^{++}_{\varphi, \text{ren}}\big(\eta, \eta^{\prime} ,k \big) = \int d^{3}\big({x}-{x}^{\prime} \big) i M^{++}_{\phi, \text{ren}}\big(x, x^{\prime} \big) e^{-i \vec{k} \cdot(\vec{x}-\vec{x}^{\prime})}\, .
\end{align}
Furthermore, spatially homogeneous contributions are proportional to delta functions in $k$-space or derivatives thereof,
\begin{align}
\int_0^{\infty} dr \, r \sin \big(k r\big) = - \pi \partial_k \delta (k)\, .
\end{align}
We will drop again such contributions.
In Appendix \ref{fourier} we establish the following Wigner transformation,
\begin{multline}
\int d^{3}\big({x}\!-\!{x}^{\prime} \big) e^{-i \vec{k} \cdot(\vec{x}-\vec{x}^{\prime})}\Bigg[\frac{1}{2}\log^2 \Big(\frac{y}{4}\Big) \!+\! f\big(\eta, \eta^{\prime} \big)\log \Big(\frac{y}{4}\Big)  \Bigg] 
\\ =
-\frac{4 \pi^2}{k^3} \Bigg[2+ \big[1+ i k |\Delta \eta | \big] \Big(  \log \Big[\frac{a a^{\prime} H^2| \Delta \eta|}{2k} \Big]+ i \frac{ \pi}{2}- \gamma_E + f\big(\eta, \eta^{\prime} \big)\Big) \Bigg] e^{-i k |\Delta \eta|} \\
+\frac{4 \pi^2}{k^3} \big(1 \!-\! i k |\Delta \eta| \big)\Bigg[ \text{ci} \big[ 2 k| \Delta \eta|  \big]  \!-\!i \,  \text{si} \big[ 2 k |\Delta \eta|  \big]  \Bigg] e^{+i k |\Delta \eta|}\, , \label{ftLog}
\end{multline}
where $\Delta \eta  = \eta - \eta^{\prime}$ and $f(\eta, \eta^{\prime})$ is some $k$-independent function.
We make use of the Wigner transform \eqref{ftLog}, rewrite the scale factor as $a = - (H \eta)^{-1}$ and obtain, after some simplifications, the self-mass in momentum space as follows,
\begin{multline}
i M^{++}_{\varphi, \text{ren}} \big(\eta, \eta^{\prime},k \big)  =  -\frac{4 \pi^2}{k^3}\frac{i h^2}{2} \frac{H^{4}}{(4 \pi)^{4}} \Bigg\lbrace\\
       \frac{\square^2_k}{H^4} \Bigg( \Bigg[2+ \big[1+ i k |\Delta \eta | \big] \Big(  \log \Big[\frac{ 2 | \Delta \eta| \mu^2}{k \eta \eta^{\prime} H^2 }\Big]+ i \frac{ \pi}{2}- \gamma_E -1 \Big) \Bigg] e^{-i k |\Delta \eta|} \\
- \big(1 - i k |\Delta \eta| \big)\Bigg[ \text{ci} \big[ 2 k| \Delta \eta|  \big]  -i \,  \text{si} \big[ 2 k |\Delta \eta|  \big]  \Bigg] e^{+i k |\Delta \eta|} \Bigg)
  \\
 +    2 \frac{\square_k}{H^2} \Bigg(\Bigg[ 2+ \big[1+ i k |\Delta \eta | \big] \Big(  \log \Big[\frac{ | \Delta \eta|H^2}{8k \eta \eta^{\prime} \mu^2} \Big]+ i \frac{ \pi}{2}- \gamma_E + 1 \Big)  \Bigg]   e^{-i k |\Delta \eta|} \\
- \big(1 - i k |\Delta \eta| \big)\Bigg[ \text{ci} \big[ 2 k| \Delta \eta|  \big]  -i \,  \text{si} \big[ 2 k |\Delta \eta|  \big]  \Bigg] e^{+i k |\Delta \eta|}  \Bigg)\\
+{ 2\big[1+ 2 \log (H^2 \eta \eta^{\prime})\big]} \frac{\square_k}{H^2} \Bigg[\big[1+ i k |\Delta \eta | \big]  e^{-i k |\Delta \eta|} \Bigg] 
  \\-8 \Bigg( \Bigg[2+ \big[1+ i k |\Delta \eta | \big] \Big(  \log \Big[ \frac{ H^2| \Delta \eta|}{2k} \Big]+ i \frac{ \pi}{2}- \gamma_E  -\frac{1}{4}  \Big) \Bigg] e^{-i k |\Delta \eta|} \\
- \big(1 - i k |\Delta \eta| \big)\Bigg[ \text{ci} \big[ 2 k| \Delta \eta|  \big]  -i \,  \text{si} \big[ 2 k |\Delta \eta|  \big]  \Bigg] e^{+i k |\Delta \eta|}   \Bigg)
\Bigg\rbrace + \text{hom}  \,,\label{M++Full}
\end{multline}
where 
\begin{equation}
\frac{\Box_k}{H^2}=-\eta^2\left(\partial_\eta^2 - \frac{2}{\eta}\partial_\eta+k^2\right)
\label{d'Alembertian in k space}
\end{equation}
 is the d'Alembertian in momentum 
space. 
For a computational convenience we shall split the self-mass \eqref{M++Full} in the following way,
\begin{align}
M^{++}_{\phi, \text{ren}}(\eta, \eta^{\prime},k) = -{ 2\big[1\!+\! 2 \log (H^2 \eta \eta^{\prime})\big]} \frac{\square_{k}}{H^2} \widehat{M}^{++}(|\Delta \eta |,k) + \sum_{n=0}^2 
\Big( \frac{\square_{k}}{H^2} \Big)^n M^{++}_{(n)}(\eta, \eta^{\prime},k)  \, , \label{boxsplit}
\end{align}
which is based on the definitions,
\begin{multline}
M^{++}_{(n)} \big(\eta, \eta^{\prime},k \big)\equiv \alpha_{(n)} \Big[  \widetilde{M}^{++}_{I} \big(|\Delta \eta|,k \big) + \widetilde{M}^{++}_{II} \big(|\Delta \eta|,k \big) \Big] + \beta_{(n)} \widehat{M}^{++} \big(|\Delta \eta|,k \big) 
\\ + \gamma_{(n)} \widehat{M}^{++} \big(| \Delta \eta |,k \big)  \log \Big[\frac{ \eta  \eta ^{\prime} H^4}{4  \mu^2} \Big] \, ,
\end{multline}
where 
\be
\alpha_{(n)} = \left\lbrace -8, 2 , 1 \right\rbrace\, ,\beta_{(n)} = \left\lbrace -2, -2\!+\!8\log\Big[\frac{2 \mu}{H}\Big] ,1 \right\rbrace\, ,\gamma_{(n)} = \left\lbrace 0, 2 ,1 \right\rbrace\, , \label{coeffsM}
\ee
and
\begin{eqnarray}
 \widehat{M}^{++} \big(| \Delta \eta |,k \big)&\equiv&  \frac{4 \pi^2}{k^3}\frac{ h^2}{2} \frac{H^{4}}{(4 \pi)^{4}}\big[1\!+\! i k |\Delta \eta | \big]  e^{-i k |\Delta \eta|} \, ,\\
 \widetilde{M}^{++}_{I} \big(|\Delta \eta|,k \big) & \equiv& -\frac{4 \pi^2}{k^3}\frac{ h^2}{2} \frac{H^{4}}{(4 \pi)^{4}}  \Bigg[2+ \big[1\!+\! i k |\Delta \eta | \big] \Big(  \log \Big[\frac{ H^2 | \Delta \eta|}{2 k} \Big]+ i \frac{ \pi}{2}- \gamma_E   \Big) \Bigg] e^{-i k |\Delta \eta|}\, ,\\
\widetilde{M}^{++}_{II} \big(| \Delta \eta|,k \big) &\equiv&   - \frac{4 \pi^2}{k^3}\frac{ h^2}{2} \frac{H^{4}}{(4 \pi)^{4}}   \big(1 \!-\! i k |\Delta \eta| \big)\text{E}_1 \big[  2 i k| \Delta \eta| \big]   e^{+i k |\Delta \eta|} \, .
\end{eqnarray} 
Here, we made use of the identity for the exponential integral function,
\be
\text{E}_1 \big[  2 i k| \Delta \eta| \big] =    i \,  \text{si} \big[ 2 k |\Delta \eta|  \big]-\text{ci} \big[ 2 k| \Delta \eta| \big]\, ,
\ee
which holds when $k>0$. The sine (si) and cosine (ci) integrals are defined in 
Eqs.~(\ref{sine integral}) and~(\ref{cos integral}), respectively.
The calculation of the other self-masses $iM_{\phi, \text{ren}}^{\pm \mp}$ and $iM_{\phi, \text{ren}}^{--}$ proceeds similarly.
By writing
\begin{align}
\log \big( \Delta x_{--}^2 \big) = \log \Big( \big|\Delta \eta^2 - \|\vec{x}\!-\!\vec{x}^{\prime}\|^2   \big|\Big) 
      - i \pi \theta \big( \Delta \eta^2 - \|\vec{x}\!-\!\vec{x}^{\prime}\|^2 \big) \, ,
\end{align}
we see that 
\begin{align}
M^{--}_{\phi, \text{ren}}  = \Big[M^{++}_{\phi, \text{ren}}\Big]^{*}\,. 
\end{align}
Moreover, due to 
\begin{align}
\log \big( \Delta x_{\mp \pm}^2 \big) = \log \Big( \big|\Delta \eta^2 - \|\vec{x}\!-\!\vec{x}^{\prime}\|^2 \big|\Big) 
 \pm  i \, \text{sign}\big(\eta,\eta^{\prime} \big) \pi \theta \big( \Delta \eta^2 - \|\vec{x}\!-\!\vec{x}^{\prime}\|^2 \big) \, , 
\end{align}
we see that,
\begin{align}
M^{ab}_{\phi, \text{ren}}(\eta, \eta^{\prime},k) = -4 \big[1\!+\! \log (\eta \eta^{\prime} H^2) \big] \frac{\square_{k}}{H^2} \widehat{M}^{ab}(|\Delta \eta |,k) + \sum_{n=0}^2 
\Big( \frac{\square_{k}}{H^2} \Big)^n M^{ab}_{(n)}(\eta, \eta^{\prime},k)  \, , 
\end{align}
where
\begin{align}
M^{\mp \pm }_{(n)}   = M^{\pm \pm}_{(n)} \theta\big(\Delta \eta \big) + M^{\mp \mp}_{(n)} \theta(- \Delta\eta )\,  , \quad \widehat{M}^{\mp \pm }_{(n)}   = \widehat{M}^{\pm \pm}_{(n)} \theta(\Delta \eta ) + \widehat{M}^{\mp \mp}_{(n)} \theta(-\Delta\eta )\,  ,
\end{align}  
where $\text{sign}\big(\eta,\eta^{\prime} \big)=\theta(\eta\!-\!\eta^{\prime})-\theta(\eta^{\prime}\!-\!\eta)$
and $\theta$ is the Heaviside step function.
It will be convenient to define
\begin{eqnarray}
M^{F}_{(n)} &\equiv& \frac{1}{2}\Big[M^{++}_{(n)}+M^{--}_{(n)} \Big] = \text{Re} \, M^{++}_{(n)}\, , \\
M^{c}_{(n)}(\eta, \eta^{\prime}) &\equiv&\frac{\text{sign} \big(\Delta\eta \big)}{i} \Big[M^{++}_{(n)}-M^{--}_{(n)}\Big](\eta, \eta^{\prime}) =2 \, \text{sign} \big(\Delta\eta \big)\,  \text{Im} \, M^{++}_{(n)}\, ,\\
\widehat{M}^{F} &\equiv& \frac{1}{2}\Big[\widehat{M}^{++} +\widehat{M}^{--} \Big] = \text{Re} \, \widehat{M}^{++}\, , \\
\widehat{M}^{c}(\eta, \eta^{\prime}) &\equiv&\frac{\text{sign} \big(\Delta\eta \big)}{i} \Big[\widehat{M}^{++}-\widehat{M}^{--}\Big](\eta, \eta^{\prime}) =2 \, \text{sign} \big(\Delta\eta \big)\,  \text{Im} \, \widehat{M}^{++} \, ,
\end{eqnarray}
and note the relations
\begin{eqnarray}
 \Big[M^{++}_{(n)}-M^{--}_{(n)} \pm \Big(M^{-+}_{(n)}-M^{+-}_{(n)} \Big)\Big](\eta, \eta^{\prime})  &= &\pm 2 \, \theta(\pm \Delta \eta )
 \, i M^{c}_{(n)}  (\eta, \eta^{\prime}) \, , \\
 \Big[M^{++}_{(n)}+M^{--}_{(n)}\Big](\eta, \eta^{\prime}) +
  \text{sign} (\tau  \!-\!  \tau^{\prime})  \Big[M^{-+}_{(n)}+M^{+-}_{(n)}\Big](\eta, \eta^{\prime})&= & 4\, \theta(\tau \!-\! \tau^{\prime})\, M^{F}_{(n)}(\eta, \eta^{\prime}) \, ,
\end{eqnarray}
which also hold for $\widehat{M}^{ab}$.

\subsection{Perturbative solution for the statistical propagator }

Let us look at the renormalized version of  equations of motion \eqref{geneom} for the Keldysh propagators $i \Delta_{\varphi}^{ab} $.
By rewriting the two-point functions in terms of real and imaginary parts, we obtain
\begin{multline}
\square_{x} F_{\varphi} (x,x^{\prime \prime})  = \frac{i}{2} \int d \eta^{\prime} d^3 x^{\prime }\big(\eta^{\prime} H \big)^{-4}\, \Big[    M^{++}_{\varphi,\text{ren}}  -   M^{--}_{\varphi,\text{ren}} +   M^{-+}_{\varphi,\text{ren}} -   M^{+-}_{\varphi,\text{ren}}   \Big](x, x^{\prime }) F_{\varphi} (x^{\prime }, x^{\prime\prime} )   \\ - \frac{1}{4} \int  d \eta^{\prime} d^3 x^{\prime }\big(\eta^{\prime} H \big)^{-4}\, \Big[\text{sign} (\eta^{\prime} - \eta^{\prime \prime} ) \Big(    M^{++}_{\varphi,\text{ren}}  +    M^{--}_{\varphi,\text{ren}}\Big) -   M^{-+}_{\varphi,\text{ren}} -   M^{+-}_{\varphi,\text{ren}}   \Big](x, x^{\prime })   \Delta_{\varphi}^c (x^{\prime }, x^{\prime\prime} )  \, . \label{statEQX}
\end{multline}
We will solve for the statistical propagator perturbatively by approximating  $ F_{\varphi}$ and $\Delta_{\varphi}^c$ on the right hand side of \eqref{statEQX} by the expressions for the Bunch-Davies vacuum \eqref{BDDelta} and \eqref{BDF}, respectively. 
Inserting the concrete expressions \eqref{boxsplit} for our model in momentum space, we find
\begin{multline}
\square_k F_{\varphi} (\eta,\eta^{\prime \prime},k)\approx    - \sum_{n=0}^2  \Big( \frac{\square_{k}}{H^2} \Big)^n \int_{- \infty}^{\eta} d \eta^{\prime} \big(\eta^{\prime} H \big)^{-4}\, M^c_{(n)}(\eta, \eta^{\prime} ,k) F_{\varphi, dS} (\eta^{\prime }, \eta^{\prime\prime},k )   \\ + \sum_{n=0}^2 \Big( \frac{\square_{k}}{H^2} \Big)^n \int_{-\infty}^{\eta^{\prime \prime}} d \eta^{\prime} \big(\eta^{\prime} H \big)^{-4}\, M^F_{(n)}(\eta, \eta^{\prime },k)  \Delta_{\varphi, dS}^c (\eta^{\prime }, \eta^{\prime\prime},k ) \\
 +2 \int_{- \infty}^{\infty} d \eta^{\prime}\frac{1+ 2\log (\eta \eta^{\prime}H^2)}{ \big(\eta^{\prime} H \big)^{4}} \frac{\square_{k}}{H^2} \Big[ \theta(\eta -\eta^{\prime}) \widehat{M}^c(\eta, \eta^{\prime} ,k) \Big]F_{\varphi, dS} (\eta^{\prime }, \eta^{\prime\prime},k )   \\ -2\int_{-\infty}^{\eta^{\prime \prime}} d \eta^{\prime} \frac{1+ 2\log (\eta \eta^{\prime}H^2)}{ \big(\eta^{\prime} H \big)^{4}} \frac{\square_{k}}{H^2} \Big[ \widehat{M}^F(\eta, \eta^{\prime} ,k) \Big] \Delta_{\varphi, dS}^c (\eta^{\prime }, \eta^{\prime\prime},k ) \, .
\end{multline}
Expanding the last two terms and rearranging the integration boundaries gives 
\begin{multline}
\square_k F_{\phi} (\eta,\eta^{\prime \prime},k)  \approx -2 \, \text{Im} \sum_{n=0}^2 \Big( \frac{\square_{k}}{H^2} \Big)^n \int_{- \infty}^{\eta} d \eta^{\prime} \big(\eta^{\prime} H \big)^{-4}\,  M^{++}_{(n)}(\eta, \eta^{\prime} ,k) i \Delta_{\varphi, dS}^{+-} (\eta^{\prime }, \eta^{\prime\prime},k )   \\ -  \sum_{n=0}^2 \Big( \frac{\square_{k}}{H^2} \Big)^n \int^{\eta}_{\eta^{\prime \prime} } d \eta^{\prime} \big(\eta^{\prime} H \big)^{-4}\,  \text{Re} \,  M^{++}_{(n)}(\eta, \eta^{\prime} ,k) \Delta_{\varphi, dS}^c (\eta^{\prime }, \eta^{\prime\prime},k ) \\
 +4 \, \text{Im} \int_{- \infty}^{\eta}  d \eta^{\prime} \frac{1+ 2\log (\eta \eta^{\prime}H^2)}{ \big(\eta^{\prime} H \big)^{4}}\Big[ \frac{\square_{k}}{H^2} \widehat{M}^{++}(\eta, \eta^{\prime} ,k) \Big] i \Delta_{\varphi, dS}^{+-} (\eta^{\prime }, \eta^{\prime\prime},k )   \\ + 2\int^{\eta}_{\eta^{\prime \prime} } d \eta^{\prime} \frac{1+ 2\log (\eta \eta^{\prime}H^2)}{ \big(\eta^{\prime} H \big)^{4}} \Big[ \frac{\square_{k}}{H^2} \text{Re} \, \widehat{M}^{++}(\eta, \eta^{\prime} ,k) \Big] \Delta_{\varphi, dS}^c (\eta^{\prime }, \eta^{\prime\prime},k )  \, , \label{eqnF}
 \end{multline}
 We will solve equation \eqref{eqnF} by using a retarded  Green's function $G_{\text{ret}}(\eta , \eta^{\prime} , k)$ (which yields no contributions of the particular solution to the initial values) for the d'Alembertian operator in momentum space 
\begin{align*}
\square_k G_{\text{ret}}(\eta , \eta^{\prime} , k) = H^2 \eta^2 \Big[  - \partial_{\eta}^2   + \frac{D-2}{\eta} \partial_{\eta} - k^2 \Big]G_{\text{ret}}(\eta , \eta^{\prime} , k) =  a^{-4}(\eta^{\prime}) \delta(\eta-\eta^{\prime}) \, ,
\end{align*}
\be
G_{\text{ret}}(\eta , \eta^{\prime} , k) = \theta(\eta - \eta^{\prime}) \frac{H^2}{k^3} \Big[ k(\eta - \eta^{\prime}) \cos \big[ k(\eta - \eta^{\prime} ) \big] -(1 + k^2 \eta \eta^{\prime} ) \sin \big[k(\eta - \eta^{\prime}) \big]\Big]\, .
\ee 
We have
\begin{multline}
F_{\varphi} (\eta,\eta^{\prime \prime},k)  =   \frac{\square_k}{H^2} \mathcal{B}_{(2)}(\eta, \eta^{\prime \prime},k) + \mathcal{B}_{(1)}(\eta, \eta^{\prime \prime},k) 
\\+ H^2 \int_{-\infty}^{\infty}d\tau \frac{G_{\text{ret}}(\eta, \tau,k)}{(\tau H)^4}\Big[ \mathcal{B}_{(0)}(\tau, \eta^{\prime \prime},k) + \mathcal{B}_{(0)}^{\text{log}}(\tau, \eta^{\prime \prime},k)\Big]+ F_{\text{hom}} (\eta ,\eta^{\prime \prime},k) \,,  \label{eqnFInv}
\end{multline}
where
\begin{eqnarray}
\mathcal{B}_{(n)}(\eta, \eta^{\prime \prime},k) &\equiv& - \frac{2}{H^2}  \, \text{Im}   \int_{- \infty}^{\eta} \frac{d \eta^{\prime}}{ (\eta^{\prime} H )^{4}}\,  M^{++}_{(n)}(\eta, \eta^{\prime} ,k) i \Delta_{\varphi, dS}^{+-} (\eta^{\prime }, \eta^{\prime\prime},k ) \nonumber \\
 && - \frac{1}{H^2}  \int^{\eta}_{ \eta^{\prime \prime}}\frac{d \eta^{\prime}}{ (\eta^{\prime} H )^{4}}\,  \text{Re} \,  M^{++}_{(n)}(\eta, \eta^{\prime} ,k) \Delta_{\varphi, dS}^c (\eta^{\prime }, \eta^{\prime\prime},k )\, , \label{BTerms}\\
 \mathcal{B}_{(0)}^{\text{log}}(\eta, \eta^{\prime \prime},k) &\equiv&4 \, \text{Im} \int_{- \infty}^{\eta}  d \eta^{\prime} \frac{1+ 2\log (\eta \eta^{\prime}H^2)}{ \big(\eta^{\prime} H \big)^{4}}\Big[ \frac{\square_{\eta}(k)}{H^2} \widehat{M}^{++}(\eta, \eta^{\prime} ,k) \Big] i \Delta_{\varphi, dS}^{+-} (\eta^{\prime }, \eta^{\prime\prime},k ) \nonumber 
 \\&& +2 \int^{\eta}_{ \eta^{\prime \prime}} d \eta^{\prime} \frac{1+2 \log (\eta \eta^{\prime}H^2)}{ \big(\eta^{\prime} H \big)^{4}} \Big[ \frac{\square_{\eta}(k)}{H^2} \text{Re} \, \widehat{M}^{++}(\eta, \eta^{\prime} ,k) \Big] \Delta_{\varphi, dS}^c (\eta^{\prime }, \eta^{\prime\prime},k )\, , \label{BLog}
\end{eqnarray}
and
\be
\square_k F_{\text{hom}} (\eta ,\eta^{\prime \prime},k)=0\,. 
\ee
Let us define
\be
\widehat{F} (\eta ,\eta^{\prime \prime},k) \equiv {F}_{\varphi} (\eta ,\eta^{\prime \prime},k) - F_{\text{hom}} (\eta ,\eta^{\prime \prime},k)\, .
\ee
The homogeneous solution has to be chosen in such a way that the symmetry properties of the statistical two-point function are satisfied
\be
F_{\text{hom}} (\eta ,\eta^{\prime \prime},k)- F_{\text{hom}} (\eta^{\prime \prime} ,\eta,k) = \widehat{F} (\eta^{\prime \prime} ,\eta,k)-\widehat{F} (\eta ,\eta^{\prime \prime},k)\, ,
\ee
and the full solution reads
\be
F_{\varphi} (\eta ,\eta^{\prime \prime},k) = \frac{1}{2} \Big[\widehat{F} (\eta ,\eta^{\prime \prime},k)+\widehat{F} (\eta^{\prime \prime} ,\eta,k)\Big] + \frac{1}{2} \Big[F_{\text{hom}} (\eta ,\eta^{\prime \prime},k)+ F_{\text{hom}} (\eta^{\prime \prime} ,\eta,k)\Big]\, .
\ee
We immediately get the consistency requirement
\be
\square_k \square_k^{{\prime \prime}}\Big[\widehat{F} (\eta ,\eta^{\prime \prime},k)-\widehat{F} (\eta^{\prime \prime} ,\eta,k) \Big]=0\, , \label{consitencyReq}
\ee
which can be used as a non-trivial check of the result of the calculation.
Let us also fix a common prefactor for the subsequent integrals
\be
\lambda 
         \equiv  \frac{h^2}{256\pi^2k^3} 
\, ,
\ee
which gives the statistical two-point function $F_{\varphi}$  correct dimensions in momentum space if all other factors and ratios are dimensionless.

Let us proceed with the calculation of \eqref{eqnFInv}. The integrals with logarithms $\mathcal{B}_{(0)}^{\text{log}}$ in  \eqref{BLog} combine to the following expression
\begin{multline}
\mathcal{B}_{(0)}^{\text{log}}(\eta, 
 \eta^{\prime \prime}, k) = -8 {\lambda} \Bigg\lbrace
\cos [k(\eta - \eta^{\prime \prime})]\Big(2 + \log [H^2 \eta^2] \Big)\\+
\sin [k(\eta - \eta^{\prime \prime})]\Big(2 k (\eta-\eta^{\prime \prime}) - k\eta^{\prime \prime} \log [H^2 \eta^2] + k \eta \log [H^2 \eta \eta^{\prime \prime}]\Big)\\ +
k \eta \Big( \text{ci} [-2 k \eta ]+\text{ci} [-2 k \eta^{\prime \prime}]\Big) \Big(k \eta^{\prime \prime } \cos [k (\eta + \eta^{\prime \prime})] - \sin[k(\eta + \eta^{\prime \prime})]  \Big)\\
+
k \eta \Big( \pi + \text{si} [-2 k \eta ]+\text{si} [-2 k \eta^{\prime \prime}]\Big) \Big(\cos [k (\eta + \eta^{\prime \prime})] +k \eta^{\prime \prime } \sin[k(\eta + \eta^{\prime \prime})]  \Big) \Bigg\rbrace\\
\longrightarrow  -8 {\lambda} \Big(2 + \log [H^2 \eta^2] \Big)\, ,
\end{multline}
where the arrow denotes the super-Hubble limit.
The next step is to tackle the $\mathcal{B}_{2,1,0}$ terms in \eqref{BTerms} for which we note that the integrals containing negative infinity as a boundary may be rewritten as
\begin{multline}
\int_{-\infty}^{\eta}\frac{d\tau}{ (\tau H )^{4}}\,    M^{++}_{(n)}(\eta, \tau ,k) i \Delta_{\varphi, dS}^{+ -} (\tau, \eta^{\prime\prime},k )\\
=\frac{1}{2}(1 + i k \eta^{\prime \prime}) \frac{e^{- i k( \eta^{\prime \prime} - \eta) }}{H^2} \int^{ \infty}_{0} d x
\Big[\alpha_{(n)} \Big[  \widetilde{M}^{++}_{I} \Big(\frac{x}{k},k \Big) + \widetilde{M}^{++}_{II} \Big(\frac{x}{k},k \Big) \Big] \\ + \beta_{(n)}   \widehat{M}^{++} \Big(\frac{x}{k},k \Big) 
 + \gamma_{(n)}\widehat{M}^{++} \Big(\frac{x}{k},k \Big) \log \Big[\frac{ \eta (k \eta - x) H^4}{ 4  k  \mu^2} \Big]    \Big] \frac{ 1-  i ( k \eta - x )}{ \big(k \eta -x \big)^{4}}  e^{ - i  x  }\, . \label{splitInfMIntegrals}
\end{multline}
We then have to solve the following integrals ($\eta\, , \eta^{\prime \prime}<0\, , k>0$),
\begin{eqnarray}
I_{\widetilde{R}} (\eta ,\eta^{\prime \prime}, k) &\equiv&  - \frac{1}{\lambda H^2} \int_{\eta^{\prime \prime}}^{\eta} d \tau \big(\tau H \big)^{-4}\,  \text{Re} \,  \big[ \widetilde{M}^{++}_I +\widetilde{M}^{++}_{II} \big](\eta, \tau ,k) \Delta_{\varphi,dS}^c (\tau, \eta^{\prime\prime},k )\, , \label{iR}\\
I_{\widehat{R}}(\eta ,\eta^{\prime \prime},k)&\equiv &  -\frac{1}{ \lambda H^2}\int_{\eta^{\prime \prime}}^{\eta}  d \tau (\tau H)^{-4}\text{Re} \,  \widehat{M} (\eta , \tau, k) \Delta_{\varphi,dS}^c (\tau, \eta^{\prime \prime},k)\, ,\label{iRHat}\\
I_{{R}_{\text{log}}}(\eta ,\eta^{\prime \prime},k) &\equiv& - \frac{1}{ \lambda H^2}\int_{\eta^{\prime \prime}}^{\eta}d \tau (\tau H)^{-4} \text{Re} \,  \widehat{M} (\eta , \tau, k) \text{log} \Big[\frac{\eta \tau H^4}{4 \mu^2} \Big] \Delta_{\varphi,dS}^c (\tau, \eta^{\prime \prime},k)\ ,\label{iRLog}\\
I_{\widetilde{M}} (\eta ,k)  &\equiv&- \frac{e^{i k \eta}}{2 \lambda H^4} \int^{ \infty}_{0}dx \Bigg[  \widetilde{M}^{++}_{I} \Big(\frac{x}{k},k \Big) + \widetilde{M}^{++}_{II} \Big(\frac{x}{k},k \Big) \Bigg]\frac{ 1-  i ( k \eta - x )}{ \big(k \eta -x \big)^{4}}  e^{ - i  x  }\, , \label{iM}\\ 
I_{\widehat{M}}(\eta ,k)  &\equiv& -\frac{e^{i k \eta}}{2 \lambda H^4} \int^{ \infty}_{0}dx \widehat{M}^{++} \Big(\frac{x}{k},k \Big)\frac{ 1-  i ( k \eta - x )}{ \big(k \eta -x \big)^{4}}  e^{ - i  x  } \, ,\label{iMHat}\\
I_{{M}_{\text{log}}}(\eta ,k)  &\equiv& - \frac{e^{i k \eta}}{2 \lambda H^4} \int^{ \infty}_{0}dx \log \Big[\frac{ \eta (k \eta - x) H^4}{ 4  k  \mu^2} \Big]   \widehat{M}^{++} \Big(\frac{x}{k},k \Big)  \frac{ 1-  i ( k \eta - x )}{ \big(k \eta -x \big)^{4}}  e^{ - i  x  } \label{iMLog}\, .
\end{eqnarray}
We are able to solve all integrals except for the first one in terms of finite sums of exponentials, exponential integrals and generalized hypergeometric functions. However, for the integral  $I_{\widetilde{R}} $ we have to define the function
\be
\mathcal{J}(\eta, 
 \eta^{\prime \prime}, k) \equiv \int_{0}^{1} d x 
 E_1\big[-2 i k \big(x (\eta - \eta^{\prime \prime} ) + \eta^{\prime \prime} \big)  \big]\frac{1- e^{-2 i k (\eta - \eta^{\prime \prime})(x-1)}}{x-1} \, . \label{jIntegral}
\ee
We note that \eqref{jIntegral} approaches a constant in the super-Hubble limit.
We solve the $I_R$ integrals in Appendix \ref{iRIntegrals} and the $I_M$ integrals in Appendix \ref{iMIntegrals}.
We then have
\begin{multline}
\mathcal{B}_{(n)}(\eta, 
 \eta^{\prime \prime}, k) =\lambda  \alpha_{(n)}\Big[2 \,  \text{Im} \Big((1 + i k \eta^{\prime \prime})e^{- i k \eta^{\prime \prime}} I_{\widetilde{M}} (\eta ,k) \Big)  + I_{\widetilde{R}} (\eta ,\eta^{\prime \prime},k) \Big] \\+\lambda \beta_{(n)}\Big[2 \,  \text{Im} \Big((1 + i k \eta^{\prime \prime})e^{- i k \eta^{\prime \prime}}  I_{\widehat{M}} (\eta ,k) \Big)  + I_{\widehat{R}} (\eta ,\eta^{\prime \prime},k) \Big]\\ +\lambda \gamma_{(n)}\Big[2 \, \text{Im} \Big( (1 + i k \eta^{\prime \prime})e^{- i k \eta^{\prime \prime}}   I_{{M}_{\text{log}}} (\eta ,k) \Big) + I_{{R}_{\text{log}}} (\eta ,\eta^{\prime \prime},k) \Big]\, , 
\end{multline}
where the coefficients are given in \eqref{coeffsM}.
If we now act with the de Sitter d'Alembertian on $\mathcal{B}_{(2)}$ we have
\begin{multline}
\frac{\square_k}{H^2}\mathcal{B}_{(2)}(\eta,\eta^{\prime \prime},k) = 
2 \lambda \Bigg\lbrace \cos[k(\eta - \eta^{\prime \prime})]- k(\eta - \eta^{\prime \prime}) \sin[k(\eta - \eta^{\prime \prime})]\\+
 \Big(\cos[k(\eta - \eta^{\prime \prime})]+ k(\eta - \eta^{\prime \prime}) \sin[k(\eta - \eta^{\prime \prime})]\Big)
\Big(\text{ci}(2 k |\eta - \eta^{\prime \prime}|)+ \gamma_E - \log \Big[\frac{2\mu^2|\eta -\eta^{\prime \prime}|}{H^2 k \eta \eta^{\prime \prime}} \Big] \Big) \\
+ \text{sign}(\eta - \eta^{\prime \prime})
\Big[ \pi k(\eta - \eta^{\prime \prime})\cos[k(\eta - \eta^{\prime \prime})] - \frac{1}{2}\sin[k(\eta - \eta^{\prime \prime})] \\
-\Big( k(\eta - \eta^{\prime \prime})\cos[k(\eta - \eta^{\prime \prime})] - \sin[k(\eta - \eta^{\prime \prime})]  \Big) \text{si}(2 k |\eta - \eta^{\prime \prime}|) \Big]\Bigg\rbrace \\
\longrightarrow 2 \lambda \Bigg(2 \gamma_E - 1 +  \log \Big[\frac{H^2 k^2 \eta \eta^{\prime \prime}}{\mu^2} \Big] \Bigg)
\, ,
\end{multline}
where we made us of
\be
E_1 [2 i k(\eta - \eta^{\prime\prime})] = - \text{ci}(2 k |\eta - \eta^{\prime \prime}|) + i \, \text{sign}(\eta - \eta^{\prime}) \text{si}(2 k |\eta - \eta^{\prime \prime}|) - i \frac{\pi}{2}\text{sign}(\eta - \eta^{\prime})\, .
\ee
The expression for  $\mathcal{B}_{(1)}$ is unfortunately much lengthier which is why we give here only the super-Hubble limit
\begin{multline}
\mathcal{B}_{(1)} (\eta ,\eta^{\prime \prime}, k) 
\longrightarrow \frac{2}{3} \lambda \Bigg[ \log^2 (-2 k \eta )+
 \log^2 (- 2k \eta^{\prime \prime})
+
2\log (- 2 k \eta )\log(- 2 k \eta^{\prime \prime}) \\+ \frac{4}{3} \log \Big[ 4 k^2 \eta \eta^{\prime \prime} \Big]\Big( 3 \log \Big[ \frac{2 \mu }{H}\Big]+3 \gamma_E-4 \Big)  
\\
+ \frac{17}{4} - \frac{32}{3} \gamma_E + 4 \gamma^2_E + \frac{\pi^2}{3} + 2\big( 4 \gamma_E -5\big) \log \Big[\frac{2 \mu}{H}\Big]\Bigg] \, .
\end{multline}
We see that the above expressions are already symmetric and we will not need a homogeneous solution for symmetrizing them.
Finally, we turn to the integral that involves the Green's function
\be
 \mathcal{G} (\eta , \eta^{\prime \prime} , k)\equiv H^2 \int_{-\infty}^{\infty}d\tau \frac{G_{\text{ret}}(\eta, \tau,k)}{(\tau H)^4}\Big[ \mathcal{B}_{(0)}(\tau, \eta^{\prime \prime},k) + \mathcal{B}_{(0)}^{\text{log}}(\tau, \eta^{\prime \prime},k)\Big]+ F_{\text{hom}} (\eta ,\eta^{\prime \prime},k)\, .
\ee
We realize that the integral boundary at negative infinity will lead to logarithmic divergences which is why we add a homogeneous solution to cancel them
\be
 \mathcal{G} (\eta , \eta^{\prime \prime} , k) =  H^2 \int_{\eta^{\prime \prime}}^{\infty}d\tau \frac{G_{\text{ret}}(\eta, \tau,k)}{(\tau H)^4}\Big[ \mathcal{B}_{(0)}(\tau, \eta^{\prime \prime},k) + \mathcal{B}_{(0)}^{\text{log}}(\tau, \eta^{\prime \prime},k)\Big]+ \widetilde{F}_{\text{hom}} (\eta ,\eta^{\prime \prime},k) \label{greensFunctionTerm}
\ee
We computed \eqref{greensFunctionTerm} in terms of finite sums of exponentials, exponential integrals and generalized hyper geometric functions as well as an additional integral which contains similar functions as \eqref{jIntegral} but is more complicated. We also find that the consistency condition \eqref{consitencyReq} applies which is a highly non-trivial statement with regard to how the various terms contribute. However, since the result fills pages and includes a lot of partial integration, we decided to give only the super-Hubble limit in this paper. We note that the Green's function has the super-Hubble limit
\be
G_{\text{ret}}(\eta , \eta^{\prime} , k) \longrightarrow  \theta(\eta - \eta^{\prime}) \frac{H^2}{k^3} \Big[ - \frac{1}{3} k^3(\eta - \eta^{\prime})^3 \Big] \, ,
\ee 
such that the full integral in the super-Hubble limit reduces to a rather simple expression 
\begin{multline}
 \mathcal{G} (\eta , \eta^{\prime \prime} , k) \longrightarrow -\frac{1}{3} \int_{\eta^{\prime \prime}}^{\eta}d\tau \frac{ (\eta - \tau)^3}{\tau^4}\Bigg[
\frac{16}{3} \log^2 (-2 k \tau)
+\frac{4}{3} \Big(8 \log \Big[\frac{H}{2 k} \Big] - 9  \Big) \log (-2 k \tau)\\
 + \frac{4}{3} \Big(8 \log \Big[\frac{H}{2 k}\Big] +7 - 8 \gamma_E \Big) \log (-2 k \eta^{\prime \prime})\\
- \frac{70}{3} + 40 \gamma_E - 16 \gamma^2_E 
- \frac{4}{9} \pi^2 
+ \frac{64}{3} \big(\gamma_E - 2 \big)  \log \Big[\frac{H}{2 k}\Big]
 \Bigg]+ \widetilde{F}_{\text{hom}} (\eta ,\eta^{\prime \prime},k)\,.
\end{multline}
The last step is to symmetrize the result by means of a homogeneous solutions that should also include the tree-level solution for the Bunch-Davies vacuum,
\begin{multline}
\widetilde{F}_{\text{hom}} (\eta ,\eta^{\prime \prime},k) ={F}_{\varphi, dS}(\eta ,\eta^{\prime \prime},k) \\ + \lambda\Big[h_1(\eta^{\prime \prime})+i ( k \eta^{\prime \prime})^{-3} h_2(\eta^{\prime \prime})\Big](1 + i k \eta) e^{- i k \eta}+\lambda \Big[h_1(\eta^{\prime \prime})-i( k \eta^{\prime \prime})^{-3} h_2(\eta^{\prime \prime})\Big](1 - i k \eta) e^{ i k \eta}\, ,
\end{multline}
where $h_{1,2}$ are real functions that we determine perturbatively as
\begin{multline}
h_1(\eta^{\prime \prime}) \longrightarrow
- \frac{733}{486} + \frac{1}{81}\Bigg(18 \gamma_E(3+2 \gamma_E) + 7\pi^2  + 16(6 \gamma_E - 11) \log \Big[\frac{H}{2 k}\Big] + 216(\gamma_E-2) \log \Big[\frac{2 \mu}{H}\Big] \Bigg)\\
+\frac{2}{81} \log(-2 k \eta^{\prime \prime})\Bigg(355 - 12 \gamma_E(47+ 18 \gamma_E) + 6\pi^2  - 288( \gamma_E - 2) \log \Big[\frac{H}{2 k}\Big]  \Bigg)\\
+ \frac{4}{27} \log^2(-2 k \eta^{\prime \prime})\Bigg(1 +  12 \gamma_E -24 \log \Big[\frac{H}{2 k}\Big]  \Bigg) - \frac{16}{27}\log^3(-2 k \eta^{\prime \prime})\, ,
\end{multline}
\begin{multline}
h_2(\eta^{\prime \prime}) \longrightarrow
- \frac{353}{81} + \frac{2}{27}\Bigg( 18 \gamma_E(2 \gamma_E - 5) + \pi^2  - 8( 11-6 \gamma_E) \log \Big[\frac{H}{2 k}\Big] \Bigg)\\
- \frac{4}{27}\log(-2 k \eta^{\prime \prime})\Bigg(1 - 12 \gamma_E   + 24 \log \Big[\frac{H}{2 k}\Big]  \Bigg)
- \frac{8}{9} \log^2(-2 k \eta^{\prime \prime})\, .
\end{multline}
Adding up all contributions for the statistical two-point function
\begin{multline}
F_{\varphi} (\eta,\eta^{\prime \prime},k)  =  {F}_{\varphi, dS}(\eta ,\eta^{\prime \prime},k)  + \lambda \Bigg(   \frac{\square_k}{H^2} \mathcal{B}_{(2)}(\eta, \eta^{\prime \prime},k) + \mathcal{B}_{(1)}(\eta, \eta^{\prime \prime},k) +  \mathcal{G} (\eta , \eta^{\prime \prime} , k)\\+ \Big[h_1(\eta^{\prime \prime})+i ( k \eta^{\prime \prime})^{-3} h_2(\eta^{\prime \prime})\Big](1 + i k \eta) e^{- i k \eta}+ \Big[h_1(\eta^{\prime \prime})-i( k \eta^{\prime \prime})^{-3} h_2(\eta^{\prime \prime})\Big](1 - i k \eta) e^{ i k \eta}  \Bigg)\, ,
\end{multline}
yields expression \eqref{resDeltaF} in the super-Hubble limit.


\section{Conclusion and outlook \label{discus}}

In the literature on cosmological perturbations, their properties 
 are often specified solely in terms of equal time two-point function of the comoving curvature perturbation ${\cal R}$. This picture is correct if the fields are Gaussian distributed and if the decaying mode on super-Hubble scales makes the (canonical) momentum perturbation $\pi_{\cal R}$ small and/or  stochastically dependent on ${\cal R}$, such that no useful or additional information is contained in it. 
This rationale can be extended at the linear level to include isocurvature modes stemming from additional field perturbation in a multi field inflation scenario. On the other, one can discuss the self-interactions of the inflaton perturbation.
There is another possibility that we discuss in this paper, namely, that the momentum of the comoving curvature perturbation can become significant at the end of inflation via quantum interactions with a spectator field. We study this scenario for a model with a  cubic coupling at the level of perturbations in which the inflaton perturbation couples linearly to the spectator. Quantum gravitational interactions during inflation have been 
schematically addressed in \cite{Weinberg:2005vy, Weinberg:2006ac} with the conclusion that,
in the single field inflationary models,  
corrections to the curvature perturbation grow on super-Hubble scales
at most with powers of logarithms of the scale factor.
In this paper we consider a simple two-field model of inflation~(\ref{action for 2 fields})
 in which the inflaton field couples bi-quadratically 
to a light spectator scalar field. We investigate how the spectator field affects the curvature perturbation 
by performing an explicit one-loop calculation with renormalized self-masses in the 2PI formalism and  
we find that the correlator of the curvature perturbation~(\ref{resDeltaF})
 grows with powers of logarithms on super-Hubble scales. 

However, the momentum correlators~(\ref{Rpi correlator}--\ref{pipi correlator}) 
grow as powers of the scale factor, such that they are not necessarily suppressed at the end of inflation. 
We calculate the Gaussian, von Neumann entropy of the curvature perturbation~(\ref{entropyPhi}), 
and show that during inflation and on super-Hubble scales it grows as $ \sim 6\ln(a)$. This rapid
growth of the entropy indicates a rapid classicalization of the curvature perturbation on super-Hubble scales 
during inflation, and it is a consequence of the rapid growth 
($\propto a^6$, see Eq.~(\ref{gaussPert}))
of the Gaussian invariant of the state~(\ref{gaussian invariant}), which in turn 
can be attributed to the rapid growth of momentum correlators~(\ref{Rpi correlator}--\ref{pipi correlator}).
This then implies that the {\it momentum operator of the curvature perturbation~(\ref{momentum of curvature perturbation}) should be regarded as stochastically independent from the curvature perturbation.} 

 When this work was nearing completion, we became aware that the idea of obtaining decoherence from spectators has been addressed in \cite{Hollowood:2017bil}, based on the work in \cite{Boyanovsky:2015tba,Boyanovsky:2015jen}. Stricly speaking, the theory with a cubic interaction studied in \cite{Hollowood:2017bil,Boyanovsky:2015jen,Boyanovsky:2015tba} is unstable and not the same as the bi-quadratic theory we start from in equations \eqref {action for 2 fields} to \eqref{potential}, which is a stable theory for a positive coupling $g$.
However, since the two-loop diagram in figure~\ref{fig:loop2} is suppressed, the principle source of decoherence in our theory is incidentally a diagram that is topologically the same as the diagram used in \cite{Hollowood:2017bil,Boyanovsky:2015jen,Boyanovsky:2015tba}, provided one identifies our coupling $h = g \bar{\phi}$ with their coupling $\lambda$. Since $\bar{\phi} =  \bar{\phi}(t)$, this identity is never exact and at some level the theories do differ. 
 
 We also emphasize that our approach (based on the one loop evaluation of the inflaton two point function) differs significantly from 
the reduced density matrix approach used in~\cite{Hollowood:2017bil,Boyanovsky:2015jen,Boyanovsky:2015tba}. 
Furthermore, our results qualitatively differ in that we find the leading order growth of the inflaton two-point function 
correlator to be $\log^3 (- k\eta)$, which differs by one power from the result obtained 
in~\cite{Hollowood:2017bil,Boyanovsky:2015jen,Boyanovsky:2015tba}.
Moreover, our result differs by a sign. Namely, we get that the two point function increases at late times while the above mentioned references find a suppression.
Since both calculational frameworks differ significantly and bear a lot of complexity, we leave it as an important task 
for the future to explain how this difference comes about.

Our study shows that the effects of interactions are typically {\it large} at the end of inflation, which 
can be clearly seen from Eq.~(\ref{pipiInf}) and is illustrated in figures~\ref{fig:snap1}--\ref{fig:snap3}. 
On the other hand, if 
interactions switch off rapidly after inflation, quite generically by the end of radiation era 
the momentum fluctuations will decay 
such that their effects will be too small to leave any observable imprint in the CMB or LSS, which 
is corroborated by the estimate given in 
Eqs.~(\ref{momentum contribution to CMB}--\ref{momentum contribution to CMB:2}).
This conclusion holds however, only if the inflaton-spectator interactions 
are switched off rapidly enough after inflation,
such that the post-inflationary evolution of cosmological perturbations 
on super-Hubble scales can be well approximated by 
the corresponding free, linear evolution, according to which the large curvature momentum perturbation 
from the end of inflation decays swiftly during radiation. 
One way to hinder the decay of the momentum correlators is to keep the inflaton-spectator 
interactions active during early parts of the radiation era. This can be achieved, for example, by
delaying the post-inflationary decays of the inflaton and spectator fluctuations, and by demanding that 
both fields are light enough such that, for some time during radiation, 
they remain approximately massless,
 {\it i.e} $m_\phi, m_\chi \ll H(t)$, where $H(t)\simeq 1/(2t)$ 
is the Hubble rate in radiation era.
 We leave a detailed study of decoherence on super-Hubble scales during radiation for future work.
 
Broadly speaking, investigations of quantum loop corrections 
 to cosmological perturbations in an inflationary setting, 
a simple example of which is performed in this work, 
can be used to test consistency of various inflationary 
models and can be considered as complementary to effective field theory methods,
which can be very useful for studying the internal consistency of inflationary 
models such as Higgs inflation~\cite{Fumagalli:2017cdo,Postma:2017hbk}. 
 Furthermore, since quantum loop corrections from light matter fields may leave observable imprints 
in the CMB and large scale structure, one can use the signatures imprinted in the 
CMB and large scale structure by the momentum correlators 
of cosmological perturbations as a means to study inflationary interactions, thus opening a 
{\it novel observational window} to inflationary physics.

\paragraph{Acknowledgments.}
This work is part of the research programme of the Foundation for Fundamental Research on Matter (FOM), which is part of the Netherlands Organisation for Scientific Research (NWO). This work is in part supported by the D-ITP consortium, a program of the Netherlands Organization for Scientific Research (NWO) that is funded by the Dutch Ministry of Education, Culture and Science (OCW). This paper is based on 
master's theses of Angelos Tzetzias, Daan Maarsman and Margherita Gambino. The authors are grateful for their 
contributions to different aspects of this research project.


\section{Appendices}

\appendix
\section{Definitions and conventions \label{defAndConv}}

We make use of the d'Alembert operator in de Sitter space-time
where
\begin{align}
\frac{\square}{H^2}\equiv \frac{\overline{\Box}_{{dS}}}{H^2} = \eta^2 \Big[- \partial^2_{\eta} + \frac{D-2}{\eta} \partial_{\eta} + \delta^{ij} \partial_i \partial_j \Big] \, ,
\end{align}
where the constant parameter $H$ is the Hubble rate at the beginning of inflation and $\eta$ denotes conformal time. 
We use the following general notation for the Wightman functions, causal and statistical propagators in the cosmological context,
\begin{eqnarray}
i \Delta_{\varphi}^{\mp \pm}(\eta, \eta^{\prime},k) &=& F_{\varphi}(\eta,\eta^\prime,k) \pm \frac{i}{2}\Delta_{\varphi}^c(\eta,\eta^\prime,k)\, ,\\
F_{\varphi}(\eta,\eta^\prime,k) &=& \int d^3(x-x^{\prime})\, {\rm  e}^{- i \vec k\cdot (\vec x-\vec{x}^{\prime})} F_\phi(x;x^\prime) 
\,,\\
i \Delta_{\varphi}^c(\eta,\eta^\prime,k) &=& \int d^3(x-x^{\prime}) \,{\rm  e}^{- i \vec k\cdot (\vec x-\vec{x}^{\prime}) } i \Delta_\phi^c(x;x^\prime) \, ,
\label{defintion of scalar correlators}
\end{eqnarray}
with
\begin{equation}
F_{\varphi}(x;x^\prime) =\frac12{\rm Tr}\left[\hat \rho(\eta_0)\{\hat \varphi(x),\hat \varphi(x^\prime)\}\right] 
\,, \quad
i \Delta_{\varphi}^c(x;x^\prime) =  {\rm Tr}\left[\hat \rho(\eta_0)[\hat \varphi(x^\prime),\hat \varphi(x)]\right] 
\label{defintion of scalar correlators:2}
\end{equation}
where $\hat \rho_0\equiv \hat \rho(\eta_0)$ is the initial density matrix (defined at $\eta=\eta_0$).
Moreover, we define the correlators
\be
\Delta_{X Y }(x;x^\prime) \equiv \frac12{\rm Tr}\left[\hat \rho(\eta_0)\{\hat{ X} (x),\hat{Y}(x^\prime)\}\right]\, .
\ee
We make frequently use of the following functions,
\begin{eqnarray}
\text{si} (z)&=&-\int_{z}^{\infty}\frac{\sin(t)dt}{t}=
\int_{0}^{z}\frac{\sin(t)dt}{t}-\frac{\pi}{2}
=\text{Si}(z)-\frac{\pi}{2}
\, ,
\label{sine integral}\\
\text{ci}(z)&=&-\int_{z}^{\infty}\frac{\cos(t)dt}{t}
\, ,
\label{cos integral}\\
\text{E}_1 ( i x)&=& - \gamma_E - \log (i x) - \sum_{k=1}^{\infty} \frac{(-ix)^k}{k k!}  =  i \,  \text{si} ( x)-\text{ci} ( x)\,, \quad x>0\, ,\\
\text{Ein} (z) &=& \int_0^z \frac{1-e^{-t}}{t} dt = \sum_{n=1}^{\infty} \frac{(-1)^{n-1}z^n }{n! n}  =  E_1 (z) + \log (z) + \gamma_E \, ,\\
\pFq{3}{3}{1,1,1}{2,2,2}{z}  &=&  \frac{1}{z}\int_0^z \frac{\text{Ein(t)}}{t}  d t\, .
\end{eqnarray}
\section{Photon kinetic equation \label{appphoton}}

The starting point for our recapitulation is the Boltzmann equation for the temperature perturbation of the photon fluids which takes the following form in Fourier space (we follow closely standard literature,  
see {\it e.g.}~\cite{Straumann:2005mz} or \cite{Dodelson:2003ft}),
\begin{multline}
\partial_{\eta} \Theta (k, \mu) + i k \mu \big[ \Theta(k,\mu) + \Phi (k) \big] - \partial_{\eta} \Psi(k) \\
 = - (\partial_\eta \tau) \Big[\Theta_0 (k) - \Theta (k,\mu) + \mu v_b (k) - \frac{1}{2} \mathcal{P}_2(\mu) \Sigma (k) \Big] \, , \quad \mu = \frac{\vec{k} \cdot \vec{p}}{kp} \, .
\label{kinetic equation: gi}
\end{multline}
Here, $\Theta (k, \mu)$ is the time-dependent gauge invariant (integrated) photon 
temperature perturbation that is obtained from,
\begin{equation}
\Theta (k, \mu) = \sum_{l=0}^{\infty} (-i)^l \theta_l (\eta,k) \mathcal{P}_l (\mu) \propto \int dp \,p^3\delta f (\eta, \vec{p}, \vec{k})
\, ,
\end{equation}
where $\delta f$ is the perturbed, gauge-invariant photon distribution function as defined in \cite{Straumann:2005mz},  and $P_l$ are Legendre polynomials. The Bardeen potentials $ \Phi$ and $\Psi$ (as defined in \cite{Malik:2008im}) are related to temporal and spatial metric perturbations. The time-dependent variable $\tau(\eta)$  is the optical depth related to Thomson scattering  with $v_b$ the (longitudinal) baryon velocity perturbation and $\Sigma$ the anisotropic stress which depends on the polarization and quadrupole moment $\Theta_2$, both of which are usually neglected in a first approximation \cite{Straumann:2005mz} \cite{Dodelson:2003ft}. We note that the gravitational slip is given by
\be
\Psi-\Phi = \frac{a^2}{M_p^2} \Sigma\, ,
\ee
and we can identify the two potentials once  anisotropic stress is absent or neglected. Moreover, we have to consider the speed of sound $c_s$ of the photon-baryon fluid which is defined via 
\be
\delta P = c_s^2 \delta \rho + \delta P_{\text{nad}}\, ,
\ee
where $\delta P$, $\delta \rho$ and $\delta P_{\text{nad}}$ are the pressure, density and non-adiabatic pressure perturbations, respectively. 
The speed of sound in radiation domination is related to the background density of photons $\rho_{\gamma}^{(0)}$ and baryons $\rho_{b}^{(0)}$ via
\be
{ c_s^{2}}(\eta) = \frac{1}{3(1 + R(\eta))}\, , \quad R(\eta) \equiv  \frac{3}{4} \frac{\rho_{b}^{(0)}(\eta)}{\rho_{\gamma}^{(0)}(\eta)}\, .
\ee
Since the baryon density is much smaller than the photon density in the radiation dominated phase, we can take as another approximation $c_s^2 \approx 1/3$ during this time which also determines the baryon velocity to first order through the photon dipole moment as
\be
v_b =  -3 i \Theta_1  +  \mathcal{O} \big ( R \big)\, .
\ee
Putting it all together, one can derive a second-order differential equation for the effective temperature fluctuation $\Delta T = \Theta_0 + \Phi$  and the gravitational potentials \cite{Straumann:2005mz} \cite{Dodelson:2003ft},
\be
\Big[\frac{d^2}{d \eta^2} +\frac{R}{1+R} \mathcal{H} \frac{d}{d \eta} + k^2 c_s^2 \Big] \Delta T   =  k^2\Big[ c_s^2 - \frac{1 }{3} \Big]  \Phi  +\Big[\frac{d^2}{d \eta^2} +\frac{R}{1+R} \mathcal{H} \frac{d}{d \eta} \Big]\Big[ \Psi + \Phi \Big]   \, .
\ee
We see that if we neglect the damping term by $c_s^2 \approx 1/3$, we have a forced harmonic oscillator, whose homogeneous solutions  are determined by the  monopole density $\Theta_0 (\eta_{\text{cmb}})$ and its time-derivative $\Theta_0^{\prime} (\eta_{\text{cmb}})$ as well as the gravitational potential $\Psi(\eta_{\text{cmb}})$ and its time-derivative $\Psi^{\prime}(\eta_{\text{cmb}})$ at some time within the radiation dominated phase $\eta_{\text{cmb}}$ that is close to recombination $\eta_{\text{cmb}}  \approx 10^{-1} \eta_{\text{rec}}$,
\begin{multline}
\Delta T (\eta) \approx \big[ \Theta_0 +  \Phi\big] (\eta_{\text{cmb}})\cos\big[ k r_s(\eta) \big]+ \Big[\frac{\Theta_0^{\prime}+\Phi^{\prime}}{k c_s} \Big](\eta_{\text{cmb}})\sin \big[ k r_s (\eta) \big]\\ + \frac{\sqrt{3}}{k}\int_{\eta_{\text{cmb}}}^{\eta}\big[ \Phi^{\prime \prime}(\bar{\eta}) + \Psi^{\prime \prime}(\bar{\eta}) \big]\sin\big[k r_s(\eta) - k r_s(\bar{\eta})\big]  d \bar{\eta} \, , \label{effTemp1}
\end{multline}
where we defined the sound horizon, 
\be
r_s(\eta) = \int_{\eta_{\text{cmb}}}^{\eta} c_s (\bar{\eta}) d \bar{\eta}\, ,
\ee
and keep the time-dependence of the speed of sound only in the phases.
By making use of \eqref{kinetic equation: gi} in the super-Hubble limit, in which also the gravitational slip vanishes,  we obtain $\Theta_0^{\prime}(\eta_{\text{cmb}}) = \Psi^{\prime}(\eta_{\text{cmb}})= \Phi^{\prime}(\eta_{\text{cmb}})$ and the temporal integration turns out to yield $2 \Theta_0(\eta_{\text{cmb}}) =- \Psi(\eta_{\text{cmb}})=- \Phi(\eta_{\text{cmb}})$ \cite{Straumann:2005mz} \cite{Dodelson:2003ft}. 
We also recall that the gravitational potential $\Psi$ obeys in the absence of gravitational slip the following differential equation \cite{Malik:2008im},
\begin{equation}
\partial_{\eta}^2 \Psi +3(1+c_s^2)\mathcal{H} \partial_{\eta} \Psi + \Big[ 2 \partial_{\eta} \mathcal{H} +(1+ 3 c_s^{2} )\mathcal{H}^2 + c_s^2 k^2 \Big] \Psi =\frac{1}{2 M_p^2} \delta P_{\text{nad}}\, . \label{eqPsiNoSlip}
\end{equation}
As a first approximation to the inhomogeneous solution in \eqref{effTemp1}, we can solve \eqref{eqPsiNoSlip} for vanishing non-adiabatic pressure, $\delta P_{\rm nad}\rightarrow 0$,  
with $c_s^2 \approx 1/3$ during radiation unless it will appear as phase in conjunction with the momentum $k$. Thus, we write
\begin{equation}
\partial_{\eta}^2 \Psi +4 \mathcal{H} \partial_{\eta} \Psi + c_s^2 k^2  \Psi  \approx 0\, . \label{eqPsiNoSlip2}
\end{equation}
We see that the solution will stay constant on super-Hubble scales or decay otherwise during radiation and we thus neglect the integral in \eqref{effTemp1}.  
We then have
\be
\Delta {T} (k, \eta) \approx  \frac{1}{2} {\Psi}_k (\eta_{\text{cmb}})\cos\big[ k r_s(\eta) \big]+  2 \frac{{\Psi}^{\prime}_k (\eta_{\text{cmb}})}{k c_s (\eta_{\text{cmb}})} \sin \big[ k r_s (\eta) \big]  \, . \label{effTempPreApp}
\ee
Finally, in order to make contact with the era of inflation, we would like to relate equation \eqref{effTempPreApp} to the gauge-invariant curvature perturbation $\mathcal{R}$ in the case of vanishing (linear) non-adiabatic pressure  ($\delta P_{\text{nad}}=0$).
First we note, that the gauge-invariant curvature perturbation $\mathcal{R}$ may be expressed in terms of the gauge-invariant gravitational potential $\Psi$ {\it via} \cite{Malik:2008im},
\begin{equation}
{\cal R}  \equiv \Psi  + \frac{\Phi}{\epsilon} +  \frac{   \partial_{\eta} \Psi}{\mathcal{H} \epsilon} = \Psi  + \frac{\Psi}{\epsilon} +  \frac{   \partial_{\eta} \Psi}{\mathcal{H} \epsilon}\,, \quad \epsilon = 1 - \frac{\partial_{\eta}\mathcal{H}}{\mathcal{H}^2} \,, \label{defR}
\end{equation}
where we neglected the gravitational slip in the second equality, which is justified on super-Hubble scales.

The squared adiabatic sound speed may be expressed as
\begin{align*}
c_s^2 \equiv \frac{\partial_{\eta} \overline{P}}{\partial_{\eta} \overline{\rho}} =  -1 + \frac{2}{3} \epsilon - \frac{\partial_{\eta} \epsilon}{3 \mathcal{H}  \epsilon }\, ,
\end{align*}
where $P$ and $\rho$ are background pressure and energy density, respectively. 
Taking a derivative of \eqref{defR}, using \eqref{eqPsiNoSlip2}, we find
\begin{align*}
\partial_{\eta} \mathcal{R} =  -\frac{ c_s^2 k^2}{\epsilon \mathcal{H} } \Psi\, .
\end{align*}
Note, that the latter relation holds also in an inflationary context with $c_s^2$ set equal to $1$. We can solve for $\Psi$ and $\partial_{\eta} \Psi$ in terms of $\mathcal{R}$ and $\partial_{\eta} \mathcal{R}$ which yields
\begin{align}
\Psi &= -\frac{  \epsilon \mathcal{H} }{c_s^2 k^2 }\partial_{\eta} \mathcal{R} \equiv -    \frac{   \mathcal{H}  }{ 2 M_p^2k^2a^2 }\pi_{\mathcal{R}}\, ,
\nonumber\\
\partial_{\eta} \Psi &= \epsilon  \mathcal{H}   \mathcal{R} + \big(1 + \epsilon  \big)\frac{  \mathcal{H}^2  }{ 2 M_p^2k^2a^2 }\pi_{\mathcal{R}}\, ,
\label{general formula for psi R}
\end{align}
where we defined the canonical momentum $\pi_{\mathcal{R}}$ associated to $\mathcal{R}$ as in Appendix \ref{curvlin}.
We now would like to evolve the gravitational potential on super-Hubble scales from the end of inflation 
deep into the radiation era by using linear relations. 
Therefore, we make use of Weinberg's theorem~\cite{Weinberg:2003sw} according to which there are always two solutions for the gravitational potential on super-Hubble scales which take the following form
\be
\Psi_{\text{ad}} (\eta)   = -\Big[\frac{1}{2 M_p^2 k^2} \pi_{\mathcal{R}} (\eta_e)+\frac{a^2(\eta_e)}{\mathcal{H}(\eta_e)}  \mathcal{R}(\eta_e) \Big]\frac{\mathcal{H}(\eta)}{a^2(\eta)} + \mathcal{R}(\eta_e) \Big[ 1 - \frac{\mathcal{H}(\eta)}{a^2(\eta)} \int_{\eta_e}^{\eta} a^2(\bar{\eta}) d \bar{\eta}\Big]\, ,
\ee
where the time $\eta_e$ signals some time shortly before the end of inflation such that we still have that $\epsilon(\eta_e) \ll 1$.
In order to set initial conditions for the CMB spectrum at time $\eta_{\text{cmb}}  \approx 10^{-1} \eta_{\text{rec}}$ close to recombination, we stick to a simplified scenario in which we neglect small contributions due to the transition from inflation to radiation and keep only leading order terms in each variable,
\begin{eqnarray}
\Psi_{\text{ad}} (\eta_{\text{cmb}})   &\approx&  -\frac{1}{2 M_p^2k^2} \frac{H a^2(\eta_e)}{a^3(\eta_{\text{cmb}})} \pi_{\mathcal{R}} (\eta_e)+ \frac{2}{3} \mathcal{R}(\eta_e) \, ,
\\
\Psi_{\text{ad}} ^{\prime} (\eta_{\text{cmb}})   &\approx& 3 \frac{H^2 a^3(\eta_e)}{a^4(\eta_{\text{cmb}})}  \Big[ \frac{1}{2 M_p^2k^2} \pi_{\mathcal{R}}(\eta_e) + \frac{a^2(\eta_e)}{H} \mathcal{R}(\eta_e)\Big]
\,.
\end{eqnarray}
Inserting the above super-Hubble initial conditions into the approximate solution for the effective 
CMB temperature perturbation \eqref{effTempPreApp} and making the stochastic character of the involved operators manifest, we have
\begin{multline}
\Delta\hat{ T} \big(\eta,\vec{k} \big) \approx   \frac{1}{2} \Big[  \frac{2}{3} \hat{\mathcal{R}}(\eta_e,\vec{k} ) - \frac{a^3(\eta_e)}{a^3(\eta_{\text{cmb}})} \frac{ H }{2 M_p^2k^2a (\eta_e)}\hat{\pi}_{\mathcal{R}} (\eta_e,\vec{k})  \Big] \cos  \big[ k r_s(\eta) \big]\\+ \frac{ 6 H}{k c_s(\eta_{\text{cmb}})}  \frac{ a^4(\eta_e)}{a^4(\eta_{\text{cmb}})}  \Big[ \frac{ H}{2 M_p^2k^2}\hat{\pi}_{\mathcal{R}} (\eta_e,\vec{k}) +    {a(\eta_e)}\hat{\mathcal{R}}(\eta_e,\vec{k} ) \Big] \sin \big[ k r_s (\eta) \big]  \, . \label{finalCMBTemp}
\end{multline}
This relation is used in see Eqs.~(\ref{effTemp2}--\ref{momentum contribution to CMB:2}) 
of section~\ref{Growing curvature momentum from quantum interactions}
 to estimate the size of the photon temperature fluctuations 
induced by the enhanced inflationary momentum perturbation.

\section{Linear evolution of curvature perturbation  \label{curvlin}}

The gauge-invariant curvature perturbation can be defined in terms of the metric perturbation $\psi$ and the perturbation of the velocity potential $\varphi_v$~\cite{Mukhanov:1990me} 
(in single field inflationary models $\varphi_v$ reduces to the inflaton field perturbation) as,
\begin{equation}
{\cal R} = \psi+\frac{H}{\sqrt{\rho+P}}\varphi_v
\,,
\label{curvature perturbation: definition}
\end{equation}
where  $\psi= -{\rm Tr}[\delta g_{ij}]/(6a^2)$, 
 $\rho$ and $P$ are the background fluid density and pressure (in inflation $\rho+P\rightarrow (\dot\phi)^2$, where $\phi(t) \equiv \langle\hat\phi(x)\rangle$ is the inflaton expectation value).
 Let us solve for the curvature perturbation ${\cal R}$ in postinflationary epochs.
The quadratic (reparametrization invariant) action for $\mathcal{R}$ reads (see {\it e.g.}~\cite{Prokopec:2013zya,Mukhanov:1990me}),
\begin{equation}
S[\mathcal{R}]= (2 M_p^2) \int d^3xdt \overline N(t) a^3\epsilon
   \left(\frac1{2c_s^2} \dot{\mathcal{R}}^2 - \frac{1}{2a^2} (\partial_i \mathcal{R})^2  \right)
\,,
\label{gi quadratic action}
\end{equation}
where $\overline N=\overline N(t)$ is the lapse function of the ADM decomposition 
(defined on a global equal time hypersuface $\Sigma_t$) and
\begin{equation}
 \epsilon(t) = -\frac{\dot H}{H^2} 
\label{principal slow roll parameter}
\end{equation}
is the principal slow roll parameter and $\dot X(t) \equiv \overline{N}^{-1}\partial/\partial t$ 
is the time derivative invariant under time reparametrizations. In inflation $\epsilon\ll 1$, in radiation $\epsilon = 2$ and in matter era $\epsilon = 3/2$. 
From~(\ref{gi quadratic action}) one easily finds the canonical momentum of $\mathcal{R}$, 
\begin{equation}
 \pi_\mathcal{R}(t,\vec x) \equiv \frac{\delta S}{\delta \partial_t\mathcal{R}(t,\vec x)}
 =\frac{2 M_p^2 a^3\epsilon}{\overline N c_s^2}\partial_t\mathcal{R}
\end{equation}
and the Hamiltonian, 
\begin{equation}
H(t) = \int d^3 x\left(\frac{\overline Nc_s^2}{4  M_p^2a^3\epsilon}\pi_\mathcal{R}^2 +M_p^2 {\overline Na\epsilon}(\partial_i \mathcal{R})^2\right)
\,.
\label{Hamiltonian}
\end{equation}
From~(\ref{Hamiltonian}) one easily arrives at the Heisenberg equations, 
\begin{equation}
 \partial_t \hat{\mathcal{R}} = \frac{\overline Nc_s^2}{2 M_p^2 a^3\epsilon}\hat  \pi_\mathcal{R}
\,,\qquad 
 \partial_t \hat \pi_\mathcal{R} =2 M_p^2 {\overline Na\epsilon}\partial_i^2 \hat{\mathcal{R}}
\,,
\label{Heisenberg equations}
\end{equation}
where $\hat{\mathcal{R}}$ and $\hat \pi_\mathcal{R}$ are the canonical pair obeying, 
\begin{equation}
 \left[\hat{\mathcal{R}}(t,\vec x),\hat \pi_\mathcal{R}(t,\vec x^{\,\prime})\right] = i \hbar \delta^3(\vec x\!-\!\vec x^{\,\prime})
\,.
\label{canonical commutation}
\end{equation}

One can solve~(\ref{Heisenberg equations}) in space-times of constant $\epsilon$ as follows.  
Let us introduce a time, $ad\eta=\overline Ndt$ (notice that time $\eta$ reduces  to the usual conformal time in the gauge, $\overline N=a$),
and~(\ref{Heisenberg equations}) reduce to, 
\begin{equation}
\partial_\eta\left[a^2\partial_\eta \hat{\mathcal{R}}\right] - a^2c_s^2\nabla^2 \hat {\mathcal{R}}= 0 
\,,
\end{equation}
where we made use of $\dot\epsilon=0$ and $\dot c_s=0$. 
Since we are primarily interested in the spectra, it is convenient to perform the following mode decomposition, 
\begin{eqnarray}
\hat{\mathcal{R}}(\eta,\vec x) &=& \int \frac{d^3k}{(2\pi)^3}\left(e^{ i \vec k\cdot \vec x}\mathcal{R}(\eta,k)\hat a(\vec k)
                                           \!+\!e^{-i \vec k\cdot \vec x}\mathcal{R}^*(\eta,k)\hat a^+(\vec k)\right)
                                                 \equiv \int \frac{d^3k}{(2\pi)^3}e^{- i  \vec k\cdot \vec x} \hat{\mathcal{R}}(\eta,\vec  k\,)
\nonumber\\
\hat \pi_\mathcal{R}(\eta,\vec x)\! &=&\! \int \frac{d^3k}{(2\pi)^3}\!\left(e^{ i  \vec k\cdot \vec x}\pi_\mathcal{R}(\eta,k)\hat a(\vec k)
                                           \!+\!e^{ -i  \vec k\cdot \vec x}\pi^*_\mathcal{R}(\eta,k)\hat a^+(\vec k)\right)
                                              \equiv \int\! \frac{d^3k}{(2\pi)^3}e^{- i \vec k\cdot \vec x} \hat \pi_\mathcal{R}(\eta,\vec  k\,)
\quad\;\;
\label{mode decomposition}
\end{eqnarray}
where 
\begin{equation}
 \left[\hat a(\vec k),\hat{a}^+(\vec k^{\,\prime})\right] = (2\pi)^3\delta^3(\vec k-\vec k^{\,\prime})
\,,\quad
\mathcal{R}(\eta,k)\pi^*_\mathcal{R}(\eta,k)-\mathcal{R}^*(\eta,k)\pi_\mathcal{R}(\eta,k)=i 
\,.
\label{normalization of modes}
\end{equation}
The equation of motion for the modes $\mathcal{R}(\eta,k)$ then becomes,
\begin{equation}
\left[ \frac{d^2}{d\eta^2}+c_s^2k^2 -\left({\cal H}^2 + \partial_\eta {\cal H}\right)\right](a \mathcal{R}) = 0
\label{mode equation}
\end{equation}
where ${\cal H} = \partial_\eta \ln(a)=aH$ is conformal Hubble rate.
For inflation we have $c_s^2 =1$ and set to leading order in the slow-roll parameters $ a(\eta) = -{H}{\eta}^{-1}$ ($ H  \approx \text{const}$).
 Thus, the two fundamental solutions in inflation  are to leading order given by
\begin{equation} 
\frac{1}{\sqrt{2 \epsilon} M_P} \frac{H}{\sqrt{2 k^3 }} (1 \mp i k \eta )e^{\pm i k \eta}
\,,
\label{solutions inflation era}
\end{equation}
such that
\bea
\hat{\mathcal{R}}(\eta,\vec  k\,)  &=&  \frac{1}{\sqrt{2 \epsilon} M_P} \frac{H}{\sqrt{2 k^3 }} \Big[(1 + i k \eta )e^{- i k \eta} \hat{a}(-\vec k) +(1 - i k \eta )e^{ i k \eta} \hat{a}^+(\vec k)\Big]
\,, \\
  \hat{\pi}_{\mathcal{R}}(\eta,\vec  k\,)& =&  \frac{1}{\sqrt{2 \epsilon} M_P} \frac{H}{\sqrt{2 k^3 }} {2 M_p^2 a^2 \epsilon}  k^2 \eta \Big[ e^{ -i k \eta} \hat{a}(-\vec k)+e^{ i k \eta} \hat{a}^+(\vec k) \Big]\, . \label{solInflationRPi}
\eea
We now restrict the degrees of freedom of (lets say) a Gaussian state associated to $\mathcal{R}$ and $\pi_{\mathcal{R}}$ to the Bunch-Davies vacuum with $\hat{a}(\vec{k}) | 0\rangle=0$ by picking up only the commutator in any two-point function. However, the dynamics of single-field inflation on super-Hubble scales within the standard linear treatment reduces the effective degrees of freedom of the Gaussian state in any case to only one stochastic variable.
In other words, of we look on super-Hubble scales $|k \eta| \ll 1$, we find that $\hat{\mathcal{R}}$ and $\hat{\pi}_{\mathcal{R}}$ are effectively no more independent operators,
\bea
\hat{\mathcal{R}}(\eta,\vec  k\,)  &\longrightarrow& \frac{1}{\sqrt{2 \epsilon} M_P}  \frac{H}{\sqrt{2 k^3 }}\Big[\hat{a}(-\vec k) +\hat{a}^+(\vec k) + \mathcal{O}\big( k \eta \big)\Big]
\,, \\
  \hat{\pi}_{\mathcal{R}}(\eta,\vec  k\,)  &\longrightarrow& \frac{1}{\sqrt{2 \epsilon} M_P} \frac{H}{\sqrt{2 k^3 }} {2 M_p^2 a^2 \epsilon}  k^2 \eta \Big[\hat{a}(-\vec k) +\hat{a}^+(\vec k) + \mathcal{O}\big( k \eta \big)\Big]
  \nonumber\\ &&\qquad\qquad \qquad\qquad=-  \frac{2 \epsilon M_p^2 a^2}{ \mathcal{H}}  k^2  \Big[\hat{\mathcal{R}}(\eta,\vec  k\,)  + \mathcal{O}\big( k \eta \big)\Big]\, . \label{solPiSingleField}
\eea
 In conclusion,  we  would like to emphasize that $\hat{\mathcal{R}}$ and $\hat{\pi}_{\mathcal{R}}$ are for every $\vec{k}$ a priori independent and it is either the choice of state or the dynamics that could effectively cease this independence. 

\section{Wigner transform of logarithms \label{fourier}}

In this appendix we show that
\begin{multline}
\int d^{3}\big({x}-{x}^{\prime} \big) e^{-i \vec{k} \cdot(\vec{x}-\vec{x}^{\prime})}\Bigg[\frac{1}{2}\log^2 \Big(\frac{y}{4}\Big) + f\big(\eta, \eta^{\prime} \big)\log \Big(\frac{y}{4}\Big)  \Bigg] 
\\ =
-\frac{4 \pi^2}{k^3} \Bigg[2+ \big[1+ i k |\Delta \eta | \big] \Big(  \log \Big[\frac{a a^{\prime} H^2| \Delta \eta|}{2k} \Big]+ i \frac{ \pi}{2}- \gamma_E + f\big(\eta, \eta^{\prime} \big)\Big) \Bigg] e^{-i k |\Delta \eta|} \\
+\frac{4 \pi^2}{k^3} \big(1 - i k |\Delta \eta| \big)\Bigg[ \text{ci} \big[ 2 k| \Delta \eta|  \big]  -i \,  \text{si} \big[ 2 k |\Delta \eta|  \big]  \Bigg] e^{+i k |\Delta \eta|}\, , \label{fTlogsApp}
\end{multline}
where $\Delta \eta  = \eta - \eta^{\prime}$ and $f(\eta, \eta^{\prime})$ is some $k$-independent function. We need integrals of the following type,
\begin{align}
 \mathcal{I}_n (x)  & \equiv x^2 \int_0^{\infty} dz \, z \sin \big[x z  \big] { \log^n \Big( |1- z^2  |\Big) }
 \\
 &= x^2 \Bigg[ \frac{d^n}{db^n} \int_0^{\infty} dz \, z \sin \big[x z  \big]  |1- z^2  |^{b}  \Bigg]_{{b}=0} \, .
 \end{align}
 By using 
 \begin{multline}
  \int_0^{\infty} dz \, z \sin \big[x z  \big]  |1- z^2  |^b =  \frac{\sqrt{\pi}}{2} \Big( \frac{2}{x} \Big)^{b+ \frac{1}{2}} \Gamma \big[ b+1\big] \Big[ J_{b + \frac{3}{2}} \big(x \big) + Y_{-b - \frac{3}{2}} \big(x \big) \Big]\, , \\ x> 0\,, \quad  -1< b < 0\, ,
 \end{multline}
 with $J_{n}\, , Y_{m}$ being the Bessel functions of the first and second kind, we find  by analytically extending
 \begin{align}
\mathcal{I}_{1} \big(x \big)   =  - \pi \Big[\cos \big(x \big)+ x \sin \big(x \big)  \Big]\, ,
 \end{align}
 and 
  \begin{multline}
\mathcal{I}_{2} \big(x \big)   =   2 \pi \Bigg[-2\cos \big(x \big)+ \Big[   \cos \big(x \big)+ x \sin \big(x \big)  \Big]\Big[  \text{ci}\big(2 x\big)  + \gamma_E - \log \Big(\frac{2}{x}\Big) \Big] \\ +\Big[    \sin \big(x \big)  -x \cos \big(x \big) \Big]  \text{si}\big(2 x\big)\Bigg]\, ,
 \end{multline}
 where we used
 \begin{align}
  \text{ci}\big(x\big) = - \int_x^{\infty} \frac{\cos \big(y \big)}{y} dy\, , \quad   \text{si}\big(x\big) = - \int_x^{\infty} \frac{\sin \big(y \big)}{y} dy\, .
 \end{align}
 Remembering that
\begin{align}
\log \big( \Delta x_{++}^2 \big) = \log \Big( |\Delta \eta^2 - |\vec{x}-\vec{x}^{\prime}|^2 |\Big) + i \pi \theta \big( \Delta \eta^2 - |\vec{x}-\vec{x}^{\prime}|^2 \big) \, ,
\end{align}
we get
\begin{multline}
\label{log1}
\int d^{3}\big({x}-{x}^{\prime} \big) e^{-i \vec{k} \cdot(\vec{x}-\vec{x}^{\prime})} \log \Big( \frac{y}{4} \Big)  =  \frac{4 \pi }{k} \int_0^{\infty} dr \, r \sin \big(k r\big) \log \Big(  \frac{y}{4} \Big) \\
=   \frac{4 \pi }{k} \int_0^{\infty} dr \, r \sin \big(k r\big) \log \Big( |1- {r^2}{\Delta \eta^{-2} }|\Big)  
+  i \frac{4 \pi^2 }{k} \int_0^{\infty} dr \, r \sin \big(k r\big) \theta \big( \Delta \eta^2 - r^2  \big) + \text{hom.}
\\
=   \frac{4 \pi }{k^3} \big[ k\Delta \eta  \big]^2 \int_0^{\infty} dz \, z \sin \big[ k | \Delta \eta | z  \big] \log \Big( |z^2 - 1  |\Big)  
+  i \frac{4 \pi^2 }{k} \int_0^{| \Delta \eta | } dr \, r \sin \big(k r\big)  + \text{hom.}
\\ 
 = -\frac{4 \pi^2}{k^3} \big[1+i k |\Delta \eta|  \big]e^{- i k |\Delta \eta| }+ \text{hom.} \, .
\end{multline}
Moreover, we calculate
\begin{multline}
\int d^{3}\big({x}-{x}^{\prime} \big) e^{-i \vec{k} \cdot(\vec{x}-\vec{x}^{\prime})} \log^2 \Big( \frac{y}{4} \Big)  
=  \frac{4 \pi }{k} \int_0^{\infty} dr \, r \sin \big(k r\big) \log^2 \Big( |1- {r^2}{\Delta \eta^{-2} }  |\Big) \\
 + \frac{8 \pi }{k} \log \Big( \frac{a a^{\prime} H^2 \Delta \eta^2}{4} \Big) \int_0^{\infty} dr \, r \sin \big(k r\big) \log \Big( |1- {r^2}{\Delta \eta^{-2} } |\Big) \\ 
+  i \frac{8 \pi^2 }{k} \int_0^{\infty} dr \, r \sin \big(k r\big) \log \Big( |1- {r^2}{\Delta \eta^{-2} }   |\Big)\theta \big( \Delta \eta^2 - r^2  \big) 
-\frac{4 \pi^3 }{k}\int_0^{\infty} dr \, r \sin \big(k r\big)\theta \big( \Delta \eta^2 - r^2  \big) \\
+i \frac{8 \pi^2 }{k}\log \Big( \frac{a a^{\prime} H^2 \Delta \eta^2}{4} \Big)\int_0^{\infty} dr \, r \sin \big(k r\big)\theta \big( \Delta \eta^2 - r^2  \big)  + \text{hom.}\\
=  \frac{4 \pi }{k^3}\big[ k\Delta \eta  \big]^2 \int_0^{\infty} dz \, z \sin \big[k |\Delta \eta |\big] { \log^2 \Big( |1- z^2  |\Big) }
\\
 + \frac{8 \pi }{k^3} \log \Big( \frac{a a^{\prime} H^2 \Delta \eta^2}{4} \Big) \big[ k\Delta \eta \big]^2 \int_0^{\infty} dz \, z \sin \big[k |\Delta \eta |z \big] \log \Big( |1- z^2  |\Big) \\ 
+  i \frac{8 \pi^2 }{k^3} \big[ k\Delta \eta\big]^2 \int_0^{1} dz \, z \sin \big[k |\Delta \eta | z \big] \log \Big( |1- z^2  |\Big)  \\
-\frac{4 \pi^2 }{k} \Bigg[\pi - 2i\log \Big( \frac{a a^{\prime} H^2 \Delta \eta^2}{4} \Big)  \Bigg]\int_0^{|\Delta \eta |} dr \, r \sin \big(k r\big)  + \text{hom.}\\
=  \frac{8 \pi^2 }{k^3}  \Bigg[-2\cos \big(k |\Delta \eta | \big)+ \Big[   \cos \big(k |\Delta \eta | \big)+ k |\Delta \eta | \sin \big(k |\Delta \eta | \big)  \Big]\Big[  \text{ci}\Big(2 k |\Delta \eta |\Big)  + \gamma_E - \log \Bigg(\frac{2}{k |\Delta \eta | }\Bigg) \Big] \\ +\Big[    \sin \big(k |\Delta \eta | \big)  -k |\Delta \eta | \cos \big(k |\Delta \eta | \big) \Big]  \text{si}\Big( 2 k |\Delta \eta |\Big)\Bigg]
\\
- \frac{8 \pi^2 }{k^3} \log \Big( \frac{a a^{\prime} H^2 \Delta \eta^2}{4} \Big)  \Big[   \cos \big(k |\Delta \eta | \big)+ k |\Delta \eta | \sin \big(k |\Delta \eta | \big)  \Big] \\ 
+  i \frac{8 \pi^2 }{k^3}\Bigg[2\sin \big(k |\Delta \eta | \big)+ \Big[\sin \big(k |\Delta \eta | \big)  -    k |\Delta \eta |  \cos \big(k |\Delta \eta | \big)  \Big]\Big[  \text{ci}\Big(2k |\Delta \eta |\Big)  - \gamma_E + \log \Bigg(\frac{2}{k |\Delta \eta | }\Bigg) \Big] \\ -\Big[    \cos \big(k |\Delta \eta | \big)  +k |\Delta \eta | \sin \big(k |\Delta \eta | \big) \Big] \Big[ \text{si}\Big( 2k |\Delta \eta | \Big) + \frac{\pi}{2} \Big]\Bigg]\\
- \frac{4 \pi^2}{k^3} \Bigg[\pi - 2i\log \Big( \frac{a a^{\prime} H^2 \Delta \eta^2}{4} \Big)  \Bigg] \Big[\sin \big[ k|\Delta \eta|   \big]- k | \Delta \eta| \cos\big[ k|\Delta \eta|   \big] \Big]  + \text{hom.}\\
= -\frac{8 \pi^2}{k^3} \Bigg[2+ \big[1+ i k |\Delta \eta | \big] \Big(  \log \Big[\frac{a a^{\prime} H^2| \Delta \eta|}{2k} \Big]+ i \frac{ \pi}{2}- \gamma_E \Big) \Bigg] e^{-i k |\Delta \eta|} \\
+\frac{8 \pi^2}{k^3} \big(1 - i k |\Delta \eta| \big)\Bigg[ \text{ci} \big[ 2 k| \Delta \eta|  \big]  -i \,  \text{si} \big[ 2 k |\Delta \eta|  \big]  \Bigg] e^{+i k |\Delta \eta|}\, ,\label{log2}
\end{multline}
where $\gamma_E \approx 0.57721$ is Euler's constant. We combine the results \eqref{log1} and \eqref{log2} in order to get \eqref{fTlogsApp}.
\section{$I_R $ integrals \label{iRIntegrals}}
In this appendix we calculate the integrals \eqref{iR}, \eqref{iRHat} and \eqref{iRLog}. Let us start with 
\begin{multline}
I_{\widetilde{R}} (\eta ,\eta^{\prime \prime}, k) =- \lambda^{-1 }H^{-2}\text{sign} (\eta, \eta^{\prime \prime}) \int_{\mathcal{C}(\eta, \eta^{\prime \prime})} d \tau \big(\tau H \big)^{-4}\,  \text{Re} \,  \big[ \widetilde{M}^{++}_I +\widetilde{M}^{++}_{II} \big](\eta, \tau ,k) \Delta_{\phi,BD}^c (\tau, \eta^{\prime\prime},k )
\\
=  \text{Im} \Bigg\lbrace  \Big[(1+ i k \eta^{\prime \prime} ) e^{-i k \eta^{\prime\prime}} \Big] 
\Big[ e^{i k \eta} \widetilde{\mathcal{R}}_1 (\eta, \eta^{\prime \prime} ,k) +e^{-i k \eta} \widetilde{\mathcal{R}}_2 (\eta, \eta^{\prime \prime} ,k)\Big] \Bigg\rbrace\, ,
\end{multline}
where
\begin{multline}
\widetilde{I}_{R_1} (\eta, \eta^{\prime \prime} ,k) \equiv \frac{1}{k^3}\int_{\eta }^{\eta^{\prime \prime}} d \tau 
\Bigg[ 2   + \big[1- i k (\eta - \tau) \big]\Big(\log \Big[\frac{iH^2 (\eta - \tau)}{2 k} \Big]-i \pi\text{sign}(\eta -\eta^{\prime \prime})  \\- \gamma_E +E_1\big[2 i k (\eta - \tau) \big]\Big)\Bigg] \frac{1- i k \tau  }{\tau^4}  \, ,
\end{multline}
\begin{multline}
\widetilde{I}_{R_2} (\eta, \eta^{\prime \prime} ,k)\equiv \frac{1}{k^3} \int_{\eta }^{\eta^{\prime \prime}} d \tau 
\Bigg[ 2 +\big[1+ i k (\eta - \tau) \big]\Big(\log \Big[-i \frac{H^2 (\eta - \tau)}{2 k} \Big]+i \pi\text{sign}(\eta -\eta^{\prime \prime}) \\- \gamma_E +E_1\big[- 2i k (\eta - \tau) \big]\Big)
\Bigg]e^{  2 i k \tau}  \frac{1- i k \tau  }{\tau^4}   \, .
\end{multline}
We were able to drop the absolute value signs in the above expressions since the function that effectively appears has no branch cut as one might expect naively due to the logarithm and the exponential integral. The branch cut is exactly cancelled and we are dealing with the entire function $ \text{Ein}(z)$, the complementary exponential integral,
\be
\text{Ein} (z) = \int_0^z \frac{1-e^{-t}}{t} dt = \sum_{n=1}^{\infty} \frac{(-1)^{n-1}z^n }{n! n}  =  E_1 (z) + \log (z) + \gamma_E \, ,
\ee
converging for all finite values of $|z|$.
We define 
\be
\mathcal{J}(\eta, 
 \eta^{\prime \prime}, k) \equiv \int_{0}^{1} d x 
 E_1\big[-2 i k \big(x (\eta - \eta^{\prime \prime} ) + \eta^{\prime \prime} \big)  \big]\frac{1- e^{-2 i k (\eta - \eta^{\prime \prime})(x-1)}}{x-1} \, ,
\ee
and have the following result
\begin{multline}
{I}_{\widetilde{R}} (\eta ,\eta^{\prime \prime}, k) 
= \text{Im} \Bigg[ ( 1 + i k \eta^{\prime \prime})e^{- i k  \eta^{\prime \prime}}\Bigg\lbrace  
e^{i k \eta} \Bigg[- \frac{4}{3 k^3 \eta^3} + \frac{4i}{3 k^2 \eta^2}- \frac{4}{3 k \eta} + \frac{2}{3 k^3 (\eta^{\prime \prime})^3} 
-\frac{2i}{3 k^2 (\eta^{\prime \prime})^2}
+\frac{2}{3 k \eta^{\prime \prime}}
\\
+ \Bigg(\frac{i}{3} + \frac{1}{3 k^3 (\eta^{\prime \prime})^3} - \frac{i k \eta}{3 k^3 (\eta^{\prime \prime})^3}  
- \frac{k \eta}{2 k^2 (\eta^{\prime \prime})^2} + \frac{1}{ k \eta^{\prime \prime}}
\Bigg)\Bigg(E_1\big[2 i k (\eta - \eta^{\prime \prime})\big]+ \log \Big[\frac{i H^2(\eta -\eta^{\prime \prime})}{2 k} \Big] - \gamma_E - i \pi \,\text{sign}(\eta - \eta^{\prime \prime})   \Bigg)\\
-\Bigg(\frac{2}{3 k^3 \eta^3} - \frac{2 i}{3 k^2 \eta^2} + \frac{1}{k\eta} + \frac{2i}{3} \Bigg)\Bigg(\log \Big[\frac{H^2}{4 k^2} \Big] - 2 \gamma_E \Bigg) + \frac{i}{3}\log \Big[\frac{\eta }{\eta^{\prime \prime}}\Big] 
\Bigg] \\+ e^{- i k \eta} \Bigg[E_1 \big[ -2 i k \eta^{\prime \prime}\big] \Bigg(i + \frac{2}{3}\Big(i - k 
\eta \Big)\Big[
E_1\big[-2 i k (\eta - \eta^{\prime \prime})\big]
+\log \Bigg[\frac{H^2(\eta - \eta^{\prime \prime})}{2 i k} \Bigg] + i \pi \, \text{sign}(\eta - \eta^{\prime \prime}) - \gamma_E \Big] \Bigg)\\
-E_1 \big( -2 i k \eta\big) \Bigg(i + \frac{2}{3}\Big[i - k 
\eta \Big]\Big[\log \Bigg[\frac{H^2}{4 k^2} \Bigg] + i \pi \, \text{sign}(\eta - \eta^{\prime \prime}) - 2 \gamma_E \Big] \Bigg)
 \Bigg]
  \\- e^{-i k \eta + 2 i k \eta^{\prime \prime}} \Bigg[ 
  \frac{2}{3 k^3 (\eta^{\prime \prime})^3} -\frac{2 i}{3 k^2 (\eta^{\prime \prime})^2} + \frac{2}{3 k \eta^{\prime \prime}}
 \\ + \Bigg( \frac{1}{3 k^3 (\eta^{\prime \prime})^3} + \frac{i k\eta}{3 k^3 (\eta^{\prime \prime})^3}
  -\frac{2 i}{3 k^2(\eta^{\prime \prime})^2}
  +\frac{k \eta}{6 k^2 (\eta^{\prime \prime})^2}
  +\frac{1}{3 k \eta^{\prime \prime}}
  + \frac{i k \eta}{3 k \eta^{\prime \prime}}\Bigg)\\ \times
  \Bigg(E_1\big[-2 i k (\eta - \eta^{\prime \prime})\big] 
+\log \Bigg[\frac{H^2(\eta - \eta^{\prime \prime})}{2 i k} \Bigg] + i \pi \, \text{sign}(\eta - \eta^{\prime \prime}) - \gamma_E   
  \Bigg) \Bigg]\\
+  \frac{2}{3} i ( 1 + i k \eta)e^{- i k \eta} \mathcal{J}(\eta ,\eta^{\prime \prime},k)\Bigg\rbrace\Bigg] \, .
\end{multline}
On super-Hubble scales this simplifies to,
\be
{I}_{\widetilde{R}} (\eta ,\eta^{\prime \prime}, k) 
\longrightarrow -\frac{4}{3}\Big( \log \Big[\frac{\eta}{\eta^{\prime \prime}} \Big] - \frac{1}{3} + \frac{1}{3} \frac{(\eta^{\prime \prime})^3}{\eta^3}\Big)\Big( \gamma_E - 1 - \log\Big[\frac{H}{2 k } \Big]\Big)  \, .
\ee
The next integral we calculate is
\begin{multline}
I_{\widehat{R}}(\eta ,\eta^{\prime \prime},k) =- \lambda^{-1} H^{-2} \int_{\eta^{\prime \prime}}^{\eta} d \tau (\tau H)^{-4}\Big[ \text{Re} \,  \widehat{M} (\eta , \tau, k) \Big] \Delta_{\phi, BD}^c (\tau, \eta^{\prime \prime},k) \\
 = -\lambda^{-1}H^{-2}\frac{(2 \pi)^3H^2 h^2}{2 k^6(4\pi)^5} \text{Im} \Bigg\lbrace (1 + i k \eta^{\prime \prime})e^{-i k \eta^{\prime \prime} }  \int_{\eta}^{\eta^{\prime \prime}} d \tau \tau^{-4}\Big[ \big[1+i k (\eta - \tau) \big]e^{ i k (\tau - \eta)}\\+\big[1-i k (\eta - \tau) \big]e^{- i k (\tau - \eta)} \Big] \Big[(1- i k \tau)e^{i k  \tau)} \Big] \Bigg\rbrace\\
=- \text{Im}\Bigg[ (1 + i k \eta^{\prime \prime})e^{-i k \eta^{\prime \prime} }  \Bigg\lbrace - \frac{2}{3}\Big( k \eta - i \Big)e^{- i k \eta} E_1 \big[-2 i k \eta \big]\\
e^{i k \eta} \Bigg(\frac{i}{3} + \frac{2}{3 k^3 \eta^3} - \frac{2i}{3 k^2 \eta^2} + \frac{1}{ k \eta} -\frac{1}{3 k^3 (\eta^{\prime \prime})^3} + \frac{i k \eta}{3  k^3 (\eta^{\prime \prime})^3} + \frac{k \eta}{2 k^2 (\eta^{\prime \prime})^2} - \frac{1}{k (\eta^{\prime \prime})} \Bigg)  \\
+ e^{2 i k  \eta^{\prime \prime} - i k \eta} \Big(-\frac{1}{3 k^3 (\eta^{\prime \prime})^3} - \frac{i k \eta}{3  k^3 (\eta^{\prime \prime})^3}
+ \frac{2 i }{3 k^2 (\eta^{\prime \prime})^2} 
 - \frac{k \eta}{6 k^2 (\eta^{\prime \prime})^2} - \frac{1}{3 k \eta^{\prime \prime}}  - \frac{i k \eta}{3 k \eta^{\prime \prime}}\Big) \\+ \frac{2}{3}\Big( k \eta - i \Big)e^{- i k \eta} E_1 \big[-2 i k \eta^{\prime \prime} \big]  \Bigg\rbrace
\Bigg] \, ,
\end{multline}
where we again made use of the fact that the absolute value sign does not matter for the real part of $\widehat{M}$.
On super-Hubble scales we have here
\be
{I}_{\widehat{R}} (\eta ,\eta^{\prime \prime}, k) 
\longrightarrow -\frac{2}{3} \Big( \log \Big[\frac{\eta}{\eta^{\prime \prime}} \Big] - \frac{1}{3} + \frac{1}{3} \frac{(\eta^{\prime \prime})^3}{\eta^3}\Big)  \, .
\ee
We also have to calculate 
\begin{multline}
\widehat{\mathcal{R}}_{\text{log}}(\eta ,\eta^{\prime \prime},k) =-\lambda^{-1} H^{-2}\int_{\eta^{\prime \prime}}^{\eta} d \tau (\tau H)^{-4}\Big[ \text{Re} \,  \widehat{M} (\eta , \tau, k) \text{log} \Big[\frac{\eta \tau H^4}{4 \mu^2} \Big] \Big] \Delta_{\phi, BD}^c (\tau, \eta^{\prime \prime},k) \\
=  \lambda^{-1} H^{-2} \text{log} \Big[\frac{-\eta  H^4}{4 \mu^2}\Big]  \widehat{\mathcal{R}}(\eta ,\eta^{\prime \prime},k)\\
-\lambda^{-1} H^{-2} \frac{(2 \pi)^3H^2 h^2}{2 k^6(4\pi)^5} \text{Im} \frac{\partial}{\partial \nu}\Bigg\lbrace (1 + i k \eta^{\prime \prime})e^{-i k \eta^{\prime \prime} }  \int_{\eta}^{\eta^{\prime \prime}} d \tau (-\tau)^{-4+\nu} \Big[ \big[1+i k (\eta - \tau) \big]e^{ i k (\tau - \eta)} \\+ \big[1+i k (\tau - \eta) \big]e^{ i k (\eta - \tau)}\Big] \Big[(1- i k \tau)e^{i k  \tau)} \Big] \Bigg\rbrace_{\nu =0}\\
=  - \text{Im} \Bigg\lbrace (1 + i k \eta^{\prime \prime})e^{-i k \eta^{\prime \prime} }  \Bigg[ 
\Bigg(-\frac{14i}{9} + \frac{5 k \eta}{9} - \frac{2}{3}\Big[ i - k \eta \Big]  \text{log} \big[\frac{\eta \tau  H^4}{4 \mu^2}\big] \Bigg)e^{- i k \eta} E_1 \big[-2 i k \tau \big] \\
+\Bigg(-\frac{1}{9 k^3 \tau^3} + \frac{i k \eta}{9  k^3 \tau^3} + \frac{k \eta}{4 k^2 \tau^2} - \frac{1}{k \tau} + \Big(-\frac{1}{3 k^3 \tau^3} + \frac{i k \eta}{3  k^3 \tau^3} + \frac{k \eta}{2 k^2 \tau^2} - \frac{1}{k \tau}\Big)  \text{log} \big[\frac{\eta \tau  H^4}{4 \mu^2}\big]\Bigg) e^{i k \eta} \\
+\Bigg(-\frac{1}{9 k^3 \tau^3} - \frac{i k \eta}{9  k^3 \tau^3} + \frac{2i}{9 k^2 \tau^2} + \frac{k \eta}{36 k^2 \tau^2} - \frac{7}{9 k \tau} - \frac{5i k \eta}{18 k \tau} \\+ \Big(-\frac{1}{3 k^3 \tau^3} - \frac{i k \eta}{3  k^3 \tau^3} +\frac{2 i}{3 k^2 \tau^2}
- \frac{k \eta}{6 k^2 \tau^2} - \frac{1}{3k \tau} - \frac{i k \eta}{3 k \tau}\Big) \text{log} \big[\frac{\eta \tau  H^4}{4 \mu^2}\big]\Bigg) e^{i k (2 \tau - \eta)} \\
+\Bigg( \frac{\pi}{3} \Big( - {14} - {5 i k \eta} \Big)-\frac{\pi^2}{6} \Big( i - k \eta \Big) - 2 \gamma_E \Big(i - k \eta \Big) \text{log}\big[-2 i k \tau \big]
- {\gamma_E^2} \Big( i - k \eta \Big) \\
+ 4 k \tau \Big(1 + i k \eta  \Big) \; \pFq{3}{3}{1,1,1}{2,2,2}{2 i k \tau} 
- \Big( i - k \eta \Big)\text{log} \big[-2 i k \tau \big]^2
\Bigg) \frac{e^{-i k  \eta} }{3}
\Bigg]_{\tau = \eta}^{\tau = \eta^{\prime \prime}} \Bigg\rbrace
\, ,
\end{multline}
where $\pFq{3}{3}{1,1,1}{2,2,2}{2 i k \tau} $ is a generalized hypergeometric function.
On super-Hubble scales we get
\begin{multline}
{I}_{{R}_{log}} (\eta ,\eta^{\prime \prime}, k) 
\longrightarrow \frac{2}{3} \log (- 2 k \eta )\log(- 2 k \eta^{\prime \prime}) - \log^2 (- 2 k \eta ) + \frac{1}{3}\log^2 (-2  k \eta^{\prime \prime} )\\-\frac{4}{9} \log (- 2 k \eta )\frac{(\eta^{\prime\prime})^3}{\eta^3} +\frac{2}{9} \log (4 k^2 \eta\eta^{\prime\prime} )-\frac{4}{3}\log \Big[\frac{\eta}{\eta^{\prime \prime}} \Big]\log \Big[ \frac{H^2}{4 k \mu}\Big] \\
+\frac{4}{9}\Big(1- \frac{(\eta^{\prime\prime})^3}{\eta^3} \Big) \log \Big[ \frac{H^2}{4 k \mu}\Big]+ \frac{2}{27}\Big(1- \frac{(\eta^{\prime\prime})^3}{\eta^3} \Big)\, .
\end{multline}
\section{$I_M$ integrals \label{iMIntegrals}}
In this appendix we will calculate the integrals \eqref{iM}, \eqref{iMHat} and \eqref{iMLog}.
Let us start by calculating the following integral
\begin{multline}
I_{\widetilde{M}_{I}} (\eta,k )
  = \int^{ \infty}_{0}dx \frac{ 1-  i ( k \eta - x )}{ \big(k \eta -x \big)^{4}} \Big[2+ (1+ i x) \Big( \log \Big[\frac{H^2 x }{2 k^2} \Big] + i \frac{\pi}{2} - \gamma_E \Big)  \Big] e^{-2 i   x}\\
 =  \frac{1}{6 k^3}\Bigg[ 2+ (1 + i k \eta)\Big(  \log \Big[\frac{H^2  }{2 k^2} \Big] +  i \frac{\pi}{2} - \gamma_E \Big) \Bigg] \partial_{\eta}^3 \int^{ \infty}_{0}dx   \frac{e^{-2 i   x}}{ x-k \eta }\\ 
 +   \frac{i}{2 k^2}  \Bigg[2+ (2 + i k \eta)\Big(  \log \Big[\frac{H^2  }{2 k^2} \Big] +  i \frac{\pi}{2} - \gamma_E \Big) \Bigg] \partial_{\eta}^2 \int^{ \infty}_{0}dx   \frac{ e^{-2 i   x}}{x- k \eta }\\
  - \frac{1}{k}\Bigg[\log \Big[\frac{H^2  }{2 k^2} \Big] +  i \frac{\pi}{2} - \gamma_E \Bigg] \partial_{\eta} \int^{ \infty}_{0}dx   \frac{e^{-2 i   x}}{ x- k \eta } 
 +  \frac{1}{6 k^3 }\Big[1 + i k \eta \Big] \partial_{\eta}^3 \int_0^{\infty} dx \frac{\log \big[x \big] e^{-2 i   x}}{x- k \eta } \\  - \frac{1}{2 k^2}\Big[ k \eta-2i \Big] \partial_{\eta}^2 \int_0^{\infty} dx \frac{\log \big[x \big] e^{-2 i   x}}{x-k \eta }   -\frac{\partial_{\eta}}{k}\int_0^{\infty} dx \frac{\log \big[x \big] e^{-2 i   x}}{x- k \eta } \label{iMIStep}
 \, .
\end{multline}
In order to proceed, we will make use of the following identities,
\be
 \int^{ \infty}_{0}dx   \frac{e^{-2 i   x}}{ x- k \eta } = e^{-2 i k \eta } E_1 \big[-2 i k \eta \big]\, ,
\ee
\bea
\frac{\partial_{\eta}}{k}    \Big[  e^{-2 i k \eta } E_1 \big[-2 i k \eta \big] \Big] &= -2 i e^{-2 i k \eta } E_1 \big[-2 i k \eta \big] - \frac{1}{ k \eta} \, , \\
\frac{\partial^2_{\eta}}{k^2}    \Big[  e^{-2 i k \eta } E_1 \big[-2 i k \eta \big] \Big] &= -4 i e^{-2 i k \eta } E_1 \big[-2 i k \eta \big] + \frac{1+2 i k \eta}{ k^2 \eta^2} \, , \\
\frac{\partial_{\eta}^3}{k^3}    \Big[  e^{-2 i k \eta } E_1 \big[-2 i k \eta \big] \Big] &= 8 i e^{-2 i k \eta } E_1 \big[-2 i k \eta \big] - \frac{2+2 i k \eta - 4 k^2 \eta^2}{ k^3 \eta^3} \, , 
\eea
\begin{multline}
\int_0^{\infty} dx \frac{\log \big[x \big] e^{-2 i   x}}{x- k \eta }  = 
-\partial_{\nu} \int_0^{\infty} dx \frac{e^{-2 i   x}}{x^{\nu} (x- k \eta) } \, \Bigg|_{\nu= 0}    = 
-\partial_{\nu} \Bigg[ \frac{\Gamma(1-\nu)}{(- k \eta)^{\nu}}e^{-2 i k \eta} \Gamma\big[\nu, - 2 i k \eta\big]  \Bigg]_{\nu= 0}  \\
=-e^{-2 i k \eta}  \big(\gamma_E + i\frac{\pi}{2} + \log (2) \big) E_{1}\big[-2 i k \eta \big] \\  - e^{-2 i k \eta} \frac{1}{2}  \Bigg[ \gamma_E^2 + \frac{\pi^2}{6}+ 4 i k \eta \;  \pFq{3}{3}{1,1,1}{2,2,2}{2 i k \eta}  + 2 \gamma_E \log \big[-2 i k \eta \big] + \log^2 \big[-2 i k \eta \big]  \Bigg]
\, ,
\end{multline}
and 
\be
 \frac{d}{dx}  \Bigg[x   \; \pFq{3}{3}{1,1,1}{2,2,2}{x}   \Bigg] = - \frac{\gamma_{E}+ \log (-x) + E_1 (-x)}{x}\, .
\ee
Plugging these expressions into \eqref{iMIStep}, we find
\begin{multline}
I_{\widetilde{M}_{I}}(\eta,k ) 
=   \Bigg[ \frac{1}{3 k^3 \eta^3} + \frac{i}{3 k^2 \eta^2} + \frac{1}{2 k \eta}  +  \frac{2 i}{3} \Big(2 \gamma_E - 2 - \log \Big[ \frac{H^2}{4 k^2} \Big]\Big)+ \frac{2}{3} k \eta  \Big( \log \Big[ \frac{H^2}{4 k^2}\Big]  - 2 \gamma_E  \Big) \Bigg] e^{-2 i k \eta } E_1 \big[-2 i k \eta \big] 
  \\
+  \frac{4 \gamma_E - 1 - 2 \log \Big[ \frac{H^2}{4 k^2}\Big]}{6 k^3 \eta^3}
+ i \frac{3 - 4 \gamma_E + 2 \log \Big[ \frac{H^2}{4 k^2}\Big]}{6 k^2 \eta^2}
+ \frac{6 \gamma_E - 5 -3\log \Big[ \frac{H^2}{4 k^2}\Big] }{6 k \eta}
+\frac{i}{3} \Big(2 \gamma_E - \log \Big[ \frac{H^2}{4 k^2}\Big] \Big)
  \\
+ \frac{1}{3} (i - k \eta )e^{-2 i k \eta} \Bigg( \gamma_E^2 + \frac{\pi^2}{6} +4 i k \eta  \;  \pFq{3}{3}{1,1,1}{2,2,2}{2 i k \eta} + 2 \gamma_E \log \big[ -2 i k \eta \big] + \log^2 \big[ -2 i k \eta \big] \Bigg)
 \, .
\end{multline}
The next integral we calculate is
\begin{multline}
I_{\widetilde{M}_{II}} (\eta,k ) = \int^{ \infty}_{0} \frac{d x}{ \big(k \eta -x \big)^{4}}  (1- i x) E_1\big[i 2 x\big] \big[ 1-  i ( k \eta - x ) \big]
  \\
 = -\frac{1}{2 k^3 \eta^3} + \frac{i}{6 k^2 \eta^2}+ \frac{1}{6 k \eta}   - \frac{1}{3}  \Bigg[\frac{1}{k^3 \eta^3}+\frac{i}{ k^2 \eta^2}  + \frac{3}{2 k \eta} - i  \Bigg]{e^{-2 i k \eta }}E_1 \big[-2 i k \eta \big] \, ,
\end{multline}
where we used the indefinite integrals
\begin{align}
\int \frac{d x}{x^2} E_1 \big[ a x + b \big] = \frac{1}{b} \Bigg[a e^{-b} E_1 \big[a x \big] - \frac{1}{x} (ax+b) E_1 \big[a x + b \big] \Bigg]\, ,
\end{align}
\be
\int \frac{d x}{x^3} E_1 \big[ a x + b \big] = 
\frac{a e^{-a x -b}}{ 2 b x} - \frac{a^2 (1+b) e^{-b} E_1 \big[ax \big]}{2 b^2} + \Big(\frac{a^2}{2 b^2} - \frac{1}{2 x^2}\Big) E_1 \big[a x + b \big]\, ,
\ee
\begin{multline}
\int \frac{d x}{x^4} E_1 \big[ a x + b \big] = 
\frac{1}{3} \frac{a^3}{b^3} e^{-ax-b}\Big[- \frac{b}{ax}+ \frac{1}{2} \big(1-ax\big) \frac{b^2}{a^2 x^2} \Big] \\+ \frac{1}{3} \frac{a^3}{b^3} \Big[1+ b +\frac{b^2}{2} \Big] e^{-b} E_1 \big[ax\big]
- \frac{1}{3} \Big[ \frac{1}{x^3} + \frac{a^3}{b^3} \Big] E_1\big[ax+b\big]\,.
\end{multline}
Adding up the last two major integrals we find the integral \eqref{iM},
\begin{multline}
I_{\widetilde{M}}(\eta,k ) = -\frac{e^{i k \eta}}{2 \lambda H^4}\int^{ \infty}_{0} d x
 \Big[  \widetilde{M}^{++}_{I} \Big(\frac{x}{k},k \Big) + \widetilde{M}^{++}_{II} \Big(\frac{x}{k},k \Big) \Big]\Big[ \frac{1}{( k \eta - x  )^{4}}-   \frac{ i }{ (k \eta -x)^3} \Big] e^{ - i  x  } \\
= e^{i k \eta}\Big[I_{\widetilde{M}_{I}} + I_{\widetilde{M}_{II}}\Big] (\eta,k )
\\=  \frac{2}{3}\Bigg[   i \Big(2 \gamma_E - \frac{3}{2} - \log \Big[ \frac{H^2}{4 k^2} \Big]\Big)+  k \eta  \Big( \log \Big[ \frac{H^2}{4 k^2}\Big]  - 2 \gamma_E  \Big) \Bigg] e^{- i k \eta } E_1 \big[-2 i k \eta \big] 
  \\
+ \frac{1}{3}e^{- i k \eta}\Bigg[ \frac{2 \gamma_E - 2 - \log \Big[ \frac{H^2}{4 k^2}\Big]}{ k^3 \eta^3}
+ i \frac{2 - 2 \gamma_E +  \log \Big[ \frac{H^2}{4 k^2}\Big]}{ k^2 \eta^2}
+ \frac{6 \gamma_E - 4 - 3\log \Big[ \frac{H^2}{4 k^2}\Big] }{2 k \eta}
+i \Big(2 \gamma_E - \log \Big[ \frac{H^2}{4 k^2}\Big] \Big)\Bigg]
  \\
+  \frac{1}{3} (i - k \eta )e^{- i k \eta} \Bigg( \gamma_E^2 + \frac{\pi^2}{6} +4 i k \eta  \;  \pFq{3}{3}{1,1,1}{2,2,2}{2 i k \eta} + 2 \gamma_E \log \big[ -2 i k \eta \big] + \log^2 \big[ -2 i k \eta \big] \Bigg)    \, .
\end{multline}
The next integral we calculate is \eqref{iMHat} and we have
\begin{multline}
I_{\widehat{M}}(\eta,k ) =-\frac{e^{i k \eta}}{2 \lambda H^4} \int^{ \infty}_{0}dx \widehat{M}^{++} \Big(\frac{x}{k},k \Big)\frac{ 1-  i ( k \eta - x )}{ \big(k \eta -x \big)^{4}} e^{ - i  x  } =-{e^{i k \eta}} \int^{ \infty}_{0}dx \frac{ 1-  i ( k \eta - x )}{ \big(k \eta -x \big)^{4}} (1+ i x) e^{-2 i   x}  \\
   = -\frac{{e^{i k \eta}} }{6 k^3} \big[ 1+ i k \eta  \big] \partial_{\eta}^3 \int^{ \infty}_{0}dx \frac{ e^{-2 i   x}}{ x- k \eta}    
  -{e^{i k \eta}}  \frac{i}{2 k^2} \big[ 2+ i k \eta  \big] \partial_{\eta}^2\int^{ \infty}_{0}dx \frac{ e^{-2 i   x}}{ x- k \eta}   
  +{e^{i k \eta}}  \frac{1}{k} \partial_{\eta} \int^{ \infty}_{0}dx  \frac{ e^{-2 i   x}}{ x- k \eta}   \\
    =- \frac{{e^{i k \eta}} }{6 k^3} \big[ 1+ i k \eta  \big] \partial_{\eta}^3  \Big(e^{-2 i k \eta } E_1 \big[-2 i k \eta \big] \Big) 
  -{e^{i k \eta}} \frac{i}{2 k^2} \big[ 2+ i k \eta  \big] \partial_{\eta}^2 \Big(e^{-2 i k \eta } E_1 \big[-2 i k \eta \big] \Big) 
   +{e^{i k \eta}}  \frac{1}{k} \partial_{\eta}  \Big(e^{-2 i k \eta } E_1 \big[-2 i k \eta \big] \Big) \\
   =  {e^{i k \eta}}\Bigg(\frac{1}{3 k^3 \eta^3} - \frac{i}{3 k^2 \eta^2}+ \frac{1}{2 k \eta}-\frac{i}{3}\Bigg) + \frac{2}{3}\Big(i  - k \eta  \Big)e^{- i k \eta}E_1\big[- 2 i k \eta \big]\, .
\end{multline}

The last integral we calculate in this appendix is \eqref{iMLog}, where we use similar techniques as above,
\begin{multline}
 I_{{M}_{\text{log}}}(\eta, k)= - \frac{e^{i k \eta}}{2 \lambda H^4} \int^{ \infty}_{0}dx \log \Big[\frac{ \eta (k \eta - x) H^4}{ 4  k  \mu^2} \Big]   \widehat{M}^{++} \Big(\frac{x}{k},k \Big)  \frac{ 1-  i ( k \eta - x )}{ \big(k \eta -x \big)^{4}}  e^{ - i  x  }  \\ 
    = e^{i k \eta}\Bigg\lbrace
 \Bigg[ \frac{1}{3 k^3 \eta^3} - \frac{i}{3 k^2 \eta^2}+ \frac{1}{2 k \eta}+\frac{i}{3} + \frac{2}{3}\Big(i  - k \eta  \Big)e^{-2 i k \eta}E_1\big[-2 i k \eta \big] \Bigg]\log \Big[\frac{ H^4 \eta^2}{ 4   \mu^2} \Big] \\
  + \frac{1}{9 k^3 \eta^3}- i \frac{1}{9 k^2 \eta^2}+ \frac{3}{4 k \eta} +  i \frac{5}{18} + \frac{e^{-2 i k \eta}}{9}\big( 14 i - 5 k \eta \big) E_1\big[ -2 i k \eta \big]  \\
  + \frac{e^{-2 i k \eta}}{3}(i- k \eta) \Bigg( \gamma_E^2 + \frac{\pi^2}{6} +4 i k \eta  \;  \pFq{3}{3}{1,1,1}{2,2,2}{2 i k \eta}  + 2 \gamma_E \log \big[ -2 i k \eta \big]   + \log^2 \big[ -2 i k \eta \big]  \Bigg) \Bigg\rbrace\, .
\end{multline}
We are also interested for super-Hubble limit of the in the integrals we calculated in this appendix. However, let us multiply them with $(1+  i k \eta^{\prime \prime})e^{- i k \eta^{\prime \prime}}$ before, since these are the expressions that enter the calculation via \eqref{splitInfMIntegrals}.
Thus, on super-Hubble scales we have 
\begin{multline}
\text{Im} \Big[ (1+  i k \eta^{\prime \prime})e^{- i k \eta^{\prime \prime}}{I}_{\widetilde{M}} (\eta , k) \Big]
\longrightarrow  - \frac{1}{3}\log^2 (- 2 k \eta) + 
\Big(\frac{2}{3} \gamma_E -1 -\frac{2}{3} \log \Big[ \frac{H^2}{4 k^2} \Big] \Big)\log (- 2 k \eta)  \\
+\frac{2}{9}\frac{(\eta^{\prime \prime})^3}{\eta^3} \Big(\gamma_E - 1 - \log \Big[ \frac{H}{2 k}\Big]\Big)
\\
+ \frac{8}{9} - \frac{26}{9} \gamma_E + \gamma_E^2 + \frac{\pi^2}{36}+
\frac{1}{9} \Big( 17-12 \gamma_E\Big)\log \Big[ \frac{H}{2 k}\Big]
\, ,
\end{multline}
\begin{multline}
\text{Im} \Big[ (1+  i k \eta^{\prime \prime})e^{- i k \eta^{\prime \prime}}{I}_{\widehat{M}} (\eta , k) \Big] \longrightarrow \frac{2}{3}\log (- 2 k \eta)  + \frac{1}{18}\Big(12 \gamma_E- 17 + 2\frac{(\eta^{\prime \prime})^3}{\eta^3}  \Big) \, ,
\end{multline}
\begin{multline}
\text{Im} \Big[ (1+  i k \eta^{\prime \prime})e^{- i k \eta^{\prime \prime}}{I}_{{M}_{log}} (\eta , k) \Big]
\longrightarrow  \log^2 (- 2k \eta)
 + \frac{1}{3}\log(- 2 k \eta)\Big( 4 \log \Big[\frac{H^2}{4 k \mu} \Big]  + 2 \gamma_E -1 + \frac{2}{3} \frac{(\eta^{\prime \prime})^3}{\eta^3} \Big) \\
+\frac{1}{27}\frac{(\eta^{\prime \prime})^3}{\eta^3} \Big( 1 +6  \log \Big[\frac{H^2}{4 k \mu} \Big]\Big)
\\
- \frac{115}{108} - \frac{14}{9} \gamma_E - \frac{1}{3} \gamma_E^2 + \frac{\pi^2}{36}-
\frac{1}{9} \Big( 17-12 \gamma_E\Big)\log \Big[\frac{H^2}{4 k \mu} \Big]   \, .
\end{multline}

\bibliographystyle{JHEP}
\bibliography{Biblio}

\end{document}